\documentclass[journal]{IEEEtran}
\IEEEoverridecommandlockouts

\usepackage{amsmath}
\usepackage{amsfonts}
\usepackage{amssymb}
\usepackage{graphicx}
\usepackage{epsfig}
\usepackage[scriptsize,center]{caption}
\usepackage{subfig}
\usepackage{float}
\usepackage{epstopdf}
\usepackage{array}
\usepackage{multirow}
\usepackage{setspace}
\usepackage{color}
\usepackage[section]{placeins}
\usepackage{url}
\usepackage{algorithm}
\usepackage{algpseudocode}
\usepackage{multicol}
\usepackage{lipsum}
\usepackage{cite}
\usepackage{soul}
\usepackage{comment}
\usepackage[all]{nowidow}
\usepackage{tikzpagenodes}

\usepackage[nopostdot, style=super, nonumberlist, toc, nogroupskip]{glossaries}

\makeglossaries
\newacronym{scs}{SCS}{Subcarrier Spacing}
\newacronym{cp}{CP}{Cyclic Prefix}
\newacronym{bwp}{BWP}{Bandwidth Part}
\newacronym{prb}{PRB}{Physical Resource Block}
\newacronym{rb}{RB}{Resource Block}
\newacronym{mcs}{MCS}{Modulation and Coding Scheme}
\newacronym{tbs}{TBS}{Transport Block Size}
\newacronym{los}{LoS}{Line-of-Sight}
\newacronym{ofdm}{OFDM}{Orthogonal Frequency Division Multiplexing}
\newacronym{fdm}{FDM}{Frequency Division Multiplexing}
\newacronym{cpu}{CPU}{Central Processing Unit}
\newacronym{tcp}{TCP}{Transmission Control Protocol}
\newacronym{cca}{CCA}{Clear Channel Assessment}
\newacronym{cws}{CWS}{Congestion Window Size}
\newacronym{lbt}{LBT}{Listen-Before-Talk}
\newacronym{lbr}{LBR}{Listen-Before-Receive}
\newacronym{omnilbt}{omniLBT}{Omnidirectional LBT}
\newacronym{dirlbt}{dirLBT}{Directional LBT}
\newacronym{pairlbt}{pairLBT}{Paired LBT}
\newacronym{lbtswitch}{LBTswitch}{LBT switching}
\newacronym{lat}{LAT}{Listen-After-Talk}
\newacronym{ed}{ED}{Energy Detection}
\newacronym{pd}{PD}{Preamble Detection}
\newacronym{mcot}{MCOT}{Maximum Channel Occupancy Time}
\newacronym{cot}{COT}{Channel Occupancy Time}
\newacronym{eirp}{EIRP}{Equivalent Isotropically Radiated Power}
\newacronym{psd}{PSD}{Power Spectral Density}
\newacronym{ocb}{OCB}{Occupied Channel Bandwidth}
\newacronym{ncb}{NCB}{Nominal Channel Bandwidth}
\newacronym{fr}{FR}{Frequency Reuse}
\newacronym{dfs}{DFS}{Dynamic Frequency Selection}
\newacronym{nav}{NAV}{Network Allocation Vector}
 \newacronym{drs}{DRS}{Discovery Reference Signals}
 
\newacronym{bsr}{BSR}{Buffer Status Report}
\newacronym{qos}{QoS}{Quality of Service}
\newacronym{aqm}{AQM}{Active Queue Management}
\newacronym{ue}{UE}{User Equipment}
\newacronym{epc}{EPC}{Evolved Packet Core}
\newacronym{cn}{CN}{Core Network}
\newacronym{gnb}{gNB}{next-Generation Node B}
\newacronym{trp}{TRP}{Transmission Reception Point}
\newacronym{ran}{RAN}{Radio Access Network}
\newacronym{rat}{RAT}{Radio Access Technology}
\newacronym{3gpp}{3GPP}{3rd Generation Partnership Project}
\newacronym{5g}{5G}{5th Generation}
\newacronym{dl}{DL}{DownLink}
\newacronym{ul}{UL}{UpLink}
\newacronym{tti}{TTI}{Transmission Time Interval}
\newacronym{sr}{SR}{Scheduling Request}
\newacronym{snr}{SNR}{Signal-to-Noise Ratio}
\newacronym{sinr}{SINR}{Signal to Interference-plus-Noise Ratio}

\newacronym{ieee}{IEEE}{Institute of Electrical and Electronics Engineers}
\newacronym{etsi}{ETSI}{European Telecommunications Standards Institute}
\newacronym{lteu}{LTE-U}{LTE Unlicensed}
\newacronym{e2e}{E2E}{End-To-End}
\newacronym{embb}{eMBB}{enhanced Mobile BroadBand}
\newacronym{urllc}{URLLC}{Ultra-Reliable and Low-Latency Communications}
\newacronym{mmtc}{mMTC}{massive Machine Type Communications}
\newacronym{ev2x}{eV2X}{enhanced Vehicle to anything communications}
\newacronym{ca}{CA}{Carrier Aggregation}
\newacronym{dc}{DC}{Dual Connectivity}
\newacronym{sa}{SA}{Standalone}
\newacronym{sps}{SPS}{Semi-Persistent Scheduling}
\newacronym{comp}{CoMP}{Coordinated Multi-Point}
\newacronym{eicic}{eICIC}{enhanced Inter-Cell Interference Coordination}

\newacronym{mmwave}{mmWave}{millimeter-wave}
\newacronym{laa}{LAA}{Licensed-Assisted Access}
\newacronym{elaa}{eLAA}{enhanced LAA}
\newacronym{phy}{PHY}{Physical}
\newacronym{mac}{MAC}{Medium Access Control}
\newacronym{rlc}{RLC}{Radio Link Control}
\newacronym{rrc}{RRC}{Radio Resource Control}
\newacronym{lte}{LTE}{Long Term Evolution}
\newacronym{nr}{NR}{New Radio}
\newacronym{nru}{NR-U}{NR-based access to Unlicensed spectrum}
\newacronym{wlans}{WLANs}{Wireless Local Area Networks}
\newacronym{wpans}{WPANs}{Wireless Personal Area Networks}
\newacronym{wifi}{Wi-Fi}{Wireless Fidelity}
\newacronym{wigig}{WiGig}{Wireless Gigabit}
\newacronym{wi}{WI}{Work Item}
\newacronym{si}{SI}{Study Item}
\newacronym{harq}{HARQ}{Hybrid Automatic Repeat and reQuest}
\newacronym{harq-ack}{HARQ-ACK}{HARQ Acknowledgement}
\newacronym{ack}{ACK}{Positive Acknowledgement}
\newacronym{nack}{NACK}{Negative Acknowledgement}
\newacronym{tx}{Tx}{transmit}
\newacronym{tdma}{TDMA}{Time Division Multiple Access}
\newacronym{csma}{CSMA}{Carrier Sense Multiple Access}
\newacronym{csma-ca}{CSMA/CA}{Carrier Sense Multiple Access with Collision Avoidance}
\newacronym{tdd}{TDD}{Time Division Duplexing}
\newacronym{pdu}{PDU}{Packet Data Unit}
\newacronym{sb}{SB}{SubBand}
\newacronym{ch}{CH}{Channel}
\newacronym{lwa}{LWA}{LTE-WLAN Aggregation}
\newacronym{lwip}{LWIP}{LTE-WLAN Radio Level Integration with IPsec Tunnel}

\newacronym{mimo}{MIMO}{Multiple-Input Multiple-Output}
\newacronym{mu-mimo}{MU-MIMO}{Multi-User MIMO}
\newacronym{ofdma}{OFDMA}{Orthogonal Frequency-Division Multiple Access}
\newacronym{rbg}{RBG}{Resource Block Group}
\newacronym{dci}{DCI}{Downlink Control Information}
\newacronym{uci}{UCI}{Uplink Control Information}
\newacronym{ipat}{IPAT}{Inter-Packet Arrival Time}
\newacronym{pdsch}{PDSCH}{Physical Downlink Shared Channel}
\newacronym{pusch}{PUSCH}{Physical Uplink Shared Channel}
\newacronym{pucch}{PUCCH}{Physical Uplink Control Channel}
\newacronym{pdcch}{PDCCH}{Physical Downlink Control Channel}
\newacronym{rach}{RACH}{Random Access Channel}
\newacronym{prach}{PRACH}{Physical Random Access Channel}
\newacronym{ssb}{SSB}{Synchronization Signal Block}
\newacronym{ss}{SS}{Synchronization Signal}
\newacronym{pbch}{PBCH}{Physical Broadcast Channel}
\newacronym{pss}{PSS}{Primary Synchronization Signal}
\newacronym{sss}{SSS}{Secondary Synchronization Signal}

\newacronym{ap}{AP}{Access Point}
\newacronym{sta}{STA}{Station}
\newacronym{aul}{AUL}{Autonomous UpLink}
\newacronym{gul}{GUL}{Grant-less UpLink}

\newacronym{csat}{CSAT}{Carrier Sense Adaptive Transmission}

\title{New Radio Beam-based Access to Unlicensed Spectrum: Design Challenges and Solutions}
\author{\IEEEauthorblockN{ 
		Sandra Lagen$^+$, Lorenza Giupponi$^+$, Sanjay Goyal$^*$, Natale Patriciello$^+$, Biljana Bojovic$^+$, \\ Alpaslan Demir$^*$, Mihaela Beluri$^*$ \\
        }
	\IEEEauthorblockA{($^+$) Centre Tecnol\`ogic de Telecomunicacions de Catalunya (CTTC/CERCA), Barcelona, Spain \\ ($^*$) InterDigital Communications, Inc., Melville, New York, USA \\ 
    corresponding e-mail: sandra.lagen@cttc.es 
    }
}

\begin{document}
\maketitle
\begin{tikzpicture}[remember picture,overlay]
\footnotesize
\node[align=center,color=red] at ([yshift=3em]current page text area.north) {This is the author's version of an article that has been accepted for publication in IEEE Commun. Surveys $\&$ Tutorials.};
\node[align=center,color=red] at ([yshift=2em]current page text area.north) {Changes were made to this version by the publisher prior to publication.};
\node[align=center,color=black] at ([yshift=-2em]current page text area.south) {Copyright (c) 2019 IEEE. Personal use of this material is permitted. However, permission to use this material for any other purposes must be \\ obtained from the IEEE by sending a request to pubs-permissions@ieee.org.};
\end{tikzpicture}

\begin{abstract}
This paper elaborates on the design challenges, opportunities, and solutions for New Radio-based access to Unlicensed spectrum (NR-U) by taking into account the beam-based transmissions and the worldwide regulatory requirements. NR-U intends to expand the applicability of 5th generation New Radio access technology to support operation in unlicensed bands by adhering to Listen-Before-Talk (LBT) requirements for accessing the channel. LBT was already adopted by different variants of 4th generation Long Term Evolution (LTE) in unlicensed spectrum, i.e., Licensed-Assisted Access and MulteFire, to guarantee fair coexistence among different radio access technologies. In the case of beam-based transmissions, the NR-U coexistence framework is significantly different as compared to LTE in unlicensed spectrum due to the use of directional antennas, which enhance the spatial reuse but also complicate the interference management. In particular, beam-based transmissions are needed in the unlicensed spectrum at millimeter-wave (mmWave) bands, which is an attractive candidate for NR-U due to its large amount of allocated spectrum. As a consequence, some major design principles need to be revisited to address coexistence for beam-based NR-U. In this paper, different problems and the potential solutions related to channel access procedures, frame structure, initial access procedures, re-transmission procedures, and scheduling schemes are discussed. A simulation evaluation of different LBT-based channel access procedures for NR-U/Wi-Fi indoor mmWave coexistence scenarios is also provided.
\end{abstract}

\begin{IEEEkeywords}
NR-U, unlicensed spectrum, beam-based transmissions, coexistence, spectrum sharing, mmWave, LBT. 
\end{IEEEkeywords}

\IEEEpeerreviewmaketitle

\section{Introduction}
\label{sec:intro}
To address the rapid increase of wireless data traffic demand in the upcoming years, the wireless industry has turned its attention to the unlicensed spectrum bands as a way to aggregate additional bands and improve the capacity of future cellular systems~\cite{zhangh:15,zhan:15,labib:17}. The unlicensed spectrum that has global worldwide availability includes the 2.4 GHz, 5 GHz, and 60 GHz bands. 
In the unlicensed 60 GHz band, there has been a release of 9 GHz of spectrum in Europe and of 14 GHz in the USA~\cite{5Gamericas}, which provides 10$\times$ times (in Europe) and 16$\times$ times (in the USA) as much unlicensed spectrum as is available in sub 6 GHz bands.
Due to the large amount of spectrum available, the design of a system able to work in \gls{mmwave} carrier frequencies (30-300 GHz) is inevitable in order to achieve multi-Gigabit/s data rates for a large number of devices~\cite{pi:11,wong17}.

The \gls{3gpp} is currently in a full standardization process of \gls{nr}\footnote{The first version of \gls{nr} specification was published as a part of \gls{nr} Rel-15 in June 2018, while the remaining part of the specification is planned to be published as a part of \gls{nr} Rel-16 (in early 2020) as well as a part of subsequent releases.}, the \gls{rat} for \gls{5g} systems~\cite{TS38300,TR38912}, which has inherent support for operation at high carrier frequencies within the \gls{mmwave} spectrum region with wide-bandwidth~\cite{parkvall:17,giordani:19}.
One of the options which is being considered is to allow \gls{nr} to operate in unlicensed bands through \gls{nru}. It is similar to what was previously proposed in the case of \gls{lte} in unlicensed spectrum for the 5 GHz band, through its different variants~\cite{labib:17}, namely
\gls{laa}~\cite{kwon:17,TR36889}, \gls{lteu}~\cite{zhan:15,LTEU}, and MulteFire~\cite{rosa:18,multefire}.

The design of \gls{nru} started in a study item of \gls{nr} Rel-16 in 2018~\cite{RP-170828,TR38889}, and it is currently being developed as one of the \gls{nr} Rel-16 work items, which will enable its inclusion in future \gls{nr} specification~\cite{RP-190706}.
The primary objective of \gls{nru} is to extend the applicability of \gls{nr} to unlicensed spectrum bands as a general purpose technology that works across different bands and uses a design that allows fair coexistence across different \gls{rat}s.
Differently from \gls{laa} and \gls{lteu} that were based on carrier aggregation using the unlicensed 5 GHz band, and from MulteFire that used standalone operation in the 5 GHz band so far, \gls{nru} considers multiple bands and various deployment modes.
The frequency bands discussed for \gls{nru} include 2.4 GHz, 5 GHz, 6 GHz, 
and 60 GHz unlicensed bands\footnote{The \gls{nru} work item in \gls{nr} Rel-16 has started while focusing on sub 7 GHz bands~\cite{RP-190706}, but the extension to unlicensed \gls{mmwave} bands will probably be included in later releases, i.e., \gls{nr} Rel-17 and beyond. References to sub 7 GHz are intended to include the unlicensed bands in the 6 GHz region that have some region exceeding 7 GHz (e.g., 7.125 GHz). This differs from the classification of spectrum in \gls{nr} that considers sub 6 GHz bands and \gls{mmwave} frequency ranges.}, as well as 3.5 GHz and 37 GHz bands, which are devoted to shared access in the USA. 
As confirmed by \gls{3gpp}, the 60 GHz band is an attractive candidate for \gls{nru}, since it is currently not very crowded and can offer a large amount of contiguous bandwidth~\cite{TR38805}. 
Regarding the deployment modes, \gls{nru} supports carrier aggregation, dual connectivity, and standalone operation in unlicensed.
All in all, \gls{nru} is a milestone for 3GPP, which will allow, among others, standalone operation of NR in unlicensed spectrum including the mmWave bands with beam-based transmissions.

One of the most critical issues of allowing cellular networks to operate in unlicensed spectrum is to ensure a fair and harmonious coexistence with other unlicensed systems, such as Wi-Fi in the 5 GHz band (IEEE 802.11a/n/ac/ax) and directional multi-Gigabit Wi-Fi in the 60 GHz band (IEEE 802.11ad/ay, also known as \gls{wigig})~\cite{wigig,nitsche:14,ghasempour:17}. 
Fairness for \gls{nru} operation in the unlicensed bands is defined as the ability that \gls{nru} devices do not impact already deployed Wi-Fi services more than an additional Wi-Fi network would do on the same carrier~\cite{RP-170828}.
For a fair coexistence, any \gls{rat} that wants to operate in the unlicensed spectrum (e.g., \gls{nru}) has to be designed in accordance with the regulatory requirements of the corresponding bands. In the case of the 5 GHz and 60 GHz bands, the regulation mandates the use of \gls{lbt} in Europe and Japan~\cite{ETSI302567}. 
\gls{lbt} is a spectrum sharing mechanism by which a device senses the channel using a \gls{cca} check before accessing to it. \gls{lbt} works across different \gls{rat}s, and it is adopted by \gls{laa}, MulteFire, Wi-Fi, and \gls{wigig} to comply with the regulation (known as \gls{csma-ca} in the IEEE 802.11 context). However, even with omnidirectional communications, \gls{lbt} suffers from the hidden node and exposed node problems due to the differences in the sensing, transmission, and reception ranges\footnote{A hidden node problem arises when a node cannot hear an on-going transmission in the channel and declares the channel free to transmit but, if that node does any transmission, it collides with the on-going transmission. An exposed node problem, instead, appears when a node senses the channel as busy because it can listen to an on-going transmission but it could have transmitted simultaneously with that on-going transmission without creating any collision.}. 

Coexistence in the 5 GHz band has been well studied in recent years to let \gls{lte} in unlicensed spectrum gracefully coexist with Wi-Fi~\cite{zhangh:15, zhan:15, labib:17, lorenza:16}. 
Since \gls{lte} was initially designed to work in licensed bands on the basis of uninterrupted and synchronous operations, it was required to be later adapted to work with asynchronous protocols for operation in the unlicensed 5 GHz band. Differently, due to the on-going \gls{nr} standardization, \gls{nru} can be designed from the start with a great amount of flexibility for efficient operation in unlicensed spectrum bands.
Nevertheless, there is a major difference between \gls{nru} coexistence with other \gls{rat}s as compared to \gls{lte}/Wi-Fi coexistence in the 5 GHz band because of the use of beam-based (or directional) transmissions in \gls{nr}.
\gls{nr} has standardized beam management procedures for \gls{gnb}s and \gls{ue}s in all operational bands~\cite[Sec. 8.2.1.6.1]{TR38912}. In particular, directional communications are needed in \gls{mmwave} bands due to its characteristic propagation conditions, which require the use of beamforming to overcome propagation limits like severe pathloss, blocking, and oxygen absorption in case of the 60 GHz band~\cite{pi:11,andrews:17}. Similarly, \gls{wigig} (IEEE 802.11ad/ay) has been particularly designed to deal with these impairments by making directionality mandatory at either the transmitter or receiver~\cite{wigig}.

The beam-based transmissions envisioned in \gls{nr} potentially may cause less interference and enable spatial reuse. However, the different interference layout due to the directional transmissions also changes the coexistence framework in the unlicensed spectrum. In particular, the directionality may aggravate the hidden node and exposed node problems in the unlicensed bands~\cite{subramanian:10}. 
As such, the beam-based transmissions make the \gls{nru} coexistence framework more challenging as compared to coexistence with omnidirectional transmissions/receptions in Wi-Fi and \gls{lte} in unlicensed spectrum.

\subsection{Objective and Contribution}
The objective of this paper is to give the reader a complete overview of the major design principles and solutions for \gls{nru} operation in unlicensed bands, with an emphasis on mmWave bands, by taking into account the beam-based transmissions and the worldwide regulatory requirements. \gls{nru} technology is currently under development, and hence we focus our discussions to a set of key features and functionalities that are likely to be included in the final specification.
For that, we go through the main \gls{nr} features defined in Rel-15 and discuss the challenges in adapting them to meet the regulation for use in unlicensed spectrum and to coexist with other \gls{rat}s. We mainly focus our discussions on the design of \gls{phy} and \gls{mac} layers.

The main contributions of this paper are summarized as follows:
\begin{itemize}
\item We review the spectrum allocation and regulatory requirements for the unlicensed bands that have the most potential for \gls{nru}, i.e., 5 GHz, 6 GHz, and 60 GHz bands.
\item We outline the \gls{nru} scenarios and \gls{lbt} procedures under discussion in 3GPP and highlight the \gls{nr} features that need to  be  revisited  for \gls{nru}.
\item By considering the
regulatory  requirements  and  the  impact  of  narrow
beam transmissions, we elaborate on a variety of critical challenges that are encountered in different \gls{nru} scenarios, related to the following areas:
\begin{itemize}
\item the redefinition and implementation of \gls{lbt}-based \textbf{channel access procedures},
\item the selection of the \textbf{frame structure} 
in \gls{tdd} systems, 
\item the adaptation of \gls{nr} \textbf{initial access procedures},
\item the redesign of \gls{nr} \textbf{re-transmission procedures} based on \gls{harq} and \textbf{scheduling schemes}.
\end{itemize}
For each one of the identified challenges, we review the available literature and interesting standard contributions, and suggest innovative design solutions that can be further elaborated in future works.
\item Finally, we evaluate and compare different LBT-compliant channel access procedures with the aid of simulations in different NR-U/WiGig coexistence scenarios at the 60 GHz band. 
\end{itemize}

\begin{figure*}[!t]
	\centering
	\includegraphics[width=0.91\textwidth]{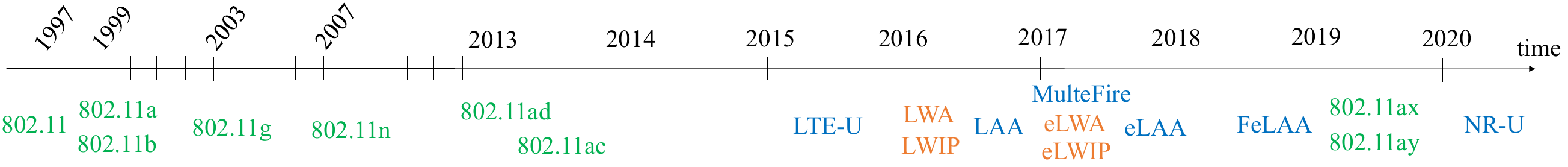}
	\caption{Standardization timeline of technologies that use unlicensed spectrum.}
	\label{fig_timing}
\end{figure*}

To the best of our knowledge, this is the first work that provides a detailed discussion on the design considerations and development process of beam-based \gls{nru}. Apart from the channel access procedures, no other work discusses other design challenges and solutions for \gls{nru}\footnote{Although we focus on beam-based transmissions, some of the discussions in this paper regarding frame structure, initial access, \gls{harq}, and scheduling, also apply to \gls{nru} with omnidirectional transmissions.}. Besides, regarding the channel access procedures, we analyze, compare, and evaluate different procedures in this paper.

Let us remark that in this paper we focus on \gls{nru}, assuming some basic knowledge from the reader about \gls{nr}. An overall description of \gls{nr} can be found in~\cite{TS38300}, and key papers are~\cite{parkvall:17,zaidi:16,liu:18}. Throughout this paper we refer the reader to specific sections of \gls{3gpp} \gls{nr} technical specification and reports when needed. 
In line with \gls{3gpp} terminology, we refer to an \gls{nr} terminal as \gls{ue} and an \gls{nr} base station as \gls{gnb}\footnote{The \gls{nr} architecture supports multiple \gls{trp}s that act as dumb antennas and are coordinated by a \gls{gnb}. Throughout this paper, we make the distinction only when needed but, in general, we refer to the \gls{nr} access point as \gls{gnb}.}. Similarly, according to IEEE 802.11 standards, Wi-Fi/\gls{wigig} terminal and base station are referred to as \gls{sta} and \gls{ap}, respectively.

\subsection{Organization}
The remainder of the paper is organized as follows. We start in Section~\ref{sec:back} by giving a review of the related work (in the areas of \gls{lte} in unlicensed spectrum (including LAA, LTE-U, and MulteFire), unlicensed IEEE-based technologies, beam-based \gls{nr} and \gls{nru}). 
Then, Section~\ref{sec:regulation} reviews the spectrum allocation and regulatory requirements for the unlicensed spectrum at 5 GHz, 6 GHz, and 60 GHz bands. Section~\ref{sec:NRU} presents the \gls{nru} scenarios and \gls{lbt} specifications, based on 3GPP discussions. Next, Section~\ref{sec:NR} introduces the different areas of the \gls{nr} system design that need to be rethought for \gls{nru}, which will be reviewed in Sections~\ref{sec:channelaccess}-\ref{sec:sched}. In Section~\ref{sec:channelaccess}, we highlight the problems and analyze potential channel access procedures for \gls{nru} to provide support for different \gls{lbt}-related problems that arise due to the beam-based transmissions and which were not present in \gls{laa} and MulteFire technologies. In Section~\ref{sec:frame}, we highlight the trade-offs in the selection of the frame structure. Section~\ref{sec:initialaccess} reviews the problems and solutions for the initial access procedure, including synchronization signal block design, random access procedure, and paging. In Section~\ref{sec:harq}, we illustrate two negative impacts of \gls{lbt} on the \gls{harq} mechanism and show how to overcome them. Section~\ref{sec:sched} elaborates on the problems related to the scheduler operation, and highlights new scheduling schemes that are suitable for beam-based \gls{nru}. After that, in Section~\ref{sec:eval}, we evaluate different \gls{lbt}-based channel access procedures in NR-U/WiGig indoor \gls{mmwave} coexistence scenarios. Finally, Section~\ref{sec:learned} summarizes the lessons learned from the discussions given in this paper, Section~\ref{sec:future} highlights future perspectives, and Section~\ref{sec:conc} concludes the paper.

\section{Background Review}
\label{sec:back}

\begin{table*}[!t]
\scriptsize
\centering
\begin{tabular}{|m{1cm}||m{1.5cm}|m{1.3cm}|m{1.5cm}|m{2.5cm}|m{1.5cm}|m{5cm}|}
\hline
 & Standardization body  & Underline \break Technology & Operational bands & Deployment capabilities & \gls{rat} in \ \  \break unlicensed  & Key features\\ \hline \hline
802.11n    & IEEE &  802.11a/g & sub 7 GHz & standalone (unlicensed) & Wi-Fi & \shortstack[l]{Unlicensed bands: 2.4, 5 GHz \\ Aggregated bandwidth: 40 MHz \\  MIMO: up to 4 streams, MU-MIMO: no \\ Modulation: up to 64-QAM \\ \gls{harq}: no \\ channel access scheme: \gls{csma-ca}} \\ \hline
802.11ad    & IEEE &  802.11 & above 7 GHz & standalone (unlicensed) & WiGig & \shortstack[l]{Unlicensed bands: 60 GHz \\ Aggregated bandwidth: 2.16 GHz \\  MIMO: up to 8 streams, MU-MIMO: no \\ Modulation: up to 64-QAM \\ \gls{harq}: no \\ channel access scheme: \gls{csma-ca}} \\ 
\hline
802.11ac    & IEEE &  802.11n & sub 7 GHz & standalone (unlicensed) & Wi-Fi & \shortstack[l]{Unlicensed bands: 5 GHz \\ Aggregated bandwidth: 160 MHz \\  MIMO: up to 8 streams, MU-MIMO: up to 4 \\ Modulation: up to 256-QAM \\ \gls{harq}: no \\ channel access scheme: \gls{csma-ca}} \\ \hline
LTE-U    & LTE-U Forum &  LTE Rel-12 & sub 7 GHz & \shortstack[l]{carrier aggregation \\ (licensed + unlicensed)} & LTE & \shortstack[l]{Unlicensed bands: 5 GHz \\ Aggregated bandwidth: 60 MHz \\  MIMO: up to 8 streams, MU-MIMO: up to 4 \\ Modulation: up to 256-QAM \\ \gls{harq}: yes \\ channel access scheme: duty-cycle} \\ 
\hline
LWA    & 3GPP &  LTE Rel-13 & sub 7 GHz & LTE + Wi-Fi integration at PDCP level & Wi-Fi & LTE Rel-13 + Wi-Fi\\
\hline
LWIP    & 3GPP &  LTE Rel-13 & sub 7 GHz & LTE + Wi-Fi integration at IP level & Wi-Fi & LTE Rel-13 + Wi-Fi\\
\hline
LAA    & 3GPP &  LTE Rel-13 & sub 7 GHz & \shortstack[l]{carrier aggregation \\ (licensed + unlicensed)} & LTE & \shortstack[l]{Unlicensed bands: 5 GHz \\ Aggregated bandwidth: 80 MHz \\  MIMO: up to 8 streams, MU-MIMO: up to 8 \\ Modulation: up to 256-QAM \\ \gls{harq}: yes \\ channel access scheme: \gls{lbt}} \\ 
\hline
MulteFire    & MulteFire Alliance &  LTE Rel-14 & sub 7 GHz & standalone (unlicensed) & LTE & \shortstack[l]{Unlicensed bands: 1.9, 2.4, 5 GHz \\ Aggregated bandwidth: 80 MHz \\  MIMO: up to 8 streams, MU-MIMO: up to 8 \\ Modulation: up to 256-QAM \\ \gls{harq}: yes \\ channel access scheme: \gls{lbt}} \\ 
\hline
eLWA    & 3GPP &  LTE Rel-14 & sub 7 GHz and above 7 GHz & LTE + Wi-Fi/WiGig integration at PDCP level & Wi-Fi/WiGig & LTE Rel-14 + Wi-Fi/WiGig\\
\hline
eLWIP    & 3GPP &  LTE Rel-14 & sub 7 GHz and above 7 GHz & LTE + Wi-Fi/WiGig integration at PDCP level & Wi-Fi/WiGig & LTE Rel-14 + Wi-Fi/WiGig\\
\hline
eLAA    & 3GPP &  LTE Rel-14 & sub 7 GHz & \shortstack[l]{carrier aggregation \\ (licensed + unlicensed), \\ dual connectivity \\ (licensed + unlicensed)}& LTE & \shortstack[l]{Unlicensed bands: 5 GHz \\ Aggregated bandwidth: 80 MHz \\  MIMO: up to 8 streams, MU-MIMO: up to 8 \\ Modulation: up to 256-QAM \\ \gls{harq}: yes \\ channel access scheme: \gls{lbt}} \\ 
\hline
FeLAA    & 3GPP &  LTE Rel-15 & sub 7 GHz & \shortstack[l]{carrier aggregation \\ (licensed + unlicensed), \\ dual connectivity \\ (licensed + unlicensed)} & LTE & \shortstack[l]{Unlicensed bands: 5 GHz \\ Aggregated bandwidth: 100 MHz \\  MIMO: up to 8 streams, MU-MIMO: up to 8 \\ Modulation: up to 256-QAM \\ \gls{harq}: yes \\ channel access scheme: \gls{lbt}} \\ 
\hline
802.11ax    & IEEE & 802.11ac  & sub 7 GHz & standalone (unlicensed) & Wi-Fi & \shortstack[l]{Unlicensed bands: 1 to 6 GHz \\ Aggregated bandwidth: 160 MHz \\  MIMO: up to 8 streams, MU-MIMO: up to 8 \\ Modulation: up to 1024-QAM \\ \gls{harq}: yes \\ channel access scheme: \gls{csma-ca}} \\ 
\hline
802.11ay    & IEEE &  802.11ad & above 7 GHz & standalone (unlicensed) & WiGig & \shortstack[l]{Unlicensed bands: 60 GHz \\ Aggregated bandwidth: 8.64 GHz \\  MIMO: up to 8 streams, MU-MIMO: up to 8 \\ Modulation: up to 64-QAM \\ \gls{harq}: no \\ channel access scheme: \gls{csma-ca}} \\ 
\hline
NR-U    & 3GPP &  NR Rel-17 & sub 7 GHz and above 7 GHz & \shortstack[l]{carrier aggregation \\ (licensed + unlicensed), \\ dual connectivity \\ (licensed + unlicensed), \\ standalone (unlicensed)} & NR & \shortstack[l]{Unlicensed bands: 2.4, 3.5, 5, 6, 37, 60 GHz \\ Aggregated bandwidth: 800 MHz \\  MIMO: up to 8 streams, MU-MIMO: up to 12 \\ Modulation: up to 1024-QAM \\ \gls{harq}: yes \\ channel access scheme: \gls{lbt}} \\ \hline
\end{tabular}
\caption{Taxonomy of technologies that use unlicensed spectrum.}
\label{table:taxonomy}
\end{table*}

In Fig.~\ref{fig_timing}, we illustrate the timeline of different \gls{rat}s that have been standardized for use in unlicensed spectrum (or are in the process of being standardized) so far. The timeline includes widely-deployed IEEE 802.11 standards (\gls{wlans}, commonly-known as Wi-Fi) with their different amendments, and the \gls{3gpp} based standards that follow different releases of \gls{lte} and \gls{nr}. In \gls{3gpp}, two main groups have been created depending on the \gls{rat} that is used to access the unlicensed spectrum: 
\begin{enumerate}
    \item technologies that are based on the integration of \gls{lte} and Wi-Fi radio links and that use Wi-Fi to access the unlicensed spectrum (i.e., \gls{lwa} and enhanced \gls{lwa} (eLWA), \gls{lwip} and enhanced \gls{lwip} (eLWIP)), and
    \item technologies that use modified versions of \gls{lte} or \gls{nr} to access and operate in the unlicensed spectrum (i.e., \gls{lteu}, \gls{laa} and its various enhancements, namely \gls{elaa} and further eLAA (FeLAA), MulteFire, and \gls{nru}).
\end{enumerate}

In Table~\ref{table:taxonomy}, we present a taxonomy of the different \gls{rat}s that use unlicensed spectrum, including the standardization body, the underline technology, the operational unlicensed spectrum bands (sub 7 GHz and/or above 7 GHz bands), the supported deployment capabilities, the \gls{rat} that is used to access the unlicensed spectrum, and the supported key features in terms of frequency bands, maximum supported bandwidth (including aggregation), \gls{mimo} support, \gls{mu-mimo} support, maximum supported modulation, \gls{harq} support for combining transmissions, and the channel access scheme that is used in the unlicensed spectrum. 

From Fig.~\ref{fig_timing} and Table~\ref{table:taxonomy}, it can be observed that IEEE 802.11 based technologies have been designed to access the unlicensed spectrum since 1997 and with the support of large bandwidth; on the other hand, 3GPP based technologies in unlicensed spectrum are more recent, and are characterized by a more sophisticated and efficient design, because they have been designed, since the very beginning, to operate in limited and expensive licensed spectrum. Nevertheless, with the latest amendments and versions (e.g., IEEE 802.11ax and NR-U), it is possible to observe that both the technologies are converging to use large bandwidth in a very efficient manner, through the support of key features such as \gls{harq}, high-order modulations, and high-order \gls{mimo}.

In this paper, we focus on the operation of cellular networks in unlicensed spectrum, i.e., the second group of 3GPP based technologies listed before, with special emphasis on the unlicensed \gls{mmwave} bands. As a result, in what follows we review only the state of the art related to the objective of this paper. Specifically, we first focus on the standardization and literature of the different variants of \gls{lte} in unlicensed spectrum. Then, we review the literature related to technologies that use directional transmissions for operation in unlicensed \gls{mmwave} bands.

\subsection{LTE in unlicensed spectrum (5 GHz band)}
To let \gls{lte} gracefully coexist with Wi-Fi in the 5 GHz band with omnidirectional transmissions and receptions, different variants of \gls{lte} in unlicensed spectrum have been proposed, widely studied in the research literature, and standardized based on modifications over \gls{lte}. The different variants are: \gls{laa}~\cite{TR36889}, \gls{lteu}~\cite{LTEU}, and MulteFire~\cite{multefire}.

\gls{3gpp} established work items on \gls{laa} in \gls{lte} Rel-13~\cite{TR36889} and on \gls{elaa} in \gls{lte} Rel-14~\cite{RP152272} to evaluate and specify \gls{dl} and \gls{ul} operations in the 5 GHz unlicensed band~\cite{TS36213}, respectively. Also, in \gls{lte} Rel-15, a work item on further enhancements to \gls{lte} operation in unlicensed spectrum (FeLAA) was concluded in 2018~\cite{RP170848}. \gls{laa} technologies (\gls{laa}/\gls{elaa}/FeLAA) operate as supplementary \gls{dl}/\gls{ul} carriers in unlicensed bands with anchor carriers in the licensed bands. As mentioned earlier, to meet worldwide regulation, a \gls{lbt}-based channel access scheme was introduced in \gls{laa} technologies to access to the unlicensed band, which is similar to the \gls{cca} procedure used in IEEE 802.11-based technologies. 
An overview of LAA technology is presented in~\cite{kwon:17}. Interested readers can also look at the comprehensive survey about \gls{laa}/Wi-Fi coexistence in the 5 GHz band in~\cite{chen:17}, and references therein. In~\cite{perf_analysis_lte_wifi}, an analytical framework based on Markov chain is developed to study the downlink throughput of \gls{laa}/Wi-Fi coexistence, for a simple \gls{lbt} with fixed contention window size and simple scenarios composed of one \gls{ap} and one \gls{laa} node. For \gls{3gpp}-based scenarios, the impact of several parameters related to the \gls{laa} \gls{lbt} mechanism on the channel access opportunities of LAA and its coexistence performance, has been assess through system-level simulations in~\cite{perf_analysis_ericsson_wireless_comm}.

In regions where the regulation does not require \gls{lbt}, as in the USA, access schemes, other than the ones standardized by \gls{3gpp}, have been designed and produced. In particular, the industrial consortium \gls{lteu} Forum specified a proprietary solution~\cite{LTEU}, known as \gls{lteu}. As for \gls{laa}, \gls{lteu} technology uses carrier aggregation of the unlicensed band with an anchor carrier in a licensed band. However, instead of relying on \gls{lbt} for accessing the channel, it basically allows coexistence by duty-cycling the \gls{lte} continuous transmission. A comprehensive overview of the \gls{lteu} technology, including implementation regulations,
principles, and typical deployment scenarios, is presented in~\cite{zhan:15}. 
A highly performing access scheme for \gls{lteu} (known as \gls{csat}) is proposed in~\cite{qualcomm_csat_algorithm}, in which the duty cycle is adapted based on the activity observed on the channel.
In~\cite{perf_analysis_lteu_stohastic_geom}, stochastic geometry is used to analyze \gls{lteu}/Wi-Fi coexistence in terms of coverage probability and throughput, as well as to perform asymptotic analysis. Resource allocation for~\gls{lteu} is studied in~\cite{7558177}, which also proposes a joint optimization of \gls{mac} and \gls{phy} layer parameters of the \gls{lteu} network.

Multiple works have focused also on modeling, analyzing, and comparing \gls{laa} and \gls{lteu}.
Authors in \cite{compare_dc_lbt_modeling_interference} derive throughput and interference models for inter-technology coexistence analysis in the 5 GHz band, considering \gls{lbt}-based as well as duty cycle-based access schemes. Through Monte Carlo simulations, they show that duty cycle (i.e., \gls{lteu}) outperforms \gls{laa} \gls{lbt} for low interference scenarios, while in high interference scenarios \gls{lbt} outperforms
duty cycle mechanisms.
Comparisons are done also through simulations in~\cite{perf_analysis_jeon_intel_globecom_ws_2014} for various indoor and outdoor setups.
In
\cite{perf_analysis_cristina_cano_douglas_leight}, a throughput model is presented to analyze LTE/Wi-Fi coexistence by focusing on the comparison of \gls{lbt} versus \gls{csat}. They conclude that, when
optimally configured, both \gls{laa} and \gls{lteu} approaches are capable of providing the same level of fairness to Wi-Fi. Authors in~\cite{perf_analysis_samsung_tcom} model, analyze, and compare different coexistence mechanisms including plain \gls{lte}, \gls{lte} with discontinuous transmission (\gls{lteu}), and \gls{lte} with \gls{lbt} (\gls{laa}). Therein, by leveraging on stochastic geometry, authors analytically derive and numerically evaluate the medium
access probability, the \gls{sinr} coverage probability, density of successful transmission, and the rate coverage probability.

In general, for the 5 GHz band, it is generally considered
that \gls{laa} is fairer to Wi-Fi than \gls{lteu}, because it uses the \gls{lbt} mechanism and so it abides similar rules as Wi-Fi. Recently, authors in~\cite{biljana:19}, have presented a detailed coexistence study and comparison of \gls{laa} and \gls{lteu} technologies through network simulations, and evaluated how the channel access procedures, besides other important aspects like the traffic patterns, simulation setup, and proprietary implementation choices, impact on the coexistence results.

Finally, the MulteFire Alliance launched the development of a new \gls{lte}-based technology capable of operating standalone in unlicensed or shared spectrum bands, also known as MulteFire~\cite{multefire, qualcomm, nokia}, without using any licensed carrier as an anchor. 
An overview of MulteFire is presented in~\cite{rosa:18}, including the main challenges due to \gls{lbt} and the standalone operation, as well as the solutions adopted in MulteFire to overcome such challenges and the attained performance benefits.
The standalone operation in unlicensed bands may open a new class of wireless private networks, e.g., for Industry 4.0 scenarios~\cite{8207346}. However, it also becomes difficult to operate without any support from licensed carriers. For example, in standalone operation, latency may be increased because of the LBT requirement for each new transmission~\cite{maldonado:18}.

A comparative analysis of the three LTE variants (LAA, LTE-U, and MulteFire) is provided in~\cite{labib:17}, including technical details of each \gls{rat} and their operational features and coexistence capabilities.

Research on the different variants of \gls{lte} in unlicensed spectrum for 5G is still on-going to improve the coexistence with Wi-Fi in sub 7 GHz bands. For example, authors in~\cite{zeng:18} propose channel selection algorithms for 5G \gls{elaa}. Recently, authors in~\cite{garcia:18} presented the massive MIMO unlicensed (mMIMO-U) technology for sub 7 GHz bands. The mMIMO-U enhances LBT by placing radiation nulls toward neighboring Wi-Fi nodes, provided that Wi-Fi nodes can be detected by the \gls{gnb}s. Authors in~\cite{song:19} present a cooperative \gls{lbt} scheme with omnidirectional transmissions/receptions, whereby neighboring \gls{gnb}s are allowed to cooperate in the sensing and transmission phases to improve the \gls{qos}. 
Let us remark that some features of \gls{laa}-based technologies and MulteFire can be reused for \gls{nru}, specially for what regards initial access from \gls{laa} and regarding \gls{harq} procedures and scheduling from MulteFire standalone operation, but they need to be adapted and/or improved for beam-based transmissions in \gls{nru} (as we will review later).

\subsection{Technologies in unlicensed mmWave bands}
\label{subsec:ieee}
One of the key features of \gls{nr}, as compared to \gls{lte}, is the wide-band support for operation at \gls{mmwave} carrier frequencies~\cite{parkvall:17}. For that, multiple procedures for beam-related operations have been defined in the \gls{nr} standard, including beam sweeping, beam measurement, beam determination, and beam reporting~\cite{giordani:19}. 
In terms of NR to  make it operate in shared/unlicensed mmWave bands, related works include~\cite{nekovee:16,nekovee:17,zhang:17,seo:18,boccardi:16}. Authors in~\cite{nekovee:16,nekovee:17} present beam scheduling solutions that are based on iterative coordination of the concurrent transmissions of different base stations by means of properly selecting their transmit beams. Also, multiple solutions based on spectrum sharing~\cite{zhang:17,seo:18} and spectrum pooling~\cite{boccardi:16} have been recently proposed, which exploit coordination among different cellular network operators to improve the spatial reuse. However, these solutions cannot ensure fair coexistence of \gls{nru} with other \gls{rat}s in the unlicensed bands because they do not employ mechanisms to avoid continuous use of the spectrum (as it is the case of \gls{lbt} or duty-cycling).

IEEE 802.11 \gls{wlans} standards have started technology development to use the unlicensed spectrum at \gls{mmwave} bands few years ago through 802.11ad specification~\cite{nitsche:14}, and its recent enhancement in 802.11ay specification~\cite{ghasempour:17} (see Fig.~\ref{fig_timing}). In this regard, both IEEE 802.11ad and 802.11ay have standardized specific beam training processes for directional transmissions~\cite{zhou:18}. However, in these specifications, \gls{cca} within \gls{csma-ca} is still defined with omnidirectional sensing. In~\cite{singh:10}, an enhanced distributed \gls{mac} protocol is proposed for CSMA-based mesh networks employing directional transmissions at 60 GHz. The proposed solution uses memory at the nodes to achieve approximate TDMA schedules without explicit coordination.

In the area of IEEE 802.15 \gls{wpans} standards (including the well-known Bluetooth and ZigBee), technology development to use the unlicensed \gls{mmwave} bands has also been considered since few years ago in 802.15.3c specification. To enhance IEEE 802.15 \gls{wpans}, multiple solutions have been proposed for beam management and time-domain coordination in \gls{mmwave} bands~\cite{an:08,pyo:09,cai:10}. A time division multiple access (TDMA) based channel allocation scheme for directional transmissions is proposed in~\cite{an:08}. An enhanced \gls{mac} with frame aggregation, unequal error protection, and block acknowledgment is defined in~\cite{pyo:09}. Authors in~\cite{cai:10} introduced the concept of an exclusive region to enable concurrent transmission with significant interference reduction in \gls{mmwave} \gls{wpans}, by considering all kinds of directional and omnidirectional transmission/reception antenna patterns. 

The coordination of the transmit beams (as proposed in~\cite{nekovee:16,nekovee:17,cai:10}) and the coordination of the channel access in time domain (as analyzed in~\cite{an:08,pyo:09}) solve hidden node problems that arise in the unlicensed spectrum. However, since these kinds of solutions require coordination between Wi-Fi/\gls{wigig} and cellular devices, they are not adequate for multi-\gls{rat} coexistence scenarios. Instead, distributed uncoordinated approaches are needed. 
For that reason, and also due to regulation mandate, \gls{lbt} was adopted to control the channel accesses in \gls{laa}/\gls{elaa}/FeLAA and MulteFire.

\subsection{Towards NR-U}
In case of directional transmissions, \gls{lbt} might not work well because of the increased hidden and exposed node problems~\cite{lagen:18d,subramanian:10}. For example, when the carrier sense is done omnidirectionally, i.e., \gls{omnilbt}, while the intended transmission is beam-based, there is a higher chance of exposed node problems (as it happens in WiGig). If the direction of the intended communication is known, directional carrier sense, i.e., \gls{dirlbt}, may help to improve the spatial reuse but it may lead to hidden node problems~\cite{R1-1713785}. This phenomenon is the so-called \gls{omnilbt}/\gls{dirlbt} trade-off. It is shown in~\cite{lagen:18b} that, for low network densities, \gls{dirlbt} performs significantly better than \gls{omnilbt}, while for high network densities, \gls{omnilbt} is a better technique.
Therefore, new regulatory-friendly and distributed channel access schemes are needed to address coexistence for \gls{nru} under beam-based transmissions. 

From the 3GPP standardization point of view, \gls{nru} for sub 7 GHz is currently being standardized by \gls{3gpp}, and \gls{nru} in \gls{mmwave} bands is planned to be addressed in future releases of 3GPP (i.e., \gls{nr} Rel-17 and beyond). In the literature, \gls{nru} with beam-based transmissions has not been discussed sufficiently. There have only been some work on the channel access procedures~\cite{lagen:18b,lagen:18d,lagen:18,R1-180xxxx,li:18}. To address the \gls{omnilbt}/\gls{dirlbt} trade-off, two distributed \gls{lbt}-based channel access procedures have been proposed by the same authors for beam-based \gls{nru}, namely \gls{pairlbt}~\cite{lagen:18b} and \gls{lbtswitch}~\cite{lagen:18d}, which we will further review, discuss, and compare throughout this paper. 
Even though, in the case of beam-based transmissions, there are interference situations that cannot be detected at the transmitter due to the significant difference in the interference dynamics at the transmitter and receiver sides. Remarkably, in some cases, it is only the receiver that can be aware of potential interference situations~\cite{lagen:18}. Therefore, \gls{lbt} at the transmitter may not be useful to detect such interference. 
In this line, a technique called \gls{lat} is introduced in~\cite[Sec. 8.2.2]{D41mmMagic}.
This approach is certainly of interest but it is not compliant with regulations regarding \gls{lbt} requirement.
This can be solved by employing receiver-assisted LBT procedures~\cite[Sec. 7.6.4]{R1-180xxxx},
or \gls{lbr}~\cite{lagen:18}, wherein the transmitter triggers a carrier sense at the receiver that is used to complement \gls{lbt}. Recently, going deeper into this issue, authors in~\cite{li:18} proposed a joint directional \gls{lbt}-\gls{lbr} and beam training for \gls{nru} in \gls{mmwave} bands.

\section{Spectrum Allocation and Regulatory Requirements}
\label{sec:regulation}
Operation in unlicensed spectrum is subject to different regulatory limitations and restrictions that are region- and band- specific. In this section, we review the spectrum allocation and the regulatory requirements for the 5 GHz and 60 GHz bands, which have common global availability and for which most major geographical areas worldwide have authorized wide unlicensed spectrum bandwidth. Also, we review the spectrum allocation for the 6 GHz band, which has been recently allocated for unlicensed use in Europe and the USA.

\subsection{Spectrum Allocation}
\begin{figure}[!t]
	\centering
	\includegraphics[width=0.41\textwidth]{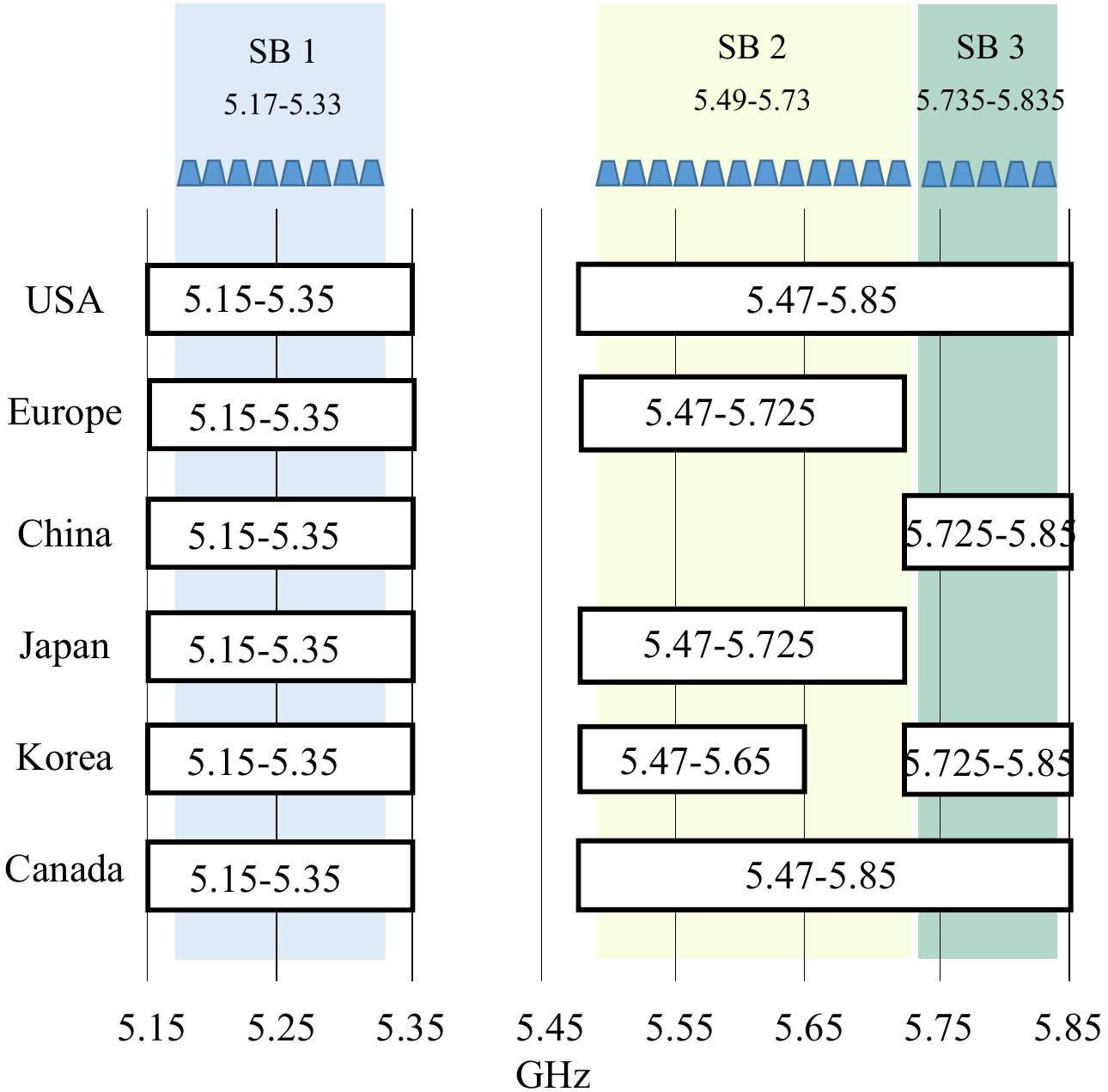}
	\caption{5 GHz unlicensed spectrum allocation in different areas of the world.}
	\label{fig_channelization5}
\end{figure}

\begin{figure}[!t]
	\centering
	\includegraphics[width=0.42\textwidth]{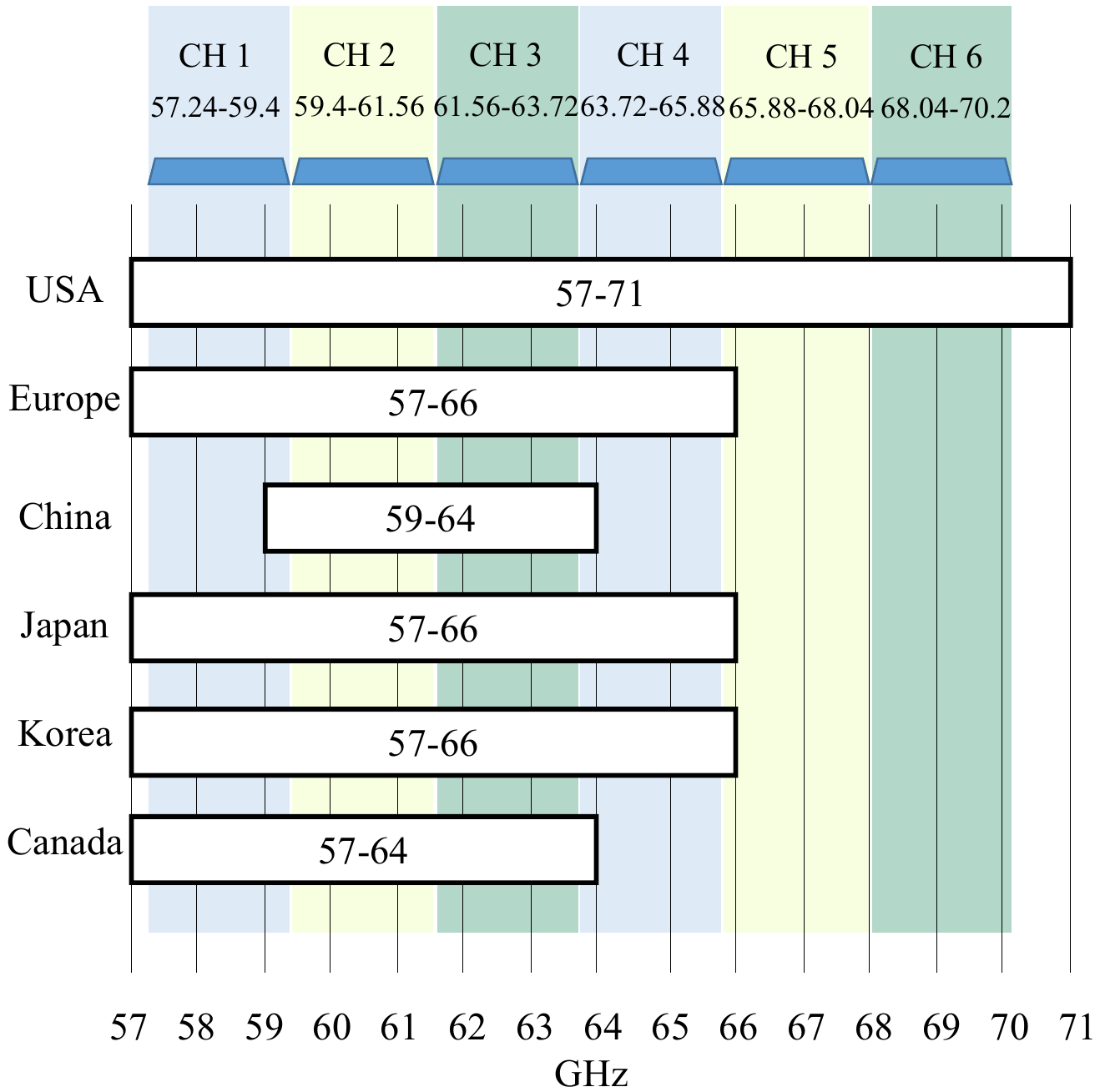}
	\caption{60 GHz unlicensed spectrum allocation in different areas of the world.}
	\label{fig_channelization}
\end{figure}

In Fig.~\ref{fig_channelization5} and Fig.~\ref{fig_channelization}, we show the unlicensed spectrum allocation of major geographic areas of the world for the 5 GHz band and the 60 GHz band, respectively, including IEEE 802.11ac channelization in Fig.~\ref{fig_channelization5} and IEEE 802.11ad channelization in Fig.~\ref{fig_channelization}. Three subbands are available in the 5 GHz band and, according to IEEE 802.11ac channelization~\cite{perahia:11}, each subband is further divided into multiple non-overlapping channels of 20 MHz bandwidth each. 
On the other hand, IEEE 802.11ad channelization in the 60 GHz band supports up to six non-overlapping channels of 2.16 GHz bandwidth each, thus having a lower number of channels but much wider channel bandwidths than Wi-Fi in the 5 GHz band.

\begin{figure}[!t]
	\centering
	\includegraphics[width=0.41\textwidth]{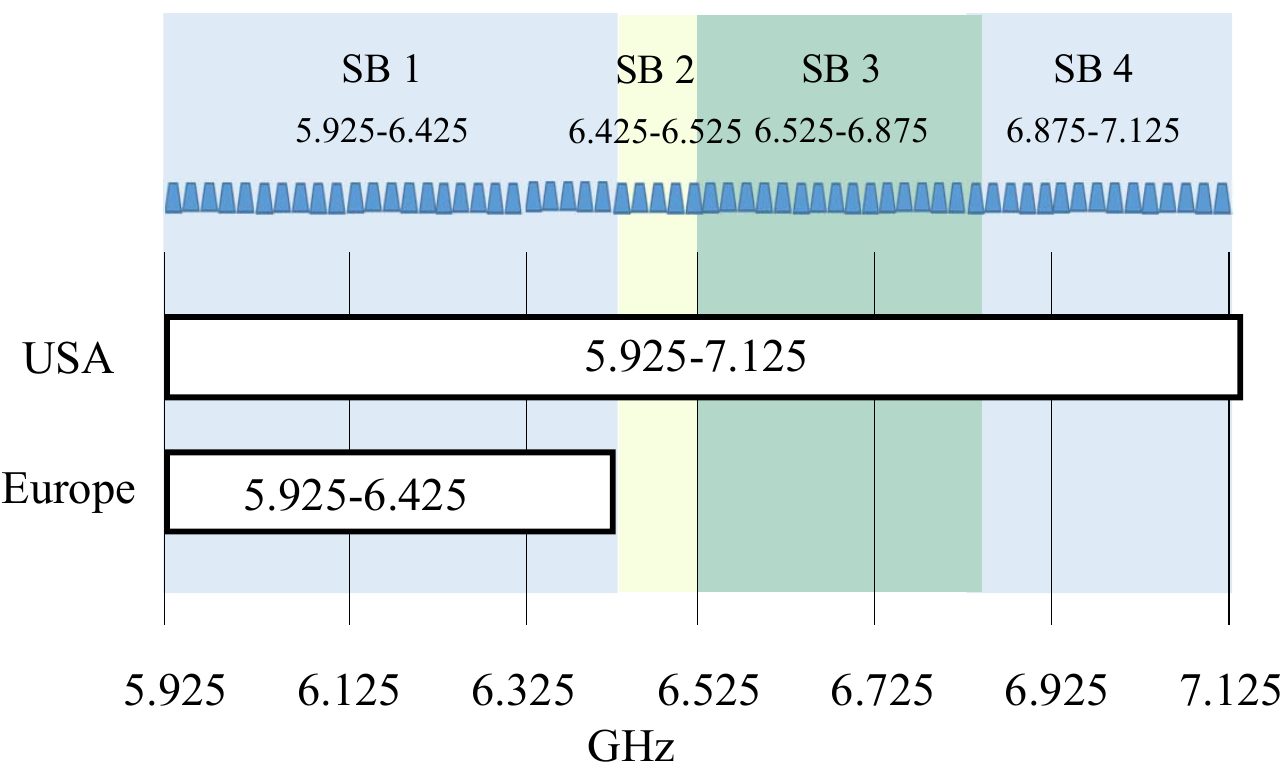}
	\caption{6 GHz potential unlicensed spectrum allocation in the USA and Europe.}
	\label{fig_channelization6}
\end{figure}

At the time of writing, the USA and Europe are analyzing the potential of the 6 GHz band for unlicensed use.
The spectrum considered in the USA (5.925-7.125 GHz) and Europe (5.925-6.425 GHz) is illustrated in Fig.~\ref{fig_channelization6}, alongside IEEE 802.11ax 20 MHz channelization.

\subsection{Regulatory Requirements}
\gls{etsi} regulation has harmonized the requirements for the 5 GHz band (5.15-5.35 GHz and 5.47-5.725 GHz) and the 60 GHz band (57-66 GHz), as included in~\cite{ETSI301893} and~\cite{ETSI302567}, respectively. To enable worldwide regulation-compliant access and satisfy a fair coexistence with the unlicensed systems (Wi-Fi, \gls{wigig}, radar) and intra-\gls{rat} services, any technology that attempts accessing to the unlicensed spectrum (like \gls{nru}) should fulfill the following regulatory requirements:

\begin{itemize}
\item \textbf{Listen-Before-Talk (LBT)}: The \gls{lbt} procedure is a mechanism by which a device should apply a \gls{cca} check (i.e., spectrum sensing
for a certain period, called the \gls{cca} period) before using the channel and which imposes certain rules after determining the channel to be busy. \gls{cca} uses \gls{ed} to detect the presence (i.e., channel is busy) or absence (i.e., channel is idle) of other signals on the channel. If the detected energy during an initial \gls{cca} period is lower than a certain threshold (the \gls{ed} threshold), the device can access the channel for a period called \gls{cot}. Otherwise, an extended \gls{cca} period starts, in which the detected energy is again compared against the \gls{ed} threshold until channel access is granted. \gls{lbt} is a mandatory procedure in Europe and Japan for the 5 GHz and 60 GHz bands but it is not required in other regions like the USA and China. The \gls{lbt} mechanism and its parameters are specified in~\cite{ETSI301893} and~\cite{ETSI302567}. Briefly, for each band, the regulation specifies the \gls{cca} slot duration ($9$ $\mu$s in the 5 GHz band, and $5$ $\mu$s in the 60 GHz band),
the initial and extended \gls{cca} check times (e.g., a multiple of $5$ $\mu$s for initial \gls{cca} and $8{+}m{\times} 5$ $\mu$s for extended \gls{cca} in the 60 GHz band, where $m$ controls the backoff), and the \gls{ed} threshold ($-72$ dBm for a 20 MHz channel bandwidth in the 5 GHz band, and $-47$ dBm for $40$ dBm of radiated power in the 60 GHz band). 

\item \textbf{\gls{mcot}}: Certain regions such as Europe and Japan prohibit continuous transmission in the unlicensed spectrum and impose limits on the \gls{cot}, i.e., the maximum continuous time a device can use the channel. The \gls{mcot} in the 5 GHz band is limited to $2$ ms, $4$ ms, or $6$ ms depending on the channel access priority class, and it may be increased up to $8$-$10$ ms in some cases~\cite{ETSI301893}. The \gls{mcot} in the 60 GHz band is $9$ ms~\cite{ETSI302567}. Besides, for the 5 GHz and 60 GHz bands, it is allowed to share the \gls{cot} with the associated devices (e.g., \gls{gnb} and \gls{ue}s), and thus enable a contiguous combination of \gls{dl} and \gls{ul} transmissions within the \gls{cot}. Sharing the COT means that once the initiating device (\gls{gnb}) gets access to the channel through \gls{lbt} and transmits, the responding devices (\gls{ue}s) are allowed to
skip the \gls{cca} check and immediately transmit in response to the received frames~\cite{ETSI302567}.

\item \textbf{\gls{eirp} and \gls{psd}}: Operation in the unlicensed spectrum is subject to power limits in all regions and bands, in terms of \gls{eirp} and \gls{psd}, to constrain the generated inter-\gls{rat} and intra-\gls{rat} interference levels. According to \gls{etsi} regulation~\cite{ETSI301893}, in the 5 GHz band, the maximum mean \gls{eirp} and \gls{psd} with transmit power control for 5.15-5.35 GHz range are limited to $23$ dBm and $10$ dBm/MHz, respectively, and for 5.47-5.725 GHz range, are limited to $30$ dBm and $17$ dBm/MHz, respectively.
In the 60 GHz band, the maximum mean \gls{eirp} and \gls{psd} are limited to $40$ dBm and $13$ dBm/MHz, respectively~\cite{ETSI302567}. Besides the \gls{etsi} power limits, more restrictive power limits are imposed in some regions~\cite{TR38805}. For example, the USA differentiates among indoor and outdoor devices with different power limits~\cite{5Gamericas,FCC}.

\item \textbf{\gls{ocb}}: The \gls{ocb} is defined as the bandwidth containing $99 \%$ of the signal power and, in certain regions, it should be larger than a percentage of the \gls{ncb} (i.e., the channel width). This enforces the unlicensed technologies to use major part of the channel bandwidth when they access the channel.
According to \gls{etsi}, for the 5 GHz band, the \gls{ocb} shall be between $70 \%$ and $100 \%$ of the \gls{ncb}~\cite{ETSI301893}. In the 60 GHz band, the \gls{ocb} shall be in between $80 \%$ and $100 \%$ of the \gls{ncb}~\cite{ETSI302567}. 

\item \textbf{\gls{fr}}: The \gls{fr} process allows reusing the same channel at the same time by different devices of the same \gls{rat}. In general, if a device is accessing the channel, then other devices in its coverage area should be muted in this channel so that it cannot be reused at the same time. This reduces the number of devices that access simultaneously (i.e., the \gls{fr} factor). The \gls{fr} mechanism is designed to allow devices of the same operator to access the channel simultaneously, and hence increase the \gls{fr} factor and improve the spectral efficiency. This is done by using different \gls{ed} thresholds for intra-\gls{rat} and inter-\gls{rat} signals, provided that the devices can distinguish between these two types of signals\footnote{For example, \gls{laa} supports \gls{fr} with an \gls{ed} threshold of $-52$ dBm for intra-\gls{rat} signals (i.e., \gls{laa} signals), as compared to $-72$ dBm of ED threshold used for inter-\gls{rat} signals (e.g., Wi-Fi signals). On the other hand, Wi-Fi is designed to avoid \gls{fr}, especially among Wi-Fi nodes. For that, Wi-Fi supports preamble detection to identify intra-\gls{rat} signals, and it uses -82 dBm of preamble detection threshold for Wi-Fi signals while a -62 dBm of \gls{ed} threshold for non-Wi-Fi signals in the 5 GHz band.}.  

\item \textbf{\gls{dfs}}: \gls{dfs} functionality is used to avoid interfering with 5 GHz and 60 GHz radar systems, as well as to uniformly spread the traffic load across the different channels in each band. The regulation states that whenever radar signals are detected, a device must switch to another channel to avoid interference. 
\end{itemize}

\section{\gls{nru} Scenarios and \gls{lbt} Specifications}
\label{sec:NRU}

\begin{figure*}[!t]
	\centering
	\includegraphics[width=0.81\textwidth]{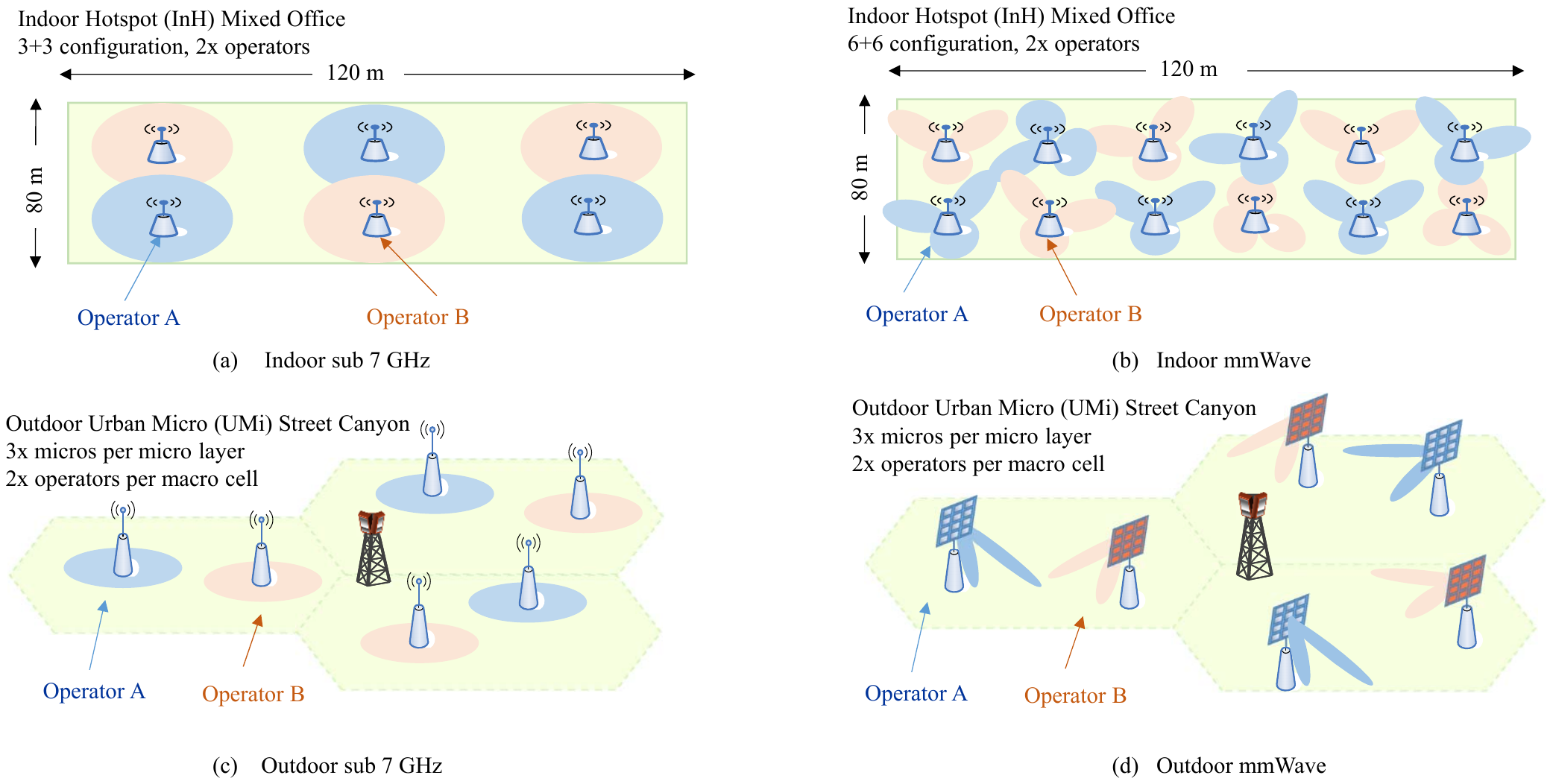}
	\caption{\gls{nru} layout scenarios.}
	\label{fig_topologies}
\end{figure*}

\begin{figure*}[!t]
	\centering
	\includegraphics[width=0.94\textwidth]{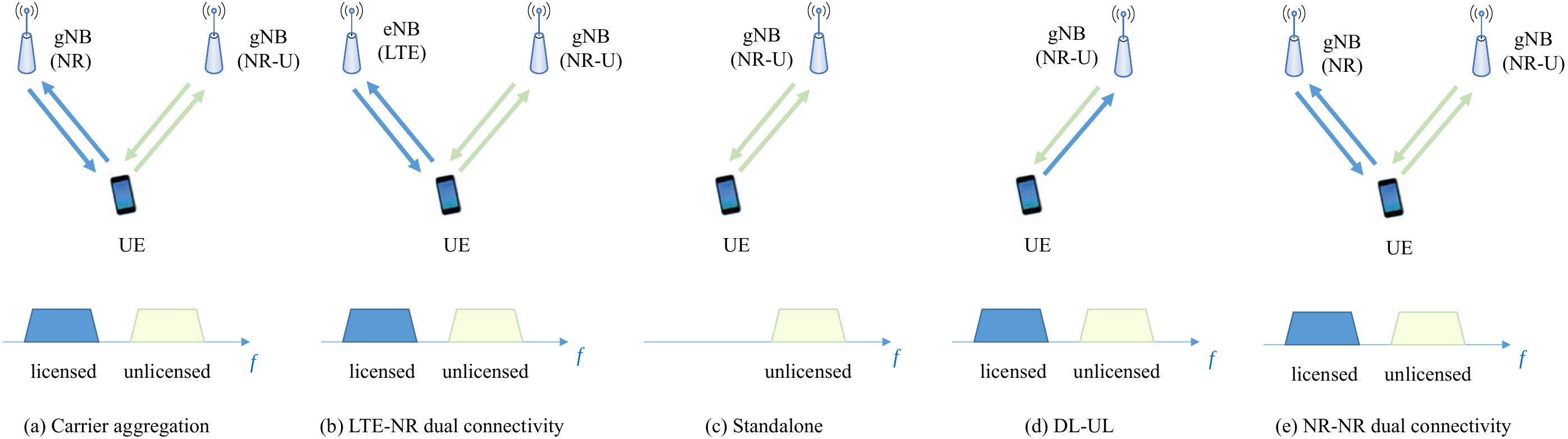}
	\caption{\gls{nru} deployment scenarios.}
	\label{fig_scenarios}
\end{figure*}

\subsection{\gls{nru} Scenarios}
\gls{laa}, \gls{lteu}, and MulteFire technologies were specifically designed to operate in the 5 GHz band. Differently, \gls{nru} considers multiple bands: 2.4 GHz (unlicensed worldwide), 3.5 GHz (shared in the USA), 5 GHz (unlicensed worldwide), 6 GHz (unlicensed in the USA and Europe), 37 GHz (shared in the USA), and 60 GHz (unlicensed worldwide). 
The \gls{3gpp} classifies these bands for \gls{nru} as sub 7 GHz and mmWave bands. Sub 7 GHz bands include the 2.4, 3.5, 5, and 6 GHz bands; meanwhile mmWave bands encompass the 37 and 60 GHz bands. First efforts in the \gls{nru} standardization focus on sub 7 GHz bands~\cite{TR38889} and mmWave bands will be addressed later. Therefore, four \textit{layout scenarios} can be defined for \gls{nru} based on the deployment and propagation environment conditions:
\begin{itemize}
\item indoor sub 7 GHz,
\item indoor \gls{mmwave},
\item outdoor sub 7 GHz,
\item outdoor \gls{mmwave}.
\end{itemize}

The \gls{nru} layout scenarios are shown in Fig.~\ref{fig_topologies}. According to standard terminology, operator A and operator B in the figure are used to denoting two different \gls{rat}s (and thus address, e.g., Wi-Fi and \gls{nru} coexistence) or two operators of the same \gls{rat}, e.g., to evaluate either Wi-Fi/Wi-Fi or \gls{nru}/\gls{nru} coexistence. The more details on the simulation methodology and parameters for indoor and outdoor sub 7 GHz can be found in the 3GPP report TR 38.889~\cite[Sec. 8.1]{TR38889}.

To assess the coexistence, five different \textit{deployment scenarios} are defined for \gls{nru} in \gls{3gpp}~\cite[Sec. 6]{TR38889}: 
\begin{itemize}
\item Carrier aggregation between licensed band \gls{nr} and unlicensed band \gls{nru},  
\item Dual connectivity between licensed band \gls{lte} and unlicensed band \gls{nru}, 
\item Standalone unlicensed band \gls{nru}, 
\item \gls{nr} with \gls{dl} in unlicensed band and \gls{ul} in licensed band, 
\item Dual connectivity between licensed band \gls{nr} and unlicensed band \gls{nru}.
\end{itemize}

The \gls{nru} deployment scenarios are illustrated in Fig.~\ref{fig_scenarios}. All of them can be applied to each of the \gls{nru} layout scenarios shown in Fig.~\ref{fig_topologies}.
The carrier aggregation scenario follows the approach in LAA technologies, with the possibility of \gls{nru} for both supplementary \gls{dl} and \gls{ul}. The standalone scenario resembles the MulteFire approach. 
Note that the \gls{nru} design is further complicated in the standalone deployment scenario because all the signals must use the unlicensed band, thus significantly affecting the initial access and scheduling procedures. 

The performance metrics for \gls{nru} coexistence evaluation are the same as in LAA~\cite{TR36889}. These include, user packet throughput and delay (mean value and value at the $5$th, $50$th, and $95$th percentiles) for low, medium and high loads, measured separately for \gls{dl} and \gls{ul}, thus leading to 48 metrics. Also, buffer occupancy is measured for \gls{nru} and Wi-Fi, separately. The coexistence evaluation scenarios include Wi-Fi/Wi-Fi, Wi-Fi/NR-U, and NR-U/NR-U~\cite{TR38889}. 

The coexistence requirement for \gls{nru} (i.e., the fairness definition) remains the same as in \gls{laa}, in which \gls{nru} devices should not impact deployed Wi-Fi/\gls{wigig} services (data, video, and voice services) more than an additional Wi-Fi/\gls{wigig} network would do on the same carrier~\cite{RP-170828}. Therefore, the standard way to evaluate the fairness is to first consider a Wi-Fi/Wi-Fi deployment (operator A/operator B) in any of the layout scenarios in Fig.~\ref{fig_topologies}, and then replace one Wi-Fi network by an \gls{nru} network to assess the Wi-Fi/\gls{nru} coexistence and determine the impact of \gls{nru} on the Wi-Fi system as compared to the Wi-Fi/Wi-Fi deployment.

\subsection{\gls{lbt} Specifications}
\gls{3gpp} has specified four \gls{lbt} categories for \gls{nru}~\cite{TR38889}:
\begin{itemize}
    \item Category 1 (Cat 1 \gls{lbt}): Immediate transmission after a short switching gap of $16$ $\mu$s.
    \item Category 2 (Cat 2 \gls{lbt}): \gls{lbt} without random back-off, in which the \gls{cca} period is deterministic (e.g., fixed to $25$ $\mu$s).
    \item Category 3 (Cat 3 \gls{lbt}): \gls{lbt} with random back-off with a contention window of fixed size, in which the extended \gls{cca} period is drawn by a random number within a fixed contention window.
    \item Category 4 (Cat 4 \gls{lbt}): \gls{lbt} with random back-off with a contention window of variable size, in which the extended \gls{cca} period is drawn by a random number within a contention window, whose size can vary based on channel dynamics.
\end{itemize} 

For different transmissions in a \gls{cot} and various channels/signals to be transmitted, different categories can be used. In brief, as in \gls{laa}, Cat 4 \gls{lbt} is used for \gls{gnb} or \gls{ue} to initiate a \gls{cot} for data transmissions, while \gls{gnb} can use Cat 2 \gls{lbt} for specific signaling like discovery reference signals (see details in~\cite{TR38889}).

The rules for shared \gls{cot} have also been defined for \gls{nru} in~\cite{TR38889}. For a \gls{gnb} initiated \gls{cot}, the responding devices are allowed to transmit without performing a \gls{cca} check (i.e., Cat 1 \gls{lbt}) if there is a gap in between \gls{dl} and \gls{ul} transmissions of less than $16$ $\mu$s. 
For a gap of more than $16$ $\mu$s but less than $25$ $\mu$s, within the \gls{cot}, only a short sensing (i.e., Cat 2 \gls{lbt}) is needed at the responding devices. Otherwise, if the gap is longer than $25$ $\mu$s, regular \gls{lbt} (i.e., Cat 4 \gls{lbt} for data) has to be done at responding devices.
Besides, differently to \gls{laa} that supported a single \gls{dl}/\gls{ul} switching point within the \gls{cot}, \gls{nru} supports multiple \gls{dl}/\gls{ul} switching points within the \gls{cot}~\cite[Sec. 7.6.2]{R1-180xxxxb}.

\section{From NR to NR-U}
\label{sec:NR}

\begin{table*}[!t]
\small
\centering
\begin{tabular}{|m{3.7cm}||m{4.3cm}|m{2.5cm}|m{2.5cm}|m{2.5cm}|}
\hline
& \textbf{NR-U}  & \textbf{LAA} & \textbf{LTE-U} & \textbf{MulteFire} \\ \hline \hline
Deployment scenario     & carrier aggregation, dual connectivity (NR-NR, LTE-NR), standalone, DL-UL & carrier aggregation & carrier aggregation & standalone \\ \hline
Operational bands     & 2.4, 3.5, 5, 6, 37, 60 GHz & 5 GHz & 5 GHz & 5 GHz \\ \hline
Duplexing mode     & FDD, semi-static TDD, \quad \quad dynamic TDD & FDD (LAA), semi-static TDD (eLAA) & FDD & semi-static TDD \\ \hline
Channel access scheme   & LBT & LBT & duty-cycle & LBT \\ \hline
Type of carrier sense    & omni/dir & omni & - & omni\\ \hline
Dimensions for carrier sense    & time, frequency (channel and bandwidth part), space & time, frequency (channel) & - & time, frequency (channel) \\ \hline 
Scheduling dimensions  & time, frequency, space & time, frequency & time, frequency & time, frequency \\ \hline
Processing delays (described in Section~\ref{sec:sched})    & 1 slot: 1, 0.5, 0.25, 0.125 ms (numerology-dependent) & 1 subframe: 1 ms & 1 subframe: 1 ms & 1 subframe: 1 ms \\ \hline 
Time-domain resource allocation granularity   & 1 OFDM symbol: 0.066, 0.033, 0.017, 0.008 ms & 1 subframe: 1 ms & 1 subframe: 1 ms & 1 subframe:1 ms \\ \hline 
Frequency-domain resource allocation granularity   & 1 \gls{rb}: 180, 260, 720, 1440 kHz (numerology-dependent) & 1 \gls{rb}: 180 kHz & 1 \gls{rb}: 180 kHz & 1 \gls{rb}: 180 kHz \\ \hline 
\end{tabular}
\caption{Comparison of NR-U and the different variants of LTE in unlicensed spectrum.}
\label{table:LTEuNRu}
\end{table*}

The \gls{nru} system should be flexible enough not only to support the different layout and deployment scenarios shown in Fig.~\ref{fig_topologies} and Fig. ~\ref{fig_scenarios} but also to follow region- and band-specific regulatory requirements (e.g., \gls{lbt}, see Section~\ref{sec:regulation}.B) to gracefully coexist with other users of the unlicensed spectrum (Wi-Fi, \gls{wigig}, radar). 
\gls{nr} has already paved the way for a fully flexible and configurable technology~\cite{TS38300}.
In particular, \gls{nr} design is highly flexible to:
\begin{itemize}
    \item support a wide range of use cases (e.g., \gls{embb}, \gls{mmtc}, \gls{urllc}, and \gls{ev2x})~\cite{TR38913},
    \item operate in a wide range of carrier frequencies (sub 6 GHz and \gls{mmwave} bands\footnote{NR in Rel-15 has been designed for up to 52.6 GHz frequencies. The frequencies above 52.6 GHz, including the unlicensed spectrum in the 60 GHz band, are expected to be part of future releases.}) with different channel bandwidths,
    \item enable different deployment options (in terms of inter-site distance, number of antennas, beamforming structures), and
    \item address a variety of architectures (non-centralized, centralized, co-sited with E-UTRA, and shared \gls{ran}).
\end{itemize}

Some of the key \gls{nr} features that enable such a flexible and configurable \gls{rat} are:
\begin{itemize}
\item a flexible \gls{ofdm} system with multiple numerologies support~\cite[Sec. 5.1]{TS38300},~\cite{zaidi:16, 5GLENA}, 
\item configurable frame and slot structures that allow fast \gls{dl}-\gls{ul} switch for bidirectional transmissions~\cite[Sec. 4.3.2]{TS38211},~\cite{qualcomm:15b}, 
\item a mini-slot-based transmission which, for the unlicensed bands, may also provide an efficient way to reduce the latency from \gls{cca} end to the start of the \gls{nru} transmission~\cite{R1-1708121,R1-1803678}, 
\item the definition of bandwidth parts and bandwidth adaptation for energy-saving purposes as well as to multiplex services with different \gls{qos} requirements~\cite[Sec. 6.10]{TS38300},~\cite{lagen:18c, biljana:18}, 
\item support for beam management procedures (including beam determination, measurement, reporting, and sweeping) at both sub 6 GHz and mmWave bands~\cite[Sec. 8.2.1.6.1]{TR38912},~\cite{giordani:19,R1-1612345,R1-1702604},
\item new dynamic \gls{nr} scheduling timing parameters~\cite{TS38213,TS38214} to flexibly govern the communication timings between \gls{gnb}s and \gls{ue}s,
and which notably reduce the high processing delays in \gls{lte}. 
\end{itemize}

In terms of NR operation in unlicensed bands, we compare it with the different variants of \gls{lte} in unlicensed spectrum, i.e., \gls{laa}, \gls{lteu}, and MulteFire, in Table~\ref{table:LTEuNRu}. 
Thanks to the flexibility inherited from \gls{nr}, the \gls{nru} system has great potential to perform well in coexistence scenarios.
As compared to \gls{lte} in unlicensed spectrum, in \gls{nru}, we may expect: 1) a lower interference generation owing to the beam-based transmissions that allow exploiting the spatial domain, and 2) a lower latency thanks to the reduced processing times as well as the better scheduling time-resource granularity provided by the \gls{nr} numerologies.

The designs of \gls{laa} and MulteFire technologies have considered the worldwide regulatory requirements of the 5 GHz band through enhancements over \gls{lte}.
For \gls{nru}, further flexibility is needed to meet the worldwide regulatory requirements of multiple operational bands, as well as to provide support for them under beam-based transmissions.
Some of the design principles that need to be rethought in beam-based \gls{nru} are~\cite{R1-180xxxx}: 
\begin{itemize}
\item the channel access procedure, 
\item the \gls{cot} structure, 
\item the initial access procedure, 
\item the \gls{harq} procedure, and
\item the \gls{mac} scheduling scheme. 
\end{itemize}

As previously highlighted, traditional \gls{lbt} might be insufficient under beam-based transmissions. As such, new regulation-compliant and distributed channel access procedures are needed. As far as the \gls{cot} structure is considered, \gls{nr} inherently includes a very flexible design due to the multiple numerologies support, but it still can be optimized for unlicensed-based access in \gls{tdd} systems to meet the \gls{mcot} limit while reducing the access delay and enabling fast \gls{dl}-\gls{ul} responses when needed. The initial access and \gls{harq} procedures that have been adopted in \gls{nr} can be reused for \gls{nru}. However, some initial access principles need to be rethought to meet the regulatory requirements (e.g., \gls{ocb}). Moreover, in case of standalone operation in unlicensed spectrum, the \gls{harq} and initial access procedures need to be improved to mitigate the negative impact that \gls{lbt} could have on the latency performance. 

In the next sections, we highlight the problems, review the available solutions, and propose new potential solutions, for each of these \gls{nru} procedures.
We would like to recall that all the procedures 
are susceptible to be standardized.

\section{Channel Access Procedures for \gls{nru}}
\label{sec:channelaccess}
\gls{nru} is required to ensure fair coexistence with other incumbent \gls{rat}s according to the regulatory requirements in the corresponding bands. 
An appropriate channel access design, including \gls{lbt}, is the key to allow a fair coexistence in all the \gls{nru} deployment scenarios shown in Fig.~\ref{fig_scenarios} (carrier aggregation, dual connectivity, and standalone), even when not mandated by the regulation~\cite[Sec. 7.6.4]{R1-180xxxxb},~\cite[Sec. 7.6.4]{R1-180xxxx}. 
The \gls{lbt} aspects that need to be designed and/or improved for beam-based \gls{nru} beyond \gls{lbt} mechanisms in \gls{laa} and MulteFire, are:
\begin{itemize}
\item \textit{\gls{lbt} for beam-based transmissions}: \gls{lbt} is a spectrum sharing mechanism that works across different \gls{rat}s. As explained in Section~\ref{sec:back}, it suffers from the hidden node and exposed node problems, which become even more likely and accentuated in case of beam-based transmissions. When an omnidirectional antenna pattern is used for carrier sense while a directional antenna pattern is used for (beam-based) transmission (as it happens in WiGig), there is a higher chance of a node being exposed. If the direction of the communication is known, directional carrier sense may help in certain situations but it may also lead to hidden node problems. 
In this line, as highlighted by 3GPP, effects of the directivity of the carrier sense, for beam-based \gls{nru}, should be studied thoroughly and improved to maximize the system performance~\cite{R1-1806761,R1-1719841}. 
\item \textit{Receiver-assisted \gls{lbt} for beam-based transmissions}: \gls{lbt} has been widely adopted in LAA and MulteFire. However, as introduced in Section~\ref{sec:back}, in case of beam-based transmissions, there are interference situations that can no longer be detected with carrier sense at the transmitting node (\gls{gnb}) because listening to the channel at the transmitter may not detect activity near to the receivers. The receivers are in a better position to assess potential interference, and thus 
the assistance from the \gls{ue} to \gls{gnb} can help to better manage interference. Therefore, as agreed among 3GPP members, interference mitigation schemes that utilize information from the \gls{ue} need to be considered for beam-based \gls{nru}~\cite{R1-1806761,R1-1804870}. 
\item \textit{Intra-\gls{rat} tight frequency reuse}: Modern cellular networks in licensed spectrum employ full frequency reuse along with interference management techniques to mitigate inter-cell interference. \gls{nru} channel access procedures could adopt similar principles within the same \gls{rat}, or at least within nodes of the same \gls{rat} that are deployed by the same operator. However, as \gls{lbt} operation based solely on \gls{ed} is uncoordinated inherently, it results in unnecessary blocking among different nodes of the same \gls{rat}, and thus it reduces the spatial reuse and efficiency as compared to full frequency reuse. Accordingly, new frequency reuse methods are needed to avoid \gls{lbt} blocking within \gls{nru} devices of the same operator, or among devices of different operators if coordination among them is permitted, as highlighted by 3GPP~\cite{R1-1806548,R1-1719841}.

\item \textit{\gls{cws} adjustment for beam-based transmissions}: The \gls{cws} is an \gls{lbt} parameter that controls the backoff period after collisions for Cat 4 \gls{lbt}, i.e., the \gls{lbt} category used for data transmissions (see Section~\ref{sec:NRU}.B). LAA-based technologies update the maximum \gls{cws} based on \gls{harq} feedback, and in particular based on the percentage of \gls{nack}s received. This procedure has some drawbacks, as \gls{nack}s do not necessarily reflect collisions and introduce delays into the \gls{cws} update procedure. Moreover, under beam-based transmission, the directionality also makes that some collisions may not be related to the transmit beam for which the \gls{cws} is being updated, e.g., collisions due to interference coming from other directions. As such, from the authors' point of view, new procedures for \gls{cws} adjustment under beam-based transmissions should be defined for \gls{nru}. 
\end{itemize}

Further in this section, we review the above challenges in more detail and discuss solutions to each of them.

\subsection{LBT for Beam-based Transmissions}
\label{sec:LBT}
Two \gls{lbt} sensing approaches are envisioned for \gls{nru} to ensure a fair multi-\gls{rat} coexistence in unlicensed bands with beam-based transmissions: \textbf{\gls{omnilbt}} and \textbf{\gls{dirlbt}}~\cite{R1-1713785}. \gls{omnilbt} senses omnidirectionally, while \gls{dirlbt} senses in a directional manner within the transmit beam towards the intended receiver. {Wi-Fi} and \gls{wigig} use \gls{omnilbt}.

\begin{figure*}[!t]
	\centering
	\includegraphics[width=0.93\textwidth]{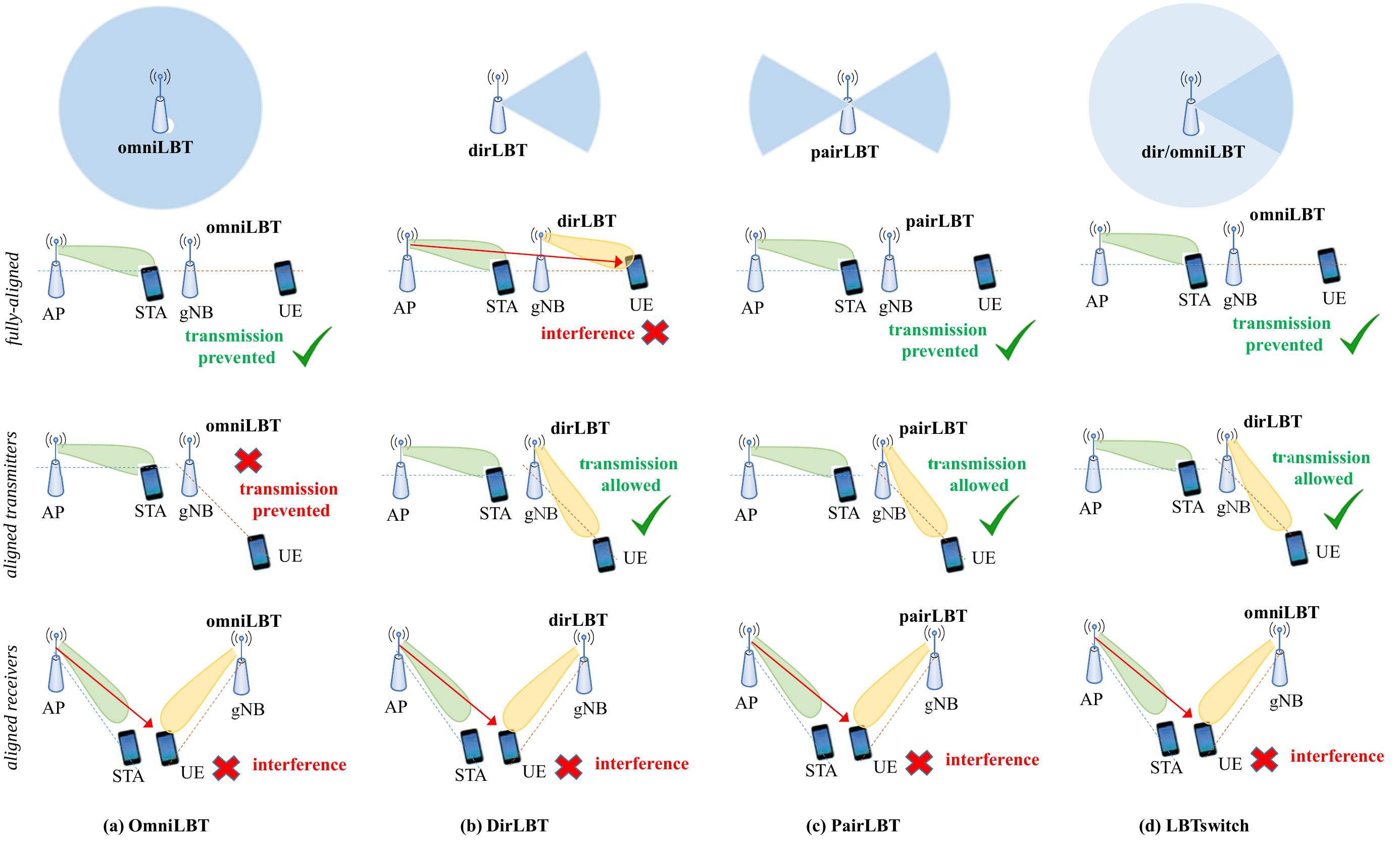}
	\caption{Behavior of (a) \gls{omnilbt}, (b) \gls{dirlbt}, (c) \gls{pairlbt}, and (d) \gls{lbtswitch} techniques, assuming beam-based transmissions and that LBT is implemented at gNB during an on-going AP-to-STA transmission, for fully-aligned (top), aligned transmitters (middle), and aligned receivers (bottom) deployment configurations.}
	\label{fig_LBT}
\end{figure*}

Under directional transmissions, \gls{omnilbt} causes overprotection because a transmission is prevented even if a signal is detected from a direction that may not create harmful interference for the intended receiver. It is an exposed node problem, as shown in Fig.~\ref{fig_LBT}.(a)-middle, for \gls{gnb}-\gls{ue}, which could have reused the spectrum but have been prevented by \gls{omnilbt} at \gls{gnb}. \gls{omnilbt} is only correct when transmissions are aligned in space, see Fig.~\ref{fig_LBT}.(a)-top. In contrast, \gls{dirlbt} does not create overprotection because it only senses the spatial direction in which the transmission will be carried out (see Fig.~\ref{fig_LBT}.(b)-middle). However, in \gls{dirlbt}, on-going nearby transmissions might not be detected, and directional hidden node problems may cause interference as shown in Fig.~\ref{fig_LBT}.(b)-top, because the transmission of \gls{ap} lies within the antenna boresight of the \gls{ue}. The above results in an \gls{omnilbt} that is overprotective and prevents spatial reuse, and a \gls{dirlbt} that enables spatial reuse with some hidden node problems. 

\begin{table*}[!t]
\small
\centering
\begin{tabular}{|m{5cm}||m{2.6cm}|m{2.6cm}|m{2.6cm}|m{2.6cm}|}
\hline
& \textbf{\gls{omnilbt}}  & \textbf{\gls{dirlbt}} & \textbf{\gls{pairlbt}} & \textbf{\gls{lbtswitch}}\\ \hline \hline
Type of carrier sense    & omnidirectional & directional & directional & omni/directional \\ \hline
Fixed or dynamic type of carrier sense  & fixed    & fixed & fixed & dynamic \\ \hline
UE-dependent carrier sense      & no  & yes & yes & yes \\ \hline
Number of carrier sense stages  & 1  & 1 & 2 or more & 1 \\ \hline
Information from UE             &  -  &  - & optional at sync, to optimize \gls{pairlbt} parameters & on-line, to switch from \gls{omnilbt} to \gls{dirlbt}, and reverse \\ \hline
\end{tabular}
\caption{Comparison of channel access procedures that use carrier sense at gNB side.}
\label{table:LBT_comparison}
\end{table*}

To properly address the \gls{omnilbt}/\gls{dirlbt} trade-off, in~\cite{lagen:18b} a distributed solution is proposed, called \textbf{\gls{pairlbt}}. The key idea of \gls{pairlbt} is to perform directional sensing in paired directions, i.e., in the transmitting direction (which is equivalent to perform legacy \gls{dirlbt}) and its opposite direction(s). The opposite directions can denote a single direction or a set of directions depending on whether the beams for carrier sense are either reconfigurable or predefined based on a set of previously configured beams, respectively. 
In this line, in~\cite{lagen:18b}, analytic expressions are derived to optimize the parameters (beam shape and \gls{ed} threshold) for \gls{lbt} in the opposite direction(s) with the objective of reducing hidden node problems. Additional extensions to the \gls{pairlbt} are also proposed, which use the sensed power during the sensing phase in the opposite direction(s) to properly adjust the transmit/receive strategy. Fig.~\ref{fig_LBT}.(c) shows how the \gls{omnilbt}/\gls{dirlbt} trade-offs are addressed by \gls{pairlbt}. 
It is shown in~\cite{lagen:18b} through simulations that the \gls{pairlbt} solution allows improving the ability to perform carrier sense by avoiding hidden node problems, which appear under \gls{dirlbt}, and by stimulating spatial reuse, which is prevented under \gls{omnilbt} (see Fig.~\ref{fig_LBT}). All in all, \gls{pairlbt} is a simple and fully distributed technique that ensures a fair indoor coexistence of different \gls{rat}s in unlicensed spectrum, and which can be properly adjusted to the network density and beamwidth configurations by optimizing the \gls{lbt} parameters. Note, however, that the procedure, as defined in~\cite{lagen:18b}, applies only to indoor scenarios (i.e., indoor \gls{mmwave} shown in Fig.~\ref{fig_topologies}.(b)), since for outdoor scenarios a new dimension (the height) should be added to the definition and optimization.

Results in~\cite{lagen:18b} also demonstrate the trade-off between \gls{omnilbt} and \gls{dirlbt}. It is shown that for low network densities, \gls{dirlbt}  performs significantly better than \gls{omnilbt}, while for high network density, \gls{omnilbt} is a good technique. The trade-off is also observed based on the beamwidth configuration (narrow versus wide beams). 
Based on that, another solution to deal with the \gls{omnilbt}/\gls{dirlbt} trade-off is to implement an \textbf{\gls{lbtswitch}} scheme~\cite{lagen:18d}. This scheme basically switches the type of carrier sense between omnidirectional and directional, based on the beamwidth configuration and density of neighboring nodes. 
Moreover, a dynamic switching method can also be implemented, where switching from \gls{dirlbt} to \gls{omnilbt} could be done based on indications like \gls{harq} feedback, \gls{ue} measurements, etc., to detect an excess of hidden node situations. To switch from \gls{omnilbt} to \gls{dirlbt}, a new procedure to measure the overprotective level of \gls{omnilbt} (i.e., an excess of exposed node situations) should be introduced, as detailed in~\cite{lagen:18d}.

The \gls{omnilbt}-\gls{dirlbt} trade-off, as well as how the \gls{pairlbt} and \gls{lbtswitch} procedures address the trade-off, is shown in Fig.~\ref{fig_LBT} for three different deployment configurations in a \gls{dl} scenario with two pairs (gNB-UE and AP-STA): 
\begin{itemize}
    \item \textit{top}: fully-aligned (i.e., AP, gNB, STA, and UE are aligned in the same spatial line),
    \item \textit{middle}: aligned transmitters (i.e., gNB is in the coverage area of AP), 
    \item \textit{bottom}: aligned receivers (i.e., UE is in the coverage area of AP).
\end{itemize}

\noindent For each configuration, we illustrate the behavior with different gNB channel access procedures and what happens when the sensing strategy fails (e.g., interference occurs or transmission is unnecessarily prevented). In the case of \gls{lbtswitch} technique, we depict the sensing strategy (\gls{dirlbt}, \gls{omnilbt}) that the gNB would use on each of the deployment configurations. The correct gNB behavior in each deployment configuration is: \textit{transmission prevented} for the fully-aligned configuration (which occurs with \gls{omnilbt}, \gls{pairlbt}, \gls{lbtswitch}), \textit{transmission allowed} for aligned transmitters configuration (which occurs with \gls{dirlbt}, \gls{pairlbt}, \gls{lbtswitch}), and \textit{transmission prevented} for aligned receivers configuration (which is not achieved with any of the methods).

In Table~\ref{table:LBT_comparison}, we provide a summary of the requirements of each LBT-based strategy to illustrate the differences in the implementation complexity. 
Note that \gls{omnilbt}, \gls{dirlbt}, \gls{pairlbt} are distributed procedures that can be implemented without UE's assistance,
while \gls{lbtswitch} is also distributed but requires information from the UE to properly adapt the type of carrier sense based on the UE's observation (see Fig.~\ref{fig_LBT}.(d)). DirLBT, \gls{pairlbt}, and \gls{lbtswitch} require knowledge of the intended beam's direction (towards UE) to perform the carrier sense, while \gls{omnilbt} does not. Indeed, \gls{pairlbt} needs at least two sensing stages before every channel access, which could increase the overhead for the sensing in case that a single radio-frequency chain could be used at a time. 

Any of the LBT schemes discussed so far cannot properly address the case of aligned receivers configuration (see Fig.~\ref{fig_LBT}-bottom). In this configuration, if the AP is transmitting towards the STA with its transmit beam (green beam) and, then, the gNB wants to access the channel to serve UE by performing LBT (either \gls{dirlbt}, \gls{omnilbt}, \gls{pairlbt}, or \gls{lbtswitch}), the gNB will sense the channel as idle. This enables the gNB to proceed with directional data transmission towards UE (yellow beam), which will generate interference onto the STA, as well as UE will receive interference from the AP. In the next section, we discuss the receiver-assisted LBT, which can help to prevent the transmission in this configuration.

\begin{figure*}[!t]
	\centering
	\includegraphics[width=0.67\textwidth]{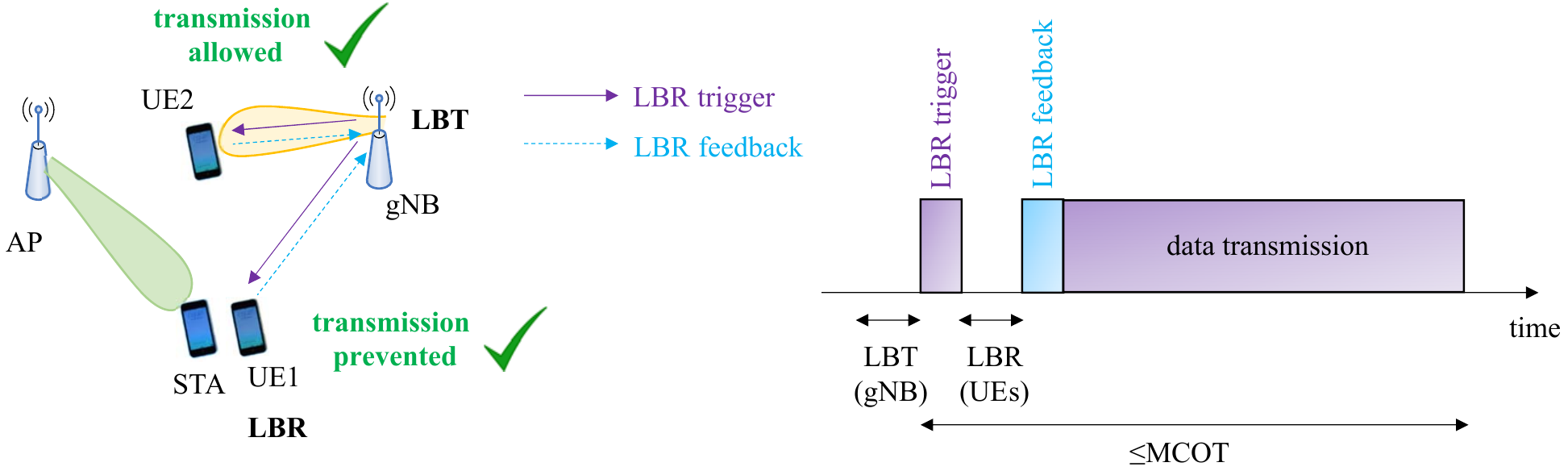}
	\caption{Receiver-assisted LBT procedure to solve the incorrectness of sensing at gNB side under beam-based transmissions, through triggering carrier sense at the receivers (UE1 and UE2). The LBR trigger and LBR feedback messages can be sent over a licensed (carrier aggregation and dual connectivity scenarios) or unlicensed carrier (standalone scenario), while carrier sense (LBR) is performed in the unlicensed band.}      
	\label{fig_lbr}
\end{figure*}

\subsection{Receiver-Assisted LBT for Beam-based Transmissions}
\label{sec:recLBT}
As shown before, there are situations (e.g., aligned receivers) in which on-going nearby beam-based transmissions cannot be detected at the gNB through any of the LBT-based schemes, thus causing hidden node problems (see Fig.~\ref{fig_LBT}-bottom). In these cases, it is the receiver (UE) which has useful information that can be properly exploited for a successful, fair, and friendly channel access in unlicensed bands with beam-based transmissions.

To address these situations, in~\cite[Sec. 8.2.2]{D41mmMagic}, a  Listen-After-Talk (\textbf{\gls{lat}}) technique based on message exchange was proposed. \gls{lat} adopts the opposite logic as compared to \gls{lbt}, in which the default mode for a transmitter is `to send data' and data is not sent only when it is confirmed that the channel is occupied by interfering transmissions. That is, the transmitter transmits when data packets arrive and then, in case that a collision is detected by the receiver, coordination signaling is used to avoid future collisions. Therefore, \gls{lat} considers involving the receiver to sense the channel directly. 
However, \gls{lat} does not use \gls{lbt}, and so it is not compliant with the regulatory requirements in the unlicensed spectrum at 5 GHz and 60 GHz bands in some regions~\cite{ETSI302567,ETSI301893}. Accordingly, it is a potential approach for the USA and China, as well as for the shared bands without the \gls{lbt} requirement, but not for Europe and Japan in 5 GHz and 60 GHz bands.

Wi-Fi and \gls{wigig} use an optional \textbf{RTS/CTS} mechanism to reduce intra-RAT collisions caused by hidden node problems. This mechanism involves physical carrier sense and virtual carrier sense but only solves intra-RAT interference problems, as IEEE 802.11 messages are not decodable by \gls{nr} devices. Note that RTS/CTS protocol is not currently adopted in LAA and MulteFire technologies. However, from the authors' point of view, it may be worth reconsidering it for \gls{nru} to deal with intra-RAT problems since the hidden node and exposed node problems become more severe under beam-based transmissions. 

Other potential solution, which is only based on the physical carrier sense of RTS/CTS, is the Listen-Before-Receive (\textbf{\gls{lbr}})~\cite{lagen:18}. According to this mechanism, the \gls{gnb} triggers the \gls{ue} to perform carrier sense, and only if the \gls{ue} responds, the \gls{gnb} can initiate the transmission. Carrier sense is used before sending the trigger and feedback messages over the unlicensed carrier, thus addressing the \gls{nru} standalone scenario. The solution is illustrated in Fig.~\ref{fig_lbr}, where the messages are referred to as LBR trigger and LBR feedback. In~\cite{lagen:18}, it is also shown how to implement LBR to complement \gls{lbt} in \gls{nr} by exploiting the \gls{nr} flexible slot structure. 
Depending on the omnidirectional/directional sensing that is performed at the \gls{gnb} (\gls{dirlbt}/\gls{omnilbt}) and at the \gls{ue} (dirLBR/omniLBR), different \gls{lbt}-LBR combinations may arise. Among all of them, it is found that \gls{dirlbt}-dirLBR is the best technique and provides significant enhancements in interference management as compared to transmitter-only based sensing approaches~\cite{lagen:18}. 

In line with the \gls{lbr} proposal, some solutions suggest sending the LBR trigger and LBR feedback (see Fig.~\ref{fig_lbr}) over the licensed carrier. This is the case of the so-called \textbf{closed-loop \gls{lbt}}, introduced in~\cite{R1-1802611}, which is useful for the carrier aggregation and dual connectivity scenarios. This way, by utilizing the licensed carrier, closed-loop \gls{lbt} procedure can become more robust to channel availability uncertainties of the unlicensed spectrum, thus resulting in lower latency.

\begin{figure*}[!t]
	\centering
	\hspace{2cm} \includegraphics[width=0.75\textwidth]{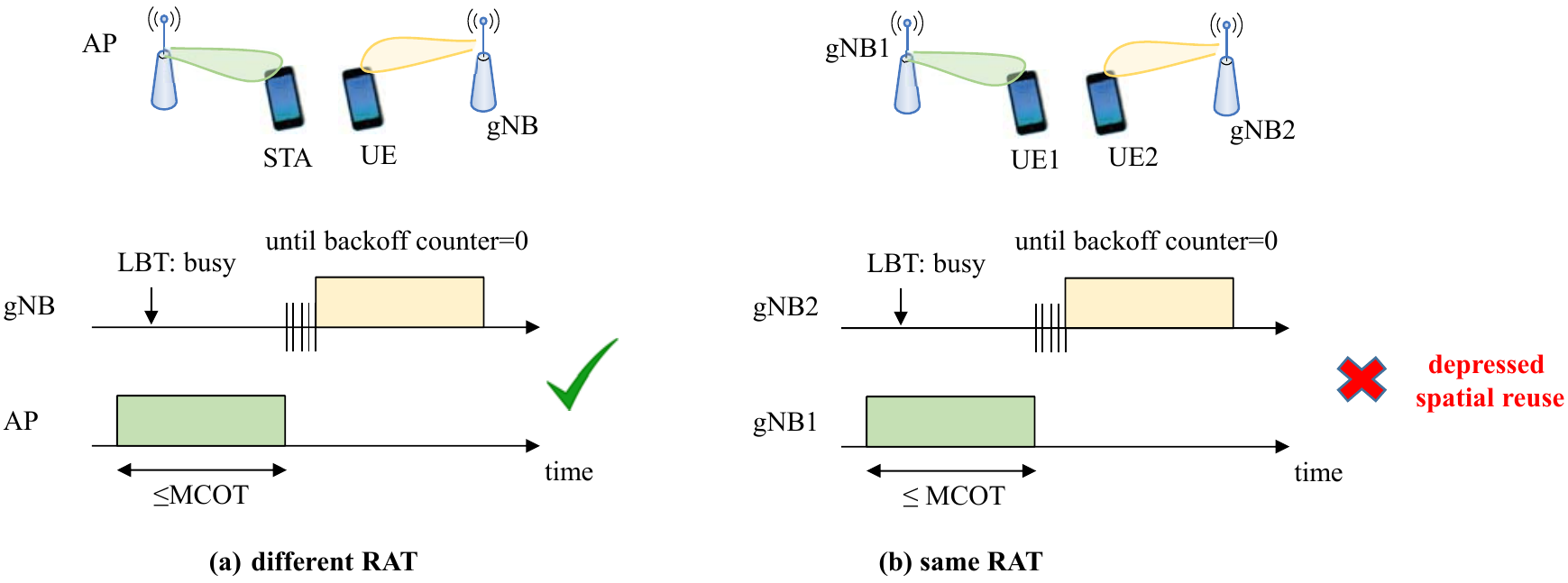}
	\caption{\gls{lbt} blocking for (a) nodes of different RATs and (b) nodes of the same RAT.}
	\label{fig_LBTcoord}
\end{figure*}

\begin{figure*}[!t]
	\centering
	\includegraphics[width=0.74\textwidth]{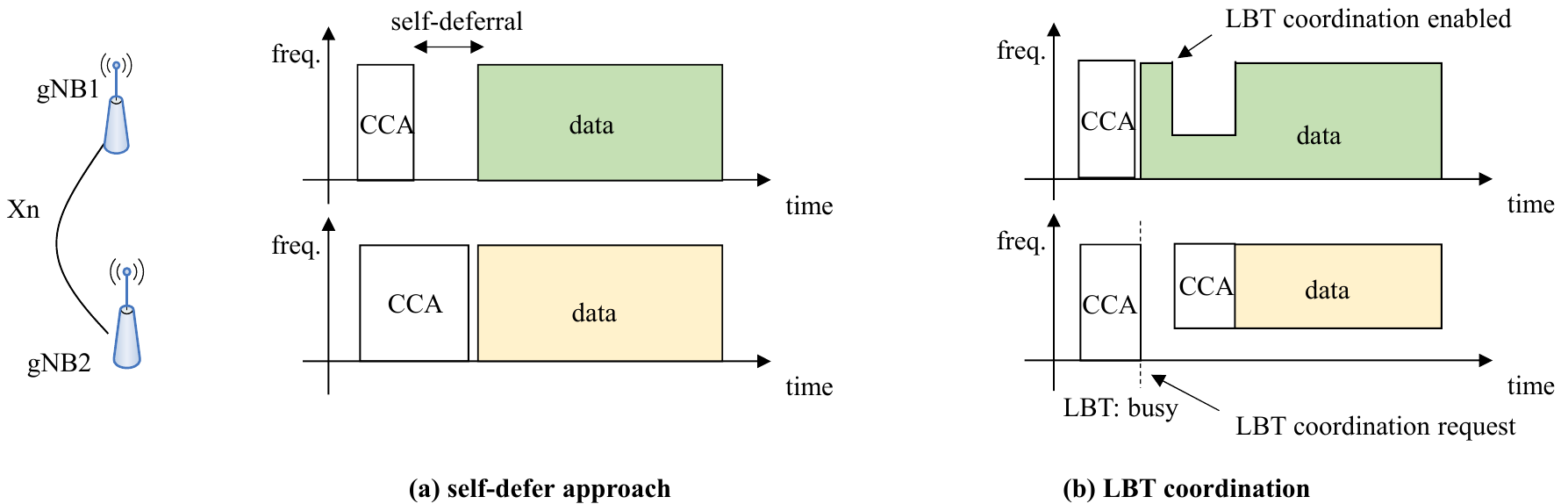} \hspace{1.4cm} 
	\caption{Solutions to avoid \gls{lbt} blocking among nodes of the same RAT/operator, through (a) self-defer approach or (b) \gls{lbt} coordination in frequency-domain.}
	\label{fig_LBTcoord2}
\end{figure*}

RTS/CTS, \gls{lbr}, and closed-loop \gls{lbt} solutions are generally referred to as \textbf{receiver-assisted \gls{lbt}}, as illustrated in Fig.~\ref{fig_lbr}. It was agreed by 3GPP to analyze whether receiver-assisted \gls{lbt} approaches, as well as on-demand receiver-assisted \gls{lbt}, enable enhancing \gls{nru} performance beyond the baseline \gls{lbt} mechanism~\cite[Sec. 7.6.4]{R1-180xxxx}.
The sensing stages for receiver-assisted \gls{lbt} procedure (i.e., \gls{lbt} at gNB and \gls{lbr} at UE) can use any of the sensing strategies that we discussed in Section~\ref{sec:LBT} (directional, omnidirectional, paired, or switching),
so that multiple \gls{lbt}-\gls{lbr} combinations can be formed. In Section~\ref{sec:eval}, we evaluate and compare different \gls{lbt}-\gls{lbr} techniques.

The efficiency of receiver-assisted \gls{lbt} is achieved at the cost of additional message exchange between gNB and UE before every channel access (see Fig.~\ref{fig_lbr}). For different \gls{nr} numerologies\footnote{Each numerology in \gls{nr} specifies a \gls{scs} and a slot length, therefore, it influences the \gls{dl}-\gls{ul} handshake timings, see~\cite[Sec. 5.1]{TS38300}.}, the overhead to implement receiver-assisted \gls{lbt} can be quantified in terms of percentage of the \gls{mcot} ($9$ ms as per ETSI for the 60 GHz band~\cite{ETSI302567}) that is used to perform the message exchange and sensing at UE side before every channel access. If we assume that one slot is required to perform a complete message handshake, which includes \gls{lbr} trigger transmission, sensing at UE side, and LBR feedback transmission, the percentage of the \gls{mcot} used for the handshake will be $11.11\%$ (\gls{scs}=$15$ kHz), $5.55\%$ (\gls{scs}=$30$ kHz), $2.77\%$ (\gls{scs}=$60$ kHz), $1.38\%$ (\gls{scs}=$120$ kHz), $0.69\%$ (\gls{scs}=$240$ kHz). This reflects the penalty in the spectral efficiency of the NR-U system.
It is observed that for high numerologies, (\gls{scs}=$60$, $120$, $240$ kHz), i.e., the ones that are used at mmWave bands, the overhead is below $3\%$.

\subsection{Intra-RAT tight Frequency Reuse}
Apart from the \gls{lbt} sensing strategies analyzed in the previous sections, another problem that may arise due to the uncoordinated \gls{lbt} among different nodes of the same \gls{rat} is unnecessary blocking of transmissions, which leads to degradation in spatial reuse.
As previously described, cellular networks have been appropriately designed to allow full frequency reuse since they have effective interference management techniques (e.g., adaptive rate control, power control, coordinated multi-point (CoMP), enhanced inter-cell interference coordination (eICIC)) to mitigate inter-cell interference within the nodes of a single \gls{rat} (e.g., \gls{nr} from a specific operator). Let us note that the transmit coordination methods, e.g., CoMP and eICIC, basically coordinate the data transmissions, which in case of NR-U occur in the unlicensed band once the channel access is obtained. Therefore, there is no need to block a transmission through \gls{lbt} among devices of the same \gls{rat} that can be coordinated for transmission in the unlicensed spectrum. 

An example of the \gls{lbt} blocking is shown in Fig.~\ref{fig_LBTcoord}, for (a) nodes of different \gls{rat}s and (b) nodes of the same \gls{rat}. In Fig.~\ref{fig_LBTcoord}.(a), the \gls{ap} has accessed the channel and then blocks transmission of the \gls{gnb}, since the \gls{gnb} senses the channel as busy with \gls{lbt}. In this case, the \gls{gnb} has to wait for the transmission of the \gls{ap} to finish and its own backoff procedure to access the channel. This behavior is correct. However, in Fig.~\ref{fig_LBTcoord}.(b), \gls{gnb}1 has accessed the channel and is blocking the transmission of \gls{gnb}2 (a node of the same \gls{rat} and operator), which detects the channel as busy. 
In this case, \gls{gnb}2 must defer the transmission, due to unnecessary \gls{lbt} blocking. Therefore, improvements can be done for \gls{nru}.

An alternative solution is presented in~\cite{R1-1719841,R1-1803679}, where a method for joint channel access using \textbf{self-defer} within a group of neighboring \gls{gnb}s/\gls{trp}s of the same operator has been proposed. The group will self-defer its transmission simultaneously after successful \gls{lbt} for joint channel access so that nodes among the group do not block each other. 
The self-defer solution is shown in Fig.~\ref{fig_LBTcoord2}.(a). Therein, once \gls{gnb}1 gets a clear channel, it communicates with the neighboring \gls{gnb}s through the Xn interface\footnote{Xn is an \gls{nr} interface through which the \gls{gnb}s may communicate with one another, similar to the X2 interface in \gls{lte}.}, and if they are performing the \gls{cca} procedure, \gls{gnb}1 would self-defer itself to avoid blocking \gls{gnb}2. \gls{gnb}1 would self-defer until \gls{gnb}2 has completed the \gls{cca} check and backoff procedure. This solution addresses simultaneous accesses. However, it does not resolve the case in which a node has already accessed the channel, and may block neighbor transmissions of the same RAT and/or operator that have not already started the \gls{cca} check. Also, during the self-deferral period, there is a risk that nodes of other RATs and/or operators do occupy the channel.

Another option that we hereby propose is to use \textbf{\gls{lbt} coordination} procedures among neighboring gNBs/TRPs of the same operator. \gls{lbt} coordination consists on coordinating the LBT processes before starting the data transmission. We foresee that \gls{lbt} coordination to finalize the backoff procedure can be either in time- or frequency-domain. A possible procedure for \gls{lbt} coordination in frequency-domain is illustrated in Fig.~\ref{fig_LBTcoord2}.(b). If a \gls{gnb} (\gls{gnb}2 in the figure) is able to detect that the node occupying the channel is a node from its own \gls{rat} and operator (\gls{gnb}1 in the figure), it could send a message over the Xn interface to request \gls{lbt} coordination to \gls{gnb}1. After receiving such request, the \gls{gnb}1 could release part of the channel bandwidth (frequency-domain \gls{lbt} coordination) and/or some slots (time-domain \gls{lbt} coordination), for the \gls{gnb}2 to complete the backoff procedure. The part of the channel bandwidth and/or the slots which will be released, as well as the starting point for transmit coordination, could be communicated through Xn so that both \gls{gnb}s, after the backoff procedure is completed, could start with transmit coordination, thus improving the spatial reuse. 

Note that, in case of time-domain \gls{lbt} coordination, the same problems as in the self-defer approach arise. That is, other nodes may occupy the channel during the request-enabled \gls{lbt} coordination process. Nevertheless, this does not happen in case of frequency-domain \gls{lbt} coordination, since \gls{gnb}1 does not release the full spectrum bandwidth and other \gls{rat}s would still detect the channel as busy. In this case, \textbf{bandwidth part-based \gls{lbt}} is needed, i.e., \gls{gnb}2 should implement \gls{cca} only in the released bandwidth part (as illustrated in Fig.~\ref{fig_LBTcoord2}.(b), second \gls{cca} block for \gls{gnb}2), and then transmit in such bandwidth part. 
To further improve the proposal and facilitate the job of detecting that channel is busy due to a \gls{gnb} of the same RAT/operator, once a \gls{gnb} gets access to the channel, it could inform nearby \gls{gnb}s over the Xn interface.

\subsection{\gls{cws} Adjustment for Beam-based Transmissions}
In \gls{laa} Cat 4 \gls{lbt}, the \gls{cws} is updated based on HARQ feedbacks.
If $80 \%$ or more of \gls{harq} feedbacks of one reference subframe are \gls{nack}, the maximum \gls{cws} is increased~\cite{TR36889}. Otherwise, it is reset. Note that this collision detection technique has some drawbacks. First, 
it is affected by the scheduler policies, e.g.,
collisions from different \gls{ue}s may affect the corrective actions of \gls{lbt} differently since they will depend on how many and which \gls{ue}s are simultaneously allocated in the reference subframe. Second, HARQ does not necessarily reflect collisions, e.g., \gls{nack} may also occur due to a sudden signal blocking. Third, since \gls{harq} is based on soft combining techniques (i.e., incremental redundancy or chase combining), an unsuccessful transmission, due to a collision, may not result in a \gls{nack} in case of successful decoding thanks to the combination of multiple transmissions. And the last, it introduces delays in the \gls{cws} update. Since LAA uses \gls{harq} feedback corresponding to the starting subframe of the most recent transmission burst, it may detect collision after at least $4$ ms, whereas, Wi-Fi detects collisions after $16$ $\mu$s. 

We would like to remark that, at the time of writing, the problems that we highlight next in this section have not been detected so far in the literature and, consequently, no solutions are available. Also, the \gls{cws} adjustment criterion for Cat 4 \gls{lbt} in \gls{nru} has not been defined yet.

\begin{figure}[!t]
	\centering
	\includegraphics[width=0.48\textwidth]{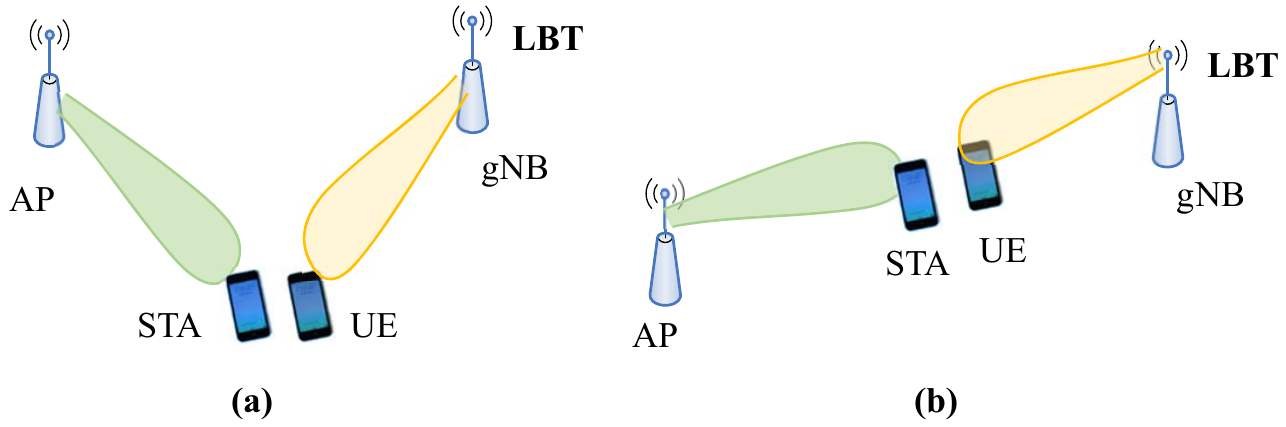}
	\caption{Situations in which \gls{lbt} and the \gls{cws} update based on HARQ feedback: (a) do not make sense, (b) make sense. }
	\label{fig_cws}
\end{figure}

For \gls{nru}, the same issues listed before for \gls{laa} will also appear if \gls{harq} feedback is used for the \gls{cws} update, except that the \gls{harq} feedback delay may be reduced due to the flexible \gls{nr} slot structure. Moreover, for beam-based \gls{nru}, the reported collision by \gls{harq} feedbacks may not be linked to the transmit beam correctly. As already mentioned in previous section, \gls{lbt} (and the extended \gls{cca} procedure) only makes sense if the beams of neighboring \gls{gnb}s are aligned. If the \gls{gnb}s/\gls{ap}s saw each other (as shown in Fig.~\ref{fig_cws}.(b)), they would backoff to each other and so randomize their accesses, taking advantage of the \gls{cws} increase. However, if beams of neighboring nodes are not aligned (see Fig.~\ref{fig_cws}.(a)), the \gls{lbt} is not effective, even if the \gls{cws} is increased. The \gls{gnb}s/\gls{ap}s never enter in the backoff phase, so the access randomization effect is not produced. In particular, in case of Fig.~\ref{fig_cws}.(a) scenario, both the \gls{gnb} and \gls{ap} listen to the channel and find it free, thus, they both access the channel and collide. Then, they increase the \gls{cws}, they listen again, find the channel free and they collide again. Therefore, in those cases where \gls{lbt} does not work properly, it is furthermore counterproductive to increase the \gls{cws} based on \gls{harq} procedure.  

To summarize, we have detected two problems that arise in beam-based transmissions when the \gls{cws} update is based on \gls{harq} feedback:
\begin{itemize}
\item \textit{The lack of correlation between a collision indicated by a \gls{nack} and the transmit beam}: \gls{harq} feedbacks may refer to collisions due to interference coming from another direction, while only collisions generated on the transmit direction line are of interest for the \gls{cws} update.
\item \textit{The inability to enter the backoff phase due to an incorrect sensing phase}: transmitters that do not see each other would never enter the backoff phase to randomize their accesses, although they increase the \gls{cws} based on \gls{harq} feedback.
\end{itemize}

Therefore, for transmitters that do not see each other, it would be beneficial that the \gls{ue} triggers the backoff procedure at its gNB to randomize its \gls{gnb}'s access to the channel since the \gls{ue} is the only one that has the knowledge about interfering nodes. In addition, it would be good to isolate the \gls{cws} update procedure from the \gls{harq} feedback because it does not properly capture the directional (and non-directional) collisions.

To solve these problems, we propose using a \gls{cws} update at the \gls{gnb} that is assisted by the \gls{ue}, i.e., \textbf{receiver-assisted \gls{cws} adjustment}. In particular, by a paired sensing at the \gls{ue}. That is, the \gls{ue} could carry out a paired sensing over the \gls{gnb} transmit beam line (receive direction and opposite direction(s)) and, if the channel is sensed as busy during some period, it could:
\begin{itemize}
\item Trigger backoff at the \gls{gnb} if it is not aligned to the source of interference.
\item Suggest the most appropriate \gls{cws} over the \gls{gnb} transmit beam line, based on, e.g., the percentage of slots sensed as busy during the paired sensing phase.
\end{itemize}

Hence, we suggest updating the \gls{cws} associated to the transmit beam based on statistical paired sensing at the \gls{ue} within the direction of the \gls{gnb} transmit beam.

\section{COT Structure for NR-U}
\label{sec:frame}

\begin{figure*}[!t]
	\centering
	\includegraphics[width=0.65\textwidth]{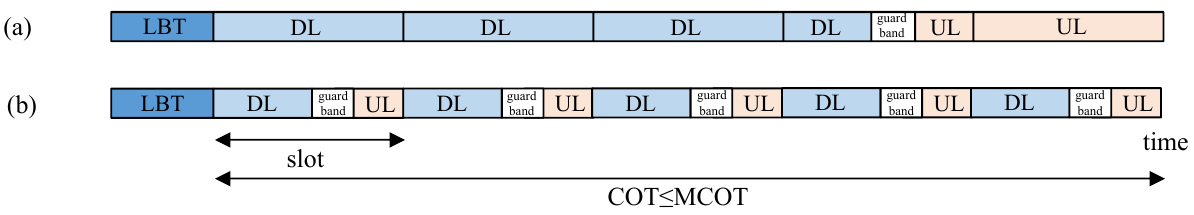}
	\caption{COT structure with (a) single \gls{dl}/\gls{ul} switch, (b) multiple \gls{dl}/\gls{ul} switches.}
	\label{fig_mcot}
\end{figure*}

After a successful \gls{lbt}, a device can access the channel at most for the duration of the \gls{mcot} ($9$ ms in the 60 GHz band). The \gls{nr} frame structure inherently allows \gls{nru} to transmit and receive in a more efficient manner compared to LTE in unlicensed spectrum technologies, thanks to the numerologies, mini-slots, and flexible slot structure~\cite{R1-1804275}. Indeed, the \gls{cot} can be shared between a \gls{gnb} and its \gls{ue}s to achieve a higher spectral efficiency and faster responses under bidirectional transmissions (see details in Section~\ref{sec:NRU}.B). In this section, we review how to define the \gls{dl} and \gls{ul} transmission periods within the \gls{cot}.

There are two options considered in 3GPP to define the structure of the \gls{cot} (as illustrated in Fig.~\ref{fig_mcot}):
\begin{itemize}
\item \gls{cot} with single \gls{dl}/\gls{ul} switch or
\item \gls{cot} with multiple \gls{dl}/\gls{ul} switches.
\end{itemize}

Note that the slot length in \gls{nr} is much shorter than the \gls{mcot}. For example, with \gls{scs}=$120$ kHz, $72$ slots fit within $9$ ms, so that multiple \gls{dl}/\gls{ul} switches could be implemented. 
Recently, support for the multiple \gls{dl}/\gls{ul} switch option within the \gls{cot} has been agreed for \gls{nru}~\cite[Sec. 7.2.1.1]{TR38889}.
Still, as highlighted by 3GPP, the number of switch points per \gls{cot} should be further studied in \gls{nru}~\cite{R1-1803678}.

Both aforementioned options have advantages and disadvantages.
A \textbf{\gls{cot} with a single \gls{dl}/\gls{ul} switch} has the advantages that: \textit{i}) there is a low overhead due to only one guard band (shown in gray color in Fig.~\ref{fig_mcot}) and \textit{ii}) avoids multiple \gls{lbt}s for successive \gls{dl}-\gls{ul} periods (in case the gaps are larger than $16$ $\mu$s and so a new \gls{lbt} has to be done at each gap\footnote{Whether \gls{lbt} before an \gls{ul} transmission that follows a \gls{dl} transmission is needed or not, depends on the gap length, as detailed in Section~\ref{sec:NRU}.B.}). The disadvantages are that: \textit{i}) it increases delays to get the \gls{harq} feedback, and \textit{ii}) the \gls{gnb} would schedule \gls{ul} far away in time, for which the channel may no longer be available in the \gls{ul} direction (in case a new \gls{lbt} has to be performed). Accordingly, this \gls{cot} configuration is suitable for high throughput situations with relaxed latency constraints, e.g., \gls{embb} traffic.

On the other hand, a \textbf{\gls{cot} with multiple \gls{dl}/\gls{ul} switches}: \textit{i}) simplifies the \gls{harq} timings related to \gls{harq} feedback and \textit{ii}) ensures channel availability in \gls{ul} (in case a new \gls{lbt} has to be done), but
\textit{i}) has a high overhead due to multiple guard bands (see Fig.~\ref{fig_mcot}) and \textit{ii}) involves multiple \gls{lbt}s for successive \gls{dl}-\gls{ul} periods at every direction switch (in case the gaps are larger than $16$ $\mu$s). This configuration is then suitable for delay-sensitive traffic, such as \gls{urllc} and \gls{ev2x}, as well as for low-load traffic categories, like \gls{mmtc}. However, it may not be suitable for applications with high throughput requirements (like \gls{embb}) as it provides a lower spectral efficiency due to the existence of multiple guard bands and the potential need for multiple \gls{lbt}s.

Based on the above advantages/disadvantages of each option, from the authors' point of view, it would be appropriate to optimize \gls{ul}/\gls{dl} structure within the \gls{cot} based on knowledge of the traffic status and patterns (e.g., \gls{bsr} and future \gls{bsr} pattern predictions), the throughput/latency requirements of the active data flows, their category type (or 5G \gls{qos} Indicator, 5QI), and the channel status at the \gls{ue}s (percentage of busy and idle slots). The \gls{gnb} could consider the information from all the active flows for the \gls{cot} period. 
In addition to the intrinsic trade-offs, it would be beneficial that the \gls{gnb} notifies the \gls{ue}s the selected \gls{cot} structure preferably at the beginning of the \gls{cot}. 
This would help the \gls{ue}s to prepare for performing \gls{lbt} ahead of time, as well as to anticipate the preparation of any potential transmission in a \gls{pucch} or \gls{pusch} resource.

\section{Initial Access Procedures for NR-U}
\label{sec:initialaccess}
The basic structure of \gls{nr} initial access is similar to
the corresponding functionality of \gls{lte}~\cite{parkvall:17}: $1$) there is a pair of \gls{dl} signals, the \gls{pss} and \gls{sss}, which are used by the \gls{ue} to find, synchronize, and identify a network, 2) there is a \gls{dl} \gls{pbch} that carries a minimum amount of system information, which is transmitted together with the \gls{pss}/\gls{sss}, and 3) there is a four-stage \gls{rach} procedure that starts with the \gls{ul} transmission
of a random access preamble~\cite{liu:18}. In NR, the combination of \gls{pss}/\gls{sss} and \gls{pbch} is referred to as an \gls{ss} block, and such signals are always sent together with the same periodicity. This section reviews the problems and solutions for the key features of the \gls{nru} initial access, which include \gls{ss} block design\footnote{In the \gls{nru} standardization, the \gls{ss} block is referred to as \gls{nru} discovery reference signal~\cite{TR38889}.}, RACH procedure, and paging.

\subsection{\gls{ss} Block Design}
\begin{figure}[!t]
	\centering
	\includegraphics[width=0.48\textwidth]{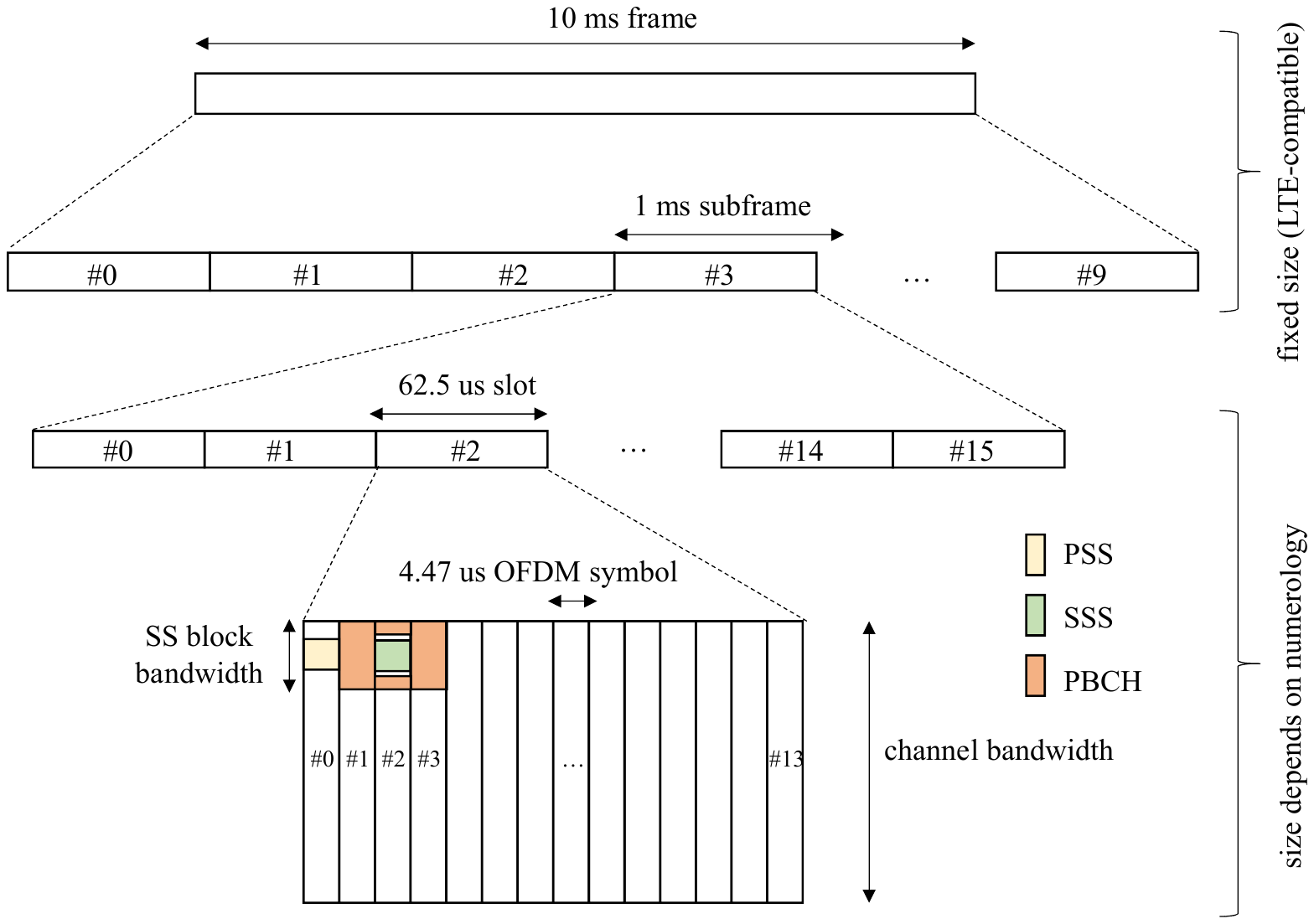}
	\caption{Location of \gls{ss} block in an \gls{nr} frame for SCS=240 kHz.}
	\label{fig_SS}
\end{figure}

\gls{ss} blocks are used in \gls{nr} to enable radio resource management measurements, synchronization, and initial access. Therefore, for \gls{nru} operation, \gls{ss} blocks should always be transmitted in all the deployment scenarios, i.e., carrier aggregation, dual connectivity, or standalone mode (see Section~\ref{sec:NRU}.A).
An \gls{ss} block spans over 240 contiguous subcarriers and 4 contiguous \gls{ofdm} symbols (as shown in Fig.~\ref{fig_SS}). The frequency location is typically not at the center of the \gls{nr} carrier (as in \gls{lte}) but shifted according to a global synchronization raster that depends on the frequency band~\cite[Sec. 5.4.3]{TS38101}. The time locations of the \gls{ss} blocks are determined by SCS and frequency range~\cite[Sec. 4.1]{TS38213}. The maximum transmission bandwidth of an \gls{ss} block has been defined to be [$5, 10, 40, 80$] MHz with [$15, 30, 120, 240$] kHz \gls{scs}, respectively. 
To support beam sweeping and \gls{ss} block repetitions, multiple \gls{ss} blocks from the same \gls{gnb} are organized in time into a burst (called \gls{ss} burst), and multiple \gls{ss} bursts further comprise an \gls{ss} burst set. The periodicity of an \gls{ss} burst set is configurable from the set of \{$5, 10, 20, 40, 80, 160$\} ms (default at $20$ ms), and each \gls{ss} burst set can contain up to $64$ \gls{ss} blocks. For more details on the \gls{pss}, \gls{sss}, and PBCH signals, see~\cite{TS38211}. The synchronization procedure for cell search is detailed in~\cite[Sec. 4.1]{TS38213}, and the time-frequency structure of the \gls{ss} block is shown in~\cite[Sec. 5.2.4]{TS38300}. 

In the following we discuss different challenges that arise in the \gls{ss} block design and transmission principles for \gls{nru}.
Note that \gls{ss} blocks must be sent by the \gls{gnb} even if there is no data, to enable the \gls{ue}s to detect and search cells, be synchronized to the \gls{gnb}, perform beam measurements, implement handover if required, and decode broadcast messages. 

The first problem is related to the transmission of the \gls{ss} blocks and \gls{lbt} requirement. Since \gls{ss} block may be interrupted due to channel occupancy, periodical \gls{ss} block transmission may not be possible~\cite[Sec. 7.6.4.2]{R1-180xxxxb}. 
This can be solved by using a solution adopted in \gls{laa} discovery reference signals, which are transmitted within a periodically occurring time window, and thus increase the chance for signal transmission~\cite{kwon:17}. Additional occasions for \gls{ss} block transmissions, over the legacy periodic \gls{ss} block transmissions, are proposed in~\cite{R1-1803977}. Also, it is shown therein how to enable multiple occasions by reusing the \gls{nr} \gls{ss} block patterns.
In addition, \gls{ss} block patterns may need to be redefined to include the \gls{lbt} resource overhead and enable \gls{ss} block transmissions through multiple beams, as discussed in~\cite{R1-1806761}. The technical contribution in~\cite{R1-1803856} describes solutions to reuse the \gls{nr} \gls{ss} block patterns while leaving enough space for \gls{lbt} as well as for switching antenna weights for beam sweeping in between the different blocks.

The second problem is related to the \gls{ss} block design in 60 GHz band, which occurs due to the \gls{ocb} requirement and the large channel bandwidth (see Section~\ref{sec:regulation}.B). The main problem is that \gls{ss} blocks occupy only a part of the NCB, as shown in Fig.~\ref{fig_SS}. For illustrative purposes, in Fig.~\ref{fig_SS}, the \gls{ss} block starts at the first OFDM symbol and is located at the upper-left corner, although its exact location is defined in~\cite[Sec. 4.1]{TS38213}. If \gls{ss} blocks are multiplexed with data, then the \gls{ocb} requirement may be met. However, if \gls{ss} blocks are not multiplexed with data, then the \gls{ocb} requirement is not met with the current \gls{ss} block design in \gls{nr}. 
Accordingly, to meet the \gls{ocb} requirement defined by ETSI, a new design of \gls{ss} blocks in frequency domain is required for \gls{nru} operation at the 60 GHz band.

In case \gls{ss} blocks are not multiplexed with data, or they are sent with data but do not fulfill the \gls{ocb} requirement, a basic solution is to send dummy non-useful data in frequency-domain to meet the \gls{ocb} requirement. However, this solution is energy-inefficient and does not add any benefit from the \gls{ue} perspective. 
Other solutions that we envision to meet the \gls{ocb} requirement are:
\begin{itemize}
\item Perform \textbf{frequency-domain \gls{ss} block repetitions}, by repeating the \gls{ss} block in multiple frequency locations within the channel bandwidth. This solution uses additional power but enhances the \gls{ue} performance, as it enables receiving the \gls{ss} block with a higher signal-to-noise ratio.
\item Redesign the \textbf{time-frequency structure} of the \gls{pss}/\gls{sss}/\gls{pbch} signals in the \gls{ss} block, by restructuring the signals placement. An example is to use a frequency-domain interlaced mapping for \gls{pss}/\gls{sss}/\gls{pbch} signals so that they span over the required channel bandwidth. This solution allows meeting the \gls{ocb} requirement without incurring additional power consumption.
\end{itemize}

\subsection{RACH Procedure}
The contention-based \gls{rach} procedure in \gls{nr} has four steps~\cite[Sec. 8]{TS38213},~\cite{liu:18}, \textit{step 1}: \gls{ue} transmits a \gls{prach} preamble to \gls{gnb}, \textit{step 2}: \gls{gnb} transmits the Random Access Response (RAR) to \gls{ue} with the \gls{pusch} resource allocation to send message 3, \textit{step 3}: \gls{ue} transmits message 3 over the allocated \gls{pusch} resource, and \textit{step 4}: \gls{gnb} transmits message 4 for contention resolution. 
In \gls{nru}, \gls{rach} procedures are needed and must be improved at least for dual connectivity and standalone deployment scenarios. Carrier sense must be performed at each step of \gls{rach} procedure, which may delay the procedure to complete if the channel is busy at any step. Therefore, high-priority channel access with Cat 2 \gls{lbt} could be preferred for \gls{rach}. Indeed, the use of two-step \gls{rach} procedures would also be of high interest to reduce the initial access delay, as proposed in~\cite{R1-1806762,R1-1803856}, and also identified by \gls{3gpp}~\cite[Sec. 7.6.4.2]{R1-180xxxxb}. Particularly, two-step \gls{rach} procedures will require fewer \gls{lbt}s than the four-step \gls{rach} procedure. Other enhancements may include the increasing transmit opportunities for each message~\cite{R1-1804405}, which is also discussed in the case of SSB transmissions. 

In addition to that, the \gls{prach} preamble format needs to fulfill the regulatory requirement of OCB, which will exclude some of the agreed \gls{nr} \gls{prach} formats. In Rel-14 eLAA~\cite{TS36213}, several types of \gls{prach} waveforms were studied, such as frequency-domain repetition of a licensed band preamble, Demodulation Reference Signals (DMRS) repetition in time domain with frequency-domain interlacing, and frequency-domain interlaced mapping of a licensed band preamble. This study in eLAA may provide a baseline for the design of \gls{nru} \gls{prach} interlace waveforms. 

\subsection{Paging}
Paging is a \gls{rrc} procedure to activate a \gls{ue} that is in idle mode. In the unlicensed context, it is needed at least for dual connectivity and standalone deployment scenarios. A paging cycle is defined to allow \gls{ue}s to wake up and listen at predefined time slots to receive possible paging messages. The paging message is scheduled through \gls{dci} and is transmitted in the associated \gls{pdsch}. 

The uncertainty of channel availability in the unlicensed bands due to \gls{lbt} makes paging \gls{dci} hard to be sent out at predefined time slots.
To solve that, a time interval composed of multiple slots for potential paging message transmission has been proposed in~\cite{US20170230933,WO2017145120}. It provides a \gls{gnb} multiple opportunities (multiple slots) to send the paging \gls{dci} as soon as \gls{lbt} allows. On the other side, \gls{ue} needs to listen for all the possible opportunities. In such solution, the probability of blocking due to channel occupancy is reduced at the cost of a higher energy consumption at \gls{ue}.

Current \gls{nr} specification already supports a paging occasion consisting of multiple slots~\cite[Sec. 9.2.5]{TS38300}, to improve the reliability of the system. Also, \gls{nr} permits the network to transmit a paging message using a different set of transmit beams or repetitions. Thus, the reliability and channel availability issues of paging for \gls{nru} can be assessed by using the already supported time- and spatial- domains for paging in \gls{nr}.

\section{HARQ Procedures for \gls{nru}}
\label{sec:harq}
In \gls{nr}, similar to \gls{lte}, after reception of data, a device has to respond with a \gls{harq} feedback to indicate whether the data transmission was successful or not. The time duration between the initial data transmission, \gls{harq} feedback, and re-transmission, as well as the way the transmitted and re-transmitted data are combined at the receiver for decoding, define the basics of the \gls{harq} procedure. \gls{harq} in \gls{nr} supports asynchronous incremental redundancy both for \gls{dl} and \gls{ul}. In \gls{dl}, the \gls{gnb} provides the \gls{harq} feedback timing configuration to \gls{ue} either dynamically using \gls{dci} or semi-statically using \gls{rrc}. In \gls{ul}, upon reception of the \gls{sr} or \gls{bsr} from \gls{ue}, the \gls{gnb} schedules each \gls{ul} transmission and re-transmission using \gls{dci}.

In \gls{nr}, the following terminologies\footnote{Let us note that, at the time of writing, only K0, K1, and K2 are included in \gls{3gpp} technical specification~\cite{TS38331}. K3 and K4 are not included, although they were mentioned in \gls{3gpp} technical discussions~\cite{R1-1719401}, and are included in this paper to illustrate the whole \gls{harq} time-line.} are defined in terms of scheduling and \gls{harq} time-line~\cite{TS38213,TS38214,TS38331,R1-1719401}:
\begin{itemize}
\item K0: Delay between \gls{dl} allocation (\gls{pdcch}) and corresponding \gls{dl} data (\gls{pdsch}) reception,~\cite[Sec. 5.1.2.1]{TS38214}, 
\item K1: Delay between \gls{dl} data (\gls{pdsch}) reception and corresponding \gls{harq} feedback transmission on \gls{ul} (\gls{pucch}),~\cite[Sec. 9.2.3]{TS38213},
\item K2: Delay between \gls{ul} grant reception in \gls{dl} (\gls{pdcch}) and \gls{ul} data (\gls{pusch}) transmission,~\cite[Sec. 6.1.2.1]{TS38214},
\item K3: Delay between \gls{harq} feedback reception in \gls{ul} (\gls{pucch}) and corresponding re-transmission of data (\gls{pdsch}) on \gls{dl},
\item K4: Delay between \gls{ul} data (\gls{pusch}) reception and corresponding \gls{harq} feedback transmission on \gls{dl} (\gls{pdcch}).
\end{itemize}

\begin{figure}[!t]
	\centering
	\includegraphics[width=0.45\textwidth]{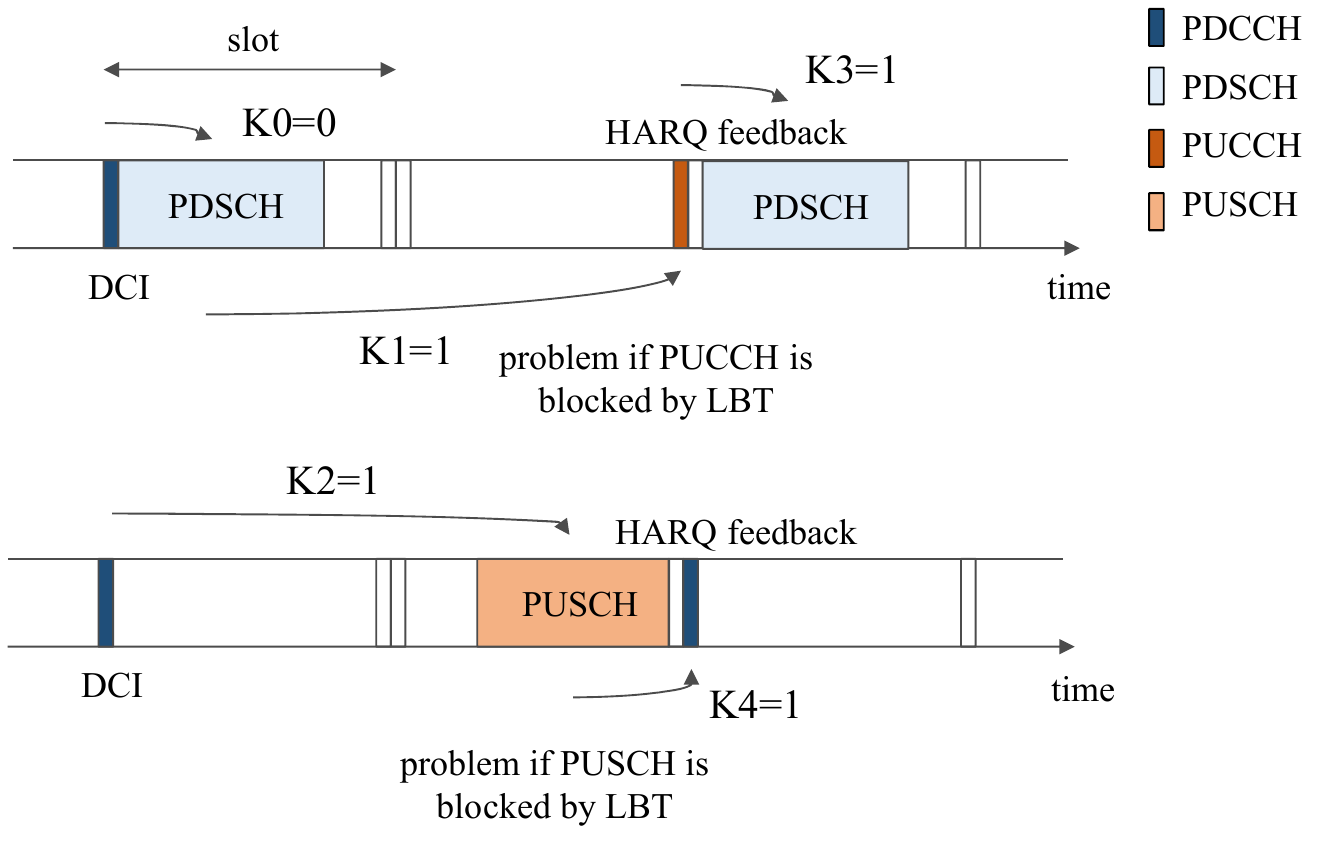}
	\caption{Problems related to scheduling and \gls{harq} due to \gls{lbt}. }
	\label{fig_harq}
\end{figure}

Fig.~\ref{fig_harq} shows an example of \gls{dl} and \gls{ul} data transmissions along with the associated \gls{harq} feedback allocation for K0=0, K1=1, K2=1, K3=1, K4=1 slots. If \gls{pdsch} is sent in slot $n$, \gls{pucch} with \gls{harq} feedback would be sent in slot $n{+}k$, where $k$ is indicated by the field \textit{\gls{pdsch}-to-HARQ-timing-indicator} (provides the value of K1) in the \gls{dci} in \gls{pdcch}. Moreover, \gls{pucch} resources, i.e., physical \gls{rb}s to be used for \gls{harq} feedback are also indicated by \gls{dci} in \gls{pdcch}~\cite[Sec. 9.2.3]{TS38213}. Similarly, in \gls{ul} transmissions, \gls{pusch} resources for \gls{ul} data transmissions and re-transmissions are configured by \gls{dci} in \gls{pdcch}, where the slot timing offset K2 is part of the \textit{Time-domain resource assignment} field in \gls{dci}~\cite[Sec. 6.1.2.1]{TS38214}.

Note that K3 and K4 need to consider the processing times at the \gls{gnb} side, while K1 and K2 have to take into account the \gls{ue} processing times. In \gls{nr}, the \gls{ue} processing time is expressed in terms of symbols, instead of slots (unlike the K parameters), for which the following terminologies are defined:
\begin{itemize}
\item N1: the number of OFDM symbols required for \gls{ue} processing from the end of \gls{pdsch} reception to the earliest possible start of the corresponding \gls{harq} feedback transmission,
\item N2: the number of OFDM symbols required for \gls{ue} processing from the end of \gls{pdcch} containing the \gls{ul} grant reception to the earliest possible start of the corresponding \gls{pusch} transmission.
\end{itemize}
More details and specific values of N1 and N2 for different configurations and numerologies can be found in~\cite{R1-1721515,R1-1719401}.

From the \gls{harq} procedure point of view, two \gls{nr} features are important: the flexible slot structure and the mini-slot-based transmissions. The flexible slot structure may reduce the \gls{harq} delay by allowing the transmission of \gls{harq} feedback in the same slot in which \gls{pdsch} was received (\textbf{self-contained \gls{harq} feedback})~\cite{R1-1806671}, and may enable re-transmissions in the subsequent slot, provided that the processing delays at \gls{ue} and \gls{gnb} are short enough to permit it. The mini-slot-based transmissions provide scheduling support with flexible transmission durations. It also reduces the delay between the time instant when the channel is found idle and the time instant when the transmission can be started. This way it reduces the need of using reservation signals to reserve the channel until the next allowed transmission time instant boundary, which was used in LAA to protect the channel until the next subframe boundary (see~\cite[Sec. 7.2.1.1]{TR36889}).

However, in case of standalone \gls{nru}, there are two important problems associated with the HARQ operation and \gls{lbt} requirement (see Fig.~\ref{fig_harq}):
\begin{itemize}
\item \textit{\gls{ul} data blocking of an \gls{ul} \gls{harq} process}: It may happen that \gls{ul} grant is transmitted through \gls{pdcch} but the corresponding \gls{ul} data in \gls{pusch} is blocked by channel occupancy (even in case of Cat 2 \gls{lbt} within a shared \gls{cot}). In such a case, the \gls{gnb} would assume it as an incorrect reception (even if there was no transmission) and so would proceed to reallocate resources for the \gls{ue} to "re-transmit". The problem is further aggravated in case of multi-slot scheduling, for which multiple slots are assigned for the \gls{ue} to transmit, through a single \gls{ul} grant. 
\item \textit{\gls{harq} feedback blocking of a \gls{dl} \gls{harq} process}: It may happen that \gls{pdcch} and \gls{pdsch} are transmitted but \gls{harq} feedback in \gls{pucch} is blocked by channel occupancy (even in case of Cat 2 \gls{lbt}).
Due to the blocking of \gls{harq} feedback transmissions, \gls{gnb} would assume a it is a \gls{nack} and additional re-transmissions would occur at \gls{gnb}.
The problem is further aggravated in case of multi-slot aggregation, for which multiple \gls{harq} feedbacks of different transport blocks are multiplexed together in a single \gls{pucch} transmission. 
\end{itemize}

The problem of \gls{ul} data blocking of an \gls{ul} \gls{harq} process has already been addressed in eLAA using triggered grant~\cite{TS36213}. The key idea is to use two step grant process instead of one. For an \gls{ul} grant, first a subset of the configuration parameters, for example, \gls{mcs}, \gls{tbs}, and assigned \gls{rb}s are sent, then, at a later point, a short triggered grant is sent on \gls{pdcch} to trigger the corresponding \gls{ul} transmission. The delay to process the triggered grant and to send the \gls{ul} transmission would be minimal at \gls{ue} side, because most of the processing has already been finished based on the configuration parameters sent earlier before the triggered grant. This allows the \gls{ue} to immediately transmit after the triggered grant without \gls{lbt}, given that the \gls{ue} transmission can be done within $16$ $\mu$s from the transmission of triggered grant, within the shared \gls{cot}. This solution can be reused for \gls{nru}.

The problem of \gls{harq} feedback blocking of a \gls{dl} HARQ process was not present in LAA technologies, because \gls{pucch} was always sent over the licensed carrier~\cite[Sec. 10]{TS36213}. In MulteFire, this problem was partially solved using new \gls{pucch} formats, i.e., an extended \gls{pucch} format (MF-ePUCCH) and a short \gls{pucch} format (MF-sPUCCH). MF-ePUCCH is sent with \gls{pusch} using interlaced configuration, while MF-sPUCCH is sent in the \gls{lte} special subframe~\cite{multefire}. 
Based on that, in MulteFire, the transmission opportunity to send HARQ feedback is defined according to the availability of either MF-sPUCCH, MF-ePUCCH (\gls{pusch} resources) for the \gls{ue}.
In addition, in case of MF-sPUCCH (if available) transmission after \gls{dl} data transmission, the \gls{lbt} for it could be avoided according to the shared \gls{cot} rule. However, \gls{lbt} blocking of \gls{harq} feedback still can arise when the MF-sPUCCH cannot be placed immediately after its \gls{dl} transmission.

One of the solutions to solve the HARQ feedback blocking of a \gls{dl} HARQ process in \gls{nru} is postponing the HARQ feedback transmission to the next available slot/symbols which are not blocked. Such a solution of postponing the HARQ feedback has also been considered in \gls{nr} for multi-slot aggregation and \gls{dl} semi-persistent scheduling. It occurs when there is a direction conflict due to \gls{dl}-\gls{ul} semi-static configuration or dynamic subframe indicator (SFI). However, in these cases, both \gls{gnb} and \gls{ue} know that there is a direction conflict, thus the \gls{gnb} postpones the reception and the \gls{ue} postpones the transmission of the HARQ feedback. In \gls{nru}, HARQ feedback can be postponed but the \gls{gnb} would not know that it was blocked in \gls{ul} and it would assume \gls{nack} instead. So, postponing the HARQ feedback is not sufficient in \gls{nru}.

A potential solution to solve the above problem can be the allocation of multiple \gls{pucch} resources for sending HARQ feedback corresponding to a \gls{pdsch} transmission within the \gls{cot} (\textbf{opportunistic HARQ feedback}). This solution has been highlighted in~\cite{R1-1806671} as a potential enhancement for \gls{nru}. The configuration of multiple \gls{pucch} resources can be given in the \gls{dci}, which requires definition of a new \gls{dci} format for \gls{nru}. The multiple \gls{pucch} resource configuration for HARQ feedback may include multiple time resources as well as various beams/TRPs. Once \gls{ue} receives \gls{pdsch} in slot $n$, the \gls{ue} will check whether the activated \gls{pucch} resources for HARQ feedback are valid. If any \gls{pucch} resource after $n{+}K1$ slots is not blocked, the HARQ feedback is transmitted. If all \gls{pucch} resources are blocked, then HARQ feedback is discarded. The \gls{gnb} must wait and check whether HARQ feedback can be decoded in any of the allocated multiple \gls{pucch} resources. As soon as the \gls{gnb} decodes the HARQ feedback, it can proceed with either re-transmissions or new data transmissions without monitoring of the remaining allocated \gls{pucch} resources. 

Another option to solve the problem is to use a \textbf{triggered HARQ feedback}~\cite{R1-1806671}. That is, to use a \gls{dl} triggered grant to trigger the transmission of HARQ feedback. This is similar to the solution adopted in eLAA that is used to resolve the \gls{ul} data blocking problem.

\section{Scheduling Methods for \gls{nru}}
\label{sec:sched}
In \gls{nr}, like in \gls{lte}, dynamic scheduled access is used for both \gls{dl} and \gls{ul}, for which the scheduling decisions are made at the \gls{gnb}. Each \gls{ue} monitors multiple \gls{pdcch}s, which, upon the detection of a valid \gls{dci}, follows the given scheduling decision and receives (transmits) its \gls{dl} (\gls{ul}) data. In \gls{nru}, the dynamic scheduler design has some challenging issues to solve because of the regulatory requirements for accessing the unlicensed bands. One of such issues arises due to \gls{mac} \& \gls{phy} processing delays and the requirement of \gls{lbt}, which we discuss in detail in Section~\ref{sec_LBTsched}. In addition to that, the scheduler needs to take the OCB and \gls{mcot} requirements into account as well. At each transmission time interval, the \gls{gnb} needs to schedule the \gls{ue}s such that the OCB requirement is full-filled. For example, multiple \gls{ue}s may be multiplexed in frequency domain in such a way that the OCB requirement is satisfied, e.g., by scheduling \gls{ue}s that are associated to the same beam in a slot. Also, the \gls{gnb} should take \gls{mcot} limitation into account while scheduling different data flows because the channel availability after \gls{mcot} cannot be ensured. 

Due to \gls{lbt} requirements, scheduling schemes other than the dynamic scheduled access might be more suitable for \gls{nru}, and particularly for \gls{ul} access. For example, autonomous \gls{ul} introduced in FeLAA~\cite[Sec. 4.2]{TS37213}, grant-less \gls{ul} in MulteFire~\cite{multefire}, or configured grant defined in \gls{nr} for \gls{ul} transmissions~\cite[Sec. 5.8.2]{TS38321}, might be good candidates for NR-U UL access. We discuss them in detail in Section~\ref{sec_alt}.

\subsection{Impact of Processing Delays and \gls{lbt} on the Scheduler}
\label{sec_LBTsched}
\begin{figure}[!t]
	\centering
	\includegraphics[width=0.3\textwidth]{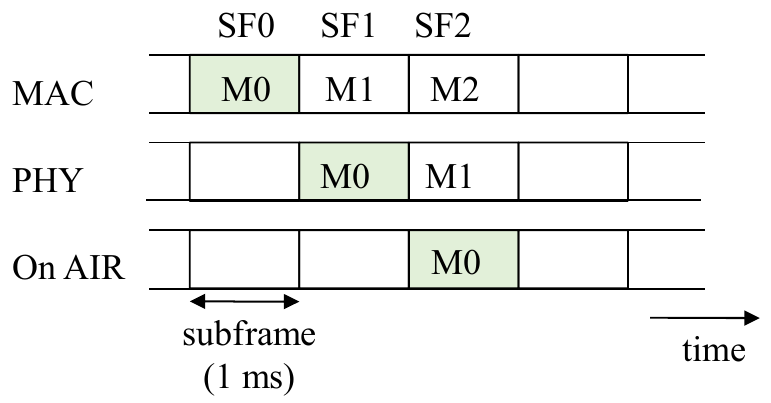}
	\caption{Processing delays in \gls{lte}.}
	\label{fig_lteprocessing}
\end{figure}

\begin{figure}[!t]
	\centering
	\includegraphics[width=0.3\textwidth]{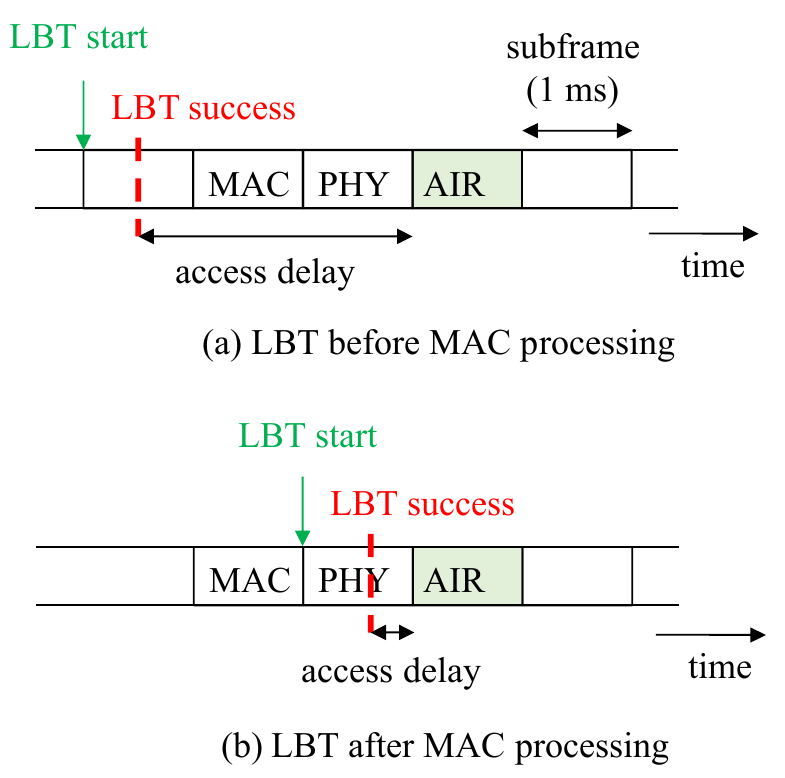}
	\caption{Options to perform \gls{lbt} in LAA. For \gls{nr}, the same options apply but the subframe would correspond to a slot or symbol (depending on the device processing capabilities) that has a numerology-dependent length.}
	\label{fig_lteprocessing2}
\end{figure}

As inherited in \gls{lte}, in \gls{laa} and MulteFire technologies there is 1 ms (one \gls{lte} subframe) of \gls{mac} processing delay and 1 ms of \gls{phy} processing delay for each transmission. For example, as shown in Fig.~\ref{fig_lteprocessing}, data scheduled in subframe number 0 (SF0) can be transmitted over the air after 2 ms in subframe number 2 (SF2). This allows two ways to perform \gls{lbt}, which are also shown in Fig.~\ref{fig_lteprocessing2}: (a) \gls{lbt} before \gls{mac} processing, (b) \gls{lbt} after \gls{mac} processing\footnote{Note that the selection of \gls{lbt} scheme out of these two options is implementation-specific and, therefore, it is not defined either in LAA-releases or MulteFire, but one of the options has to be implemented in chipsets.}. 

In the \textbf{\gls{lbt} before \gls{mac} processing} option, the delay to access the channel, given that the channel is clear, is larger than two subframes (see Fig.~\ref{fig_lteprocessing2}.(a)). In this solution, the \gls{mac}/\gls{phy} configuration of the current transmission can be modified based on the \gls{lbt} outcome (e.g., adjust the \gls{mcs} based on the sensed power during \gls{lbt}). In the \textbf{\gls{lbt} after \gls{mac} processing} option, if the channel is clear, then the delay to access the channel is lower than one subframe (see Fig.~\ref{fig_lteprocessing2}.(b)). If the channel is not clear within the duration of the \gls{phy} processing, then the \gls{mac} \gls{pdu} needs to be rescheduled, which will incur an access delay of more than three subframes to reschedule at \gls{rlc}, and then reprocess at \gls{mac} and \gls{phy}. In addition, in this case, when the channel is clear, the \gls{mac}/\gls{phy} configuration of the current transmission cannot be modified based on the \gls{lbt} outcome. In both the options, when the channel is clear, reservation signals may be needed to reserve the channel until the subframe boundary corresponding to the data transmission starts. In line with the above, \gls{lbt} before or after \gls{mac} processing solutions have clear trade-offs. The \gls{lbt} before \gls{mac} processing provides more flexibility at the scheduler but it requires the use of reservation signals during \gls{mac} and \gls{phy} processing for a long duration. On the other hand, the \gls{lbt} after \gls{mac} processing reduces the duration of use of reservation signals but requires handling rescheduling if \gls{lbt} fails which complicates the scheduler operation.

In \gls{nr}, \gls{mac}/\gls{phy} processing delays are of the order of OFDM symbol length, for which the specific values can be derived based on the device capability and the numerology~\cite{R1-1721515}. Although the processing delays are reduced in \gls{nr}, the same trade-offs of \gls{lbt} before and after \gls{mac} processing options described above will still exist for \gls{nru}. However, for \gls{lbt} after \gls{mac} processing, due to small delay in accessing the channel, i.e., less than one OFDM symbol, which can be for example 8.93 $\mu$s for \gls{scs}=120 kHz, there may not be any need for using reservation signals. 
This is an important aspect, since there are some suggestions in \gls{3gpp} to eliminate the use of reservation signals, which may also be prevented by the \gls{etsi} regulation in the future~\cite{R1-1714479}.

In case of scheduled \gls{ul} transmissions, \gls{lbt} after \gls{mac} processing is a better solution because the scheduling decision has already been made by the \gls{gnb} and it becomes important to not lose the allocated resource for \gls{ul} access. Losing the transmission opportunity in \gls{ul} may delay successful transmission. It may also affect the \gls{dl} performance for example in the case of \gls{tcp}, which requires transmission of timely \gls{tcp} \gls{ack}s in the opposite direction. 

One of the solutions that we propose here to increase the probability of channel access while performing \gls{lbt} for the beam-based transmissions is to use \textbf{multiple spatial replicas} of the same transmission. This is more suitable for the \gls{dl} transmissions, where multiple \gls{trp}s or multiple beams of the same \gls{trp} can be used to generate multiple spatial replicas for the same \gls{ue}. 
However, it also applies to the \gls{ul} in case the \gls{ue} has connectivity with multiple \gls{trp}s/beams. In this solution, we propose:
\begin{itemize}
\item preparing multiple replicas of the same \gls{mac} \gls{pdu} scheduled for a certain slot/symbol of a specific \gls{ue} with different beam-pairs or \gls{trp}s for that \gls{ue}, 
\item performing simultaneous \gls{lbt} processes on different \gls{trp}s/beams after \gls{mac} processing, and 
\item then proceeding with the best beam/\gls{trp} for which \gls{lbt} is successful (i.e., that finds the channel available on time). In case of multiple \gls{trp}s/beams get a successful \gls{lbt}, the final selection can be based on the channel conditions on the selected \gls{trp}s/beams. 
\end{itemize}

\begin{figure}[!t]
	\centering
  \includegraphics[width=0.28\textwidth]{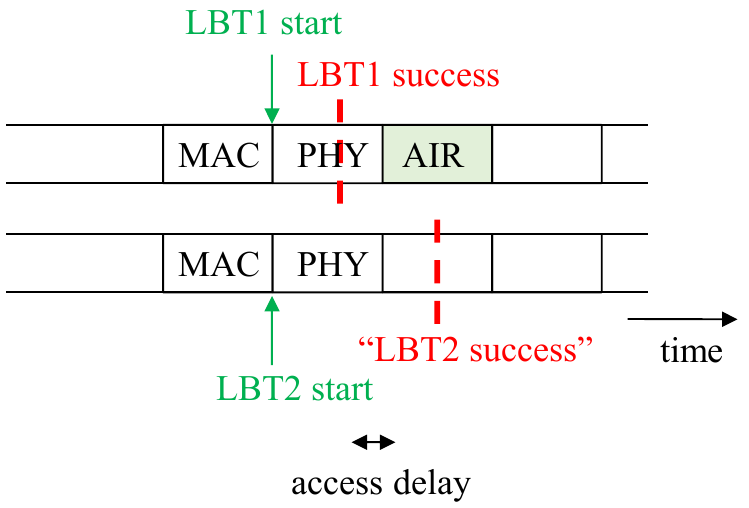}
	\caption{\gls{lbt} after \gls{mac} processing with two spatial \gls{mac} \gls{pdu} replicas and two parallel \gls{lbt} processes.}
	\label{fig_lteprocessing3}
\end{figure}

This is illustrated in Fig.~\ref{fig_lteprocessing3} for two spatial replicas. The proposed solution requires a process of selecting multiple beams/\gls{trp}s for each transmission link, as well as the capability of performing \gls{lbt} simultaneously on multiple beams/\gls{trp}s. In case different \gls{trp}s are used, the sensing for \gls{lbt} can be either directional or omnidirectional. If multiple beams of the same \gls{trp} are used, then this solution is only applied in case the \gls{gnb} uses directional \gls{lbt}.
In any case, it also requires the \gls{ue} to listen simultaneously on the multiple configured beams for data reception. 
This method would increase the reliability and reduce the impact of \gls{lbt} failure on latency. It would also reduce the access delay and improve performance in case that \gls{mac}/\gls{phy} processing delays are of the slot length order and/or the use of reservation signals is not allowed.

In FeLAA~\cite{TS36213}, a similar kind of solution was adopted by allowing multiple starting positions in the \gls{dl} and \gls{ul} special subframes, which is basically using multiple replicas of \gls{mac} \gls{pdu} in the temporal domain. Similarly, in~\cite{WO2017074498A1,WO2016122786A1,WO2017078796A1}, it was proposed that multiple \gls{pdcch}s were used to indicate different starting positions for the special subframes, whereas in~\cite{EP3}, it was suggested adjusting the \gls{mcs} according to the remaining time available for transmission which will depend on time instant at which the channel is found available by \gls{lbt}. For \gls{lbt} after \gls{mac}, these solutions involve preparing multiple replicas related to different starting temporal points~\cite{WO2017074498A1,WO2016122786A1,WO2017078796A1} and \gls{mcs}s~\cite{EP3}.

\subsection{Non-dynamic Scheduling Schemes}
\label{sec_alt}
In the case of \gls{ul} dynamic scheduling, first, a \gls{ue} has to send a \gls{sr}/\gls{bsr} to request an \gls{ul} grant (\gls{dci} in \gls{pdcch}) from its \gls{gnb}. Then, after receiving the \gls{ul} grant, the \gls{ue} performs the data transmission in \gls{pusch}. In unlicensed spectrum, this process will need multiple \gls{lbt}s (in particular, 3 \gls{lbt}s for \gls{nru} standalone scenario). This means that, if channel is occupied at any step, it will incur long delays to \gls{ul} data transmissions. Alternative (non-dynamic) scheduling schemes may be more suitable for \gls{ul} \gls{nru} to reduce the message exchange overhead of dynamic scheduled \gls{ul}.

In Rel-14 eLAA, it was found that scheduled \gls{ul} transmission has disadvantages in terms of throughput and latency, compared to contention-based transmissions used in other coexisting \gls{rat}s, such as Wi-Fi. 
To compensate for that, Rel-15 FeLAA introduced the \textbf{autonomous \gls{ul}} transmissions~\cite[Sec. 4.2]{TS37213} and MulteFire defined \textbf{grant-less \gls{ul}}~\cite{multefire}, which have a high resemblance. Both in autonomous \gls{ul} and grant-less \gls{ul}, there is a predefined set of radio resources, which are
configured on a per-cell basis and are for contention-based access. A \gls{ue} is allowed, after a successful \gls{lbt}, to transmit its \gls{pusch} on such resources without an \gls{ul} grant. 
Therefore, autonomous \gls{ul} and grant-less \gls{ul} eliminate the handshake of \gls{sr}, \gls{bsr}, and dynamic \gls{ul} grant for \gls{ul} 
access~\cite{R1-1804313}. 
However, losses due to collisions and blocking owing to channel occupancy may occur in autonomous \gls{ul} and grant-less \gls{ul}. To solve that, if a similar approach is followed for \gls{nru}, then the multi-\gls{trp} deployment and multi-beam operation could be exploited to configure \gls{ul} transmissions to multiple \gls{trp}s by following the same approach as in the spatial replicas based solution that we described in Section~\ref{sec_LBTsched}. 

Non-dynamic scheduling schemes have also been introduced in \gls{nr} to reduce the latency of dynamic scheduled \gls{ul}.
\gls{nr} defines a new non-dynamic scheduling for \gls{ul} data transmission~\cite[Sec. 5.8.2]{TS38321}, called \textbf{configured grant}. In configured grant, the \gls{ul} data transmissions follow a semi-statically configured resource allocation corresponding to a UE-specific configured grant. The configured grant may either be provided by \gls{rrc} (Type 1) or via \gls{dci} (Type 2). 
Due to the semi-static and periodic configuration of resources, configured scheduling requires less control signaling as compared to dynamic scheduling. 
This is convenient for \gls{nru} \gls{ul} to simplify the \gls{sr}/\gls{bsr}/\gls{ul} grant handshake and reduce the number of required \gls{lbt}s that are needed before a \gls{ue} can successfully access the unlicensed channel~\cite{TR38889}. 
Therefore, it is a potential scheme to reduce the access delay in \gls{nru} \gls{ul}, provided that its parameters: size $\gamma$ (i.e., amount of data, in bits, which is given by the number of assigned resources and \gls{mcs}) and periodicity $p$ (in number of slots) 
are properly configured for the available traffic pattern. For example, consider a \gls{ue} which needs to download some data from a remote host, in that case, configured grant can be used to reserve space for TCP \gls{ack}s every $p$ slots for an amount of data $\gamma$. Moreover, to avoid blocking of \gls{ul} transmissions on configured resources due to \gls{lbt}, the \gls{gnb} can also use the triggered grants (described in Section~\ref{sec:harq}) to enable the UE transmit immediately after the triggered grant.

\section{Evaluation}
\label{sec:eval}
In this section, by using simulation of an \textbf{NR-U/WiGig coexistence} scenario, we evaluate the performance of different \gls{lbt}-based channel access procedures discussed in Section~\ref{sec:channelaccess}. The evaluation of other open design improvements (like \gls{cot} structure, initial access, HARQ, and scheduler analyzed in Sections~\ref{sec:frame},~\ref{sec:initialaccess},~\ref{sec:harq},~\ref{sec:sched}, respectively) is left for future works. 
The details of the deployment scenario and the simulation results are given in the following sections.

\subsection{Deployment Scenario}
\label{sec:scen}
A dense indoor network deployment, composed of $K$ pairs that are randomly deployed in a $25 {\times} 25$ m$^2$ area is considered. 
We consider an NR-U/WiGig coexistence scenario, for which half of the pairs ($K/2$) are NR-U pairs (gNB-UE) and the other half of the pairs ($K/2$) are WiGig pairs (AP-STA). The minimum distance among gNBs/APs is set to $1$ meter, and UEs/STAs are deployed in a random distance between $3$ and $8$ meters from the serving gNB/AP.
Performance of the downlink transmission is assessed, assuming that gNBs/APs operate at carrier frequency $60$ GHz with $1$ GHz channel bandwidth and transmit power of $10$ dBm. The channel models of IEEE 802.11ad are used. The noise power spectral density and the noise figure are set to ${-}174$ dBm/Hz and $7$ dB, respectively. 

According to WiGig specification, we assume that APs perform \gls{omnilbt}. For NR-U gNBs, different channel access procedures described in Section~\ref{sec:LBT}, i.e., \gls{omnilbt}, \gls{dirlbt}, \gls{pairlbt}, and \gls{lbtswitch} are considered. We also combine each of these strategies with LBR (i.e., receiver-assisted LBT, as detailed in Section~\ref{sec:recLBT}), which are denoted by \gls{omnilbt}+LBR, \gls{dirlbt}+LBR, \gls{pairlbt}+LBR, and \gls{lbtswitch}+LBR, respectively. In addition to these schemes, we introduce a dummy design in which gNBs do not perform any LBT before a transmission, denoted as no-LBT. The no-LBT option is not compliant with ETSI regulation~\cite{ETSI302567} but it is just included as a benchmark in the simulations.
For the LBR-based options, the additional time required to perform LBR handshake given in Section~\ref{sec:recLBT} is taken into account. For NR-U, we consider \gls{scs}=$120$ kHz, since it is a common numerology in \gls{mmwave} bands.

Directional transmissions are assumed at gNBs/APs. The transmit beam gain at gNBs/APs is fixed to $10$ dB with a transmit main lobe beamwidth of $30^o$, and ideal antenna radiation efficiency is assumed. For data reception, two configurations for the UEs/STAs' antennas are considered:
\begin{itemize}
\item \textbf{Omnidirectional reception}: UEs/STAs receive data omnidirectionally. In this case, for the LBR scheme, sensing at UE side will also be performed omnidirectionally.
\item \textbf{Quasi-omnidirectional reception}: the receive beam gain at UEs/STAs is fixed to $7$ dB with a receive main lobe beamwidth of $90^o$ while assuming ideal antenna radiation efficiency. In this case, LBR will be implemented through directional sensing (in the receive beam) at UE side.
\end{itemize}

\begin{figure*}[!t]
	\centering
	\includegraphics[width=0.88\textwidth]{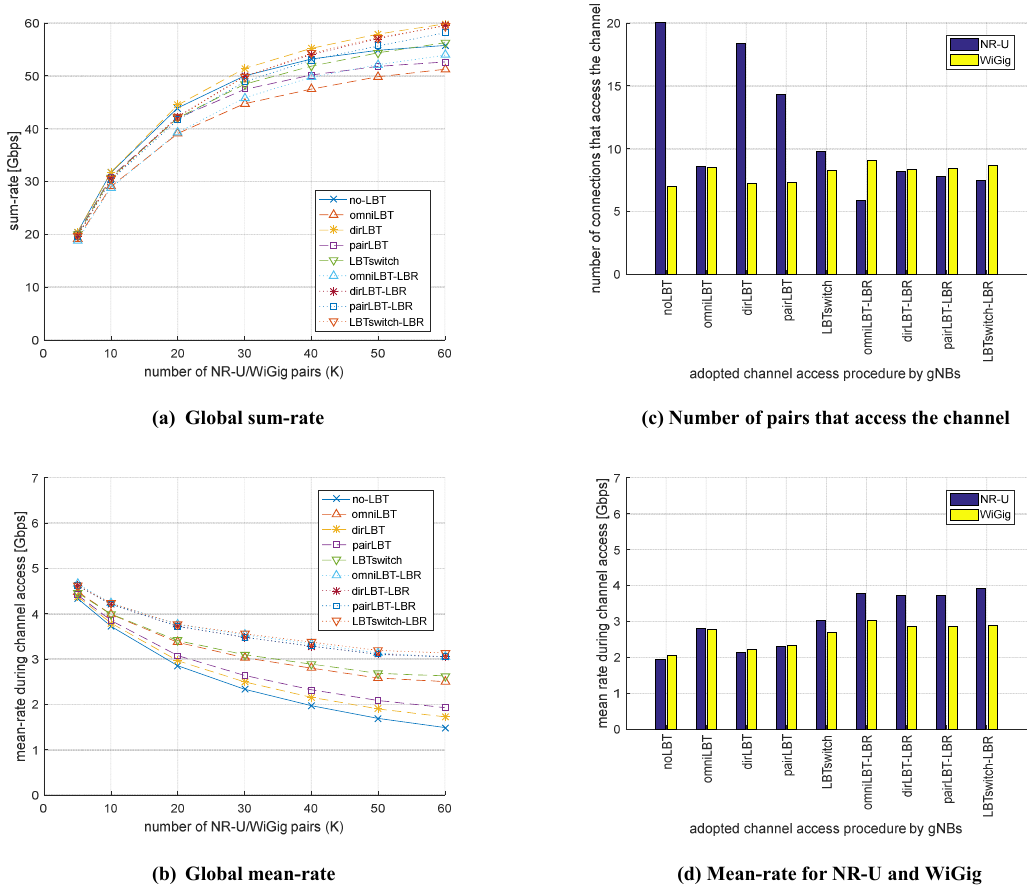}
	\caption{Performance evaluation of different NR-U channel access procedures, for omnidirectional reception at UEs/STAs. The WiGig channel access is kept as per IEEE 802.11ad standard, i.e., \gls{omnilbt}. (a) Sum-rate (Gbps) vs $K$. (b) Mean-rate during channel access (Gbps) vs $K$. (c) Number of pairs that get access to the channel when $K{=}40$, for NR-U and WiGig, separately. (d) Mean-rate during channel access (Gbps) when $K{=}40$, for NR-U and WiGig, separately.}      
	\label{fig_res1}
\end{figure*}

\begin{figure*}[!t]
	\centering
	\includegraphics[width=0.88\textwidth]{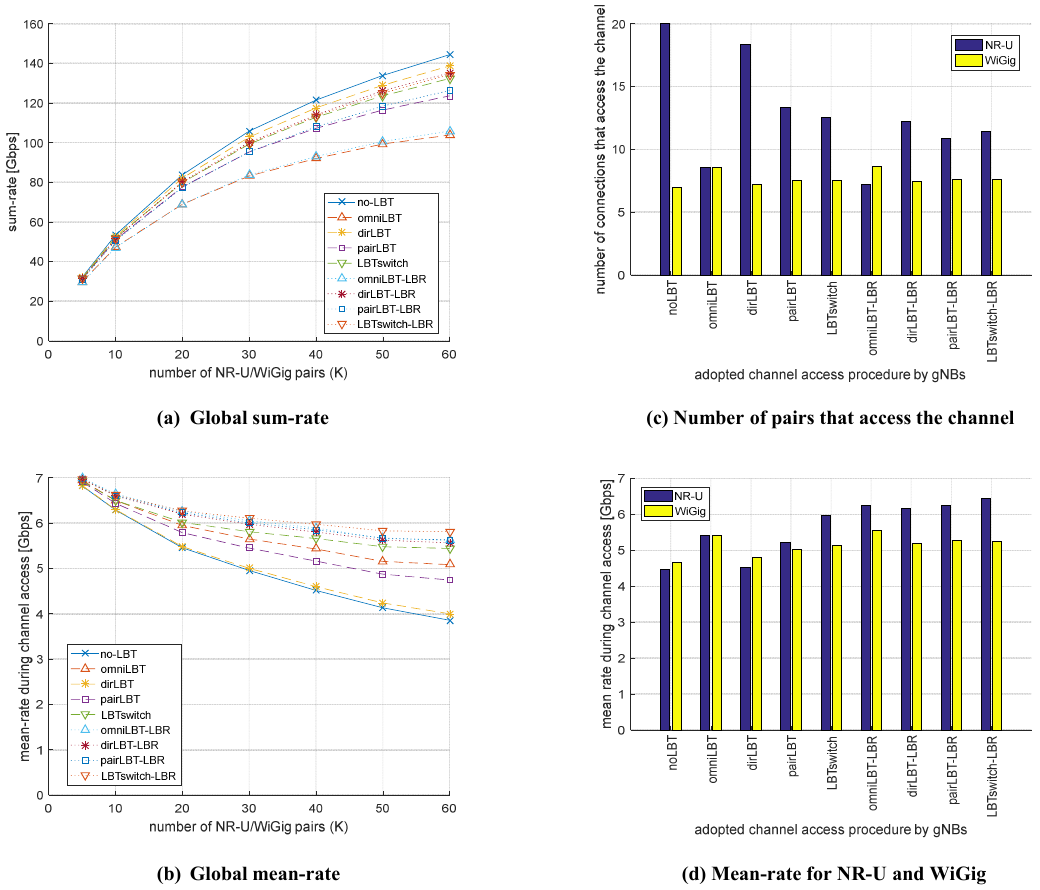}
	\caption{Performance evaluation of different NR-U channel access procedures, for quasi-omnidirectional reception at UEs/STAs. The WiGig channel access is kept as per IEEE 802.11ad standard, i.e., \gls{omnilbt}. (a) Sum-rate (Gbps) vs $K$. (b) Mean-rate during channel access (Gbps) vs $K$. (c) Number of pairs that get access to the channel when $K{=}40$, for NR-U and WiGig, separately. (d) Mean-rate during channel access (Gbps) when $K{=}40$, for NR-U and WiGig, separately.}
	\label{fig_res2}
\end{figure*}

The \gls{ed} threshold for LBT, normalized by the maximum antenna gain for sensing, is set to -74 dBm\footnote{Note that directional transmissions are considered but the sensing stage can be performed either directionally or omnidirectionally. Thus, a normalized \gls{ed} threshold of -74 dBm is considered by taking into account the receive gain used for sensing, which corresponds to an \gls{ed} threshold of -74 dBm for \gls{omnilbt} and -64 dBm for \gls{dirlbt}. Similarly, in case of LBR, it depends on the data reception configuration; for omnidirectional reception, the \gls{ed} threshold is -74 dBm, while it is -67 dBm for quasi-omnidirectional reception. Recall that the noise power for $W{=}1$ GHz with noise power spectral density of $-174$ dBm/Hz results in ${-}84$ dBm, and thus we consider -74 dBm as the \gls{ed} in \gls{omnilbt} to account for the noise figure.}. 
We do not emulate backoff processes for both WiGig \gls{cca} and \gls{nru} \gls{lbt}, and simply consider how many pairs (connections) can reuse the spectrum according to the different channel access procedures. Simulation results are averaged among $1000$ random deployments.

For the performance metrics, we collect sum-rate and mean-rate during channel access. The sum-rate is the sum of data rates of all the pairs that can simultaneously access the channel. On the other hand, the mean-rate corresponds to the average of the rates over the connections that get access to the channel. This may be a useful metric to measure the \gls{qos} obtained by the different RATs. In addition, to account for fairness, we also evaluate the average number of connections that get access to the channel for both NR-U and WiGig systems.

\subsection{Results and Comparison}
\label{sec:res}
We categorize the results based on the reception type implemented at the UEs and STAs sides. For the omnidirectional reception at UEs/STAs, the collected results are shown in Fig.~\ref{fig_res1}, and for the quasi-omnidirectional reception, the results are shown in Fig.~\ref{fig_res2}.
Within the figures, subfigures (a) and (b) show the sum-rate and mean-rate with different number of total pairs ($K$), respectively. Subfigures (c) and (d) depict the number of pairs that get access to the channel and their attained mean-rate in each of the systems, i.e., NR-U and WiGig, respectively, with $K{=}40$.

For omnidirectional reception, we observe that:
\begin{itemize}
\item No-LBT provides the lowest mean-rate for all $K$. It is worse than \gls{omnilbt} for coexistence since it reduces the number of WiGig connections and their attained rate (see Fig.~\ref{fig_res1}.(c)-(d)). Also, as $K$ increases, the sum-rate gets saturated due to the interference (see Fig.~\ref{fig_res1}.(a)).
\item LBT strategies at gNB side (\gls{omnilbt}, \gls{dirlbt}, \gls{pairlbt}, \gls{lbtswitch}): 

\begin{itemize}
\item The \gls{omnilbt}-\gls{dirlbt} trade-off is observed. OmniLBT is overprotective (low number of NR-U connections access), so that it obtains a lower sum-rate but a higher mean-rate than \gls{dirlbt} (see Fig.~\ref{fig_res1}.(a)-(b)). DirLBT enables spatial reuse at gNBs (high number of NR-U connections) but hidden nodes arise, which also impacts WiGig performance negatively since more NR-U nodes access and interfere (see Fig.~\ref{fig_res1}.(d)). 
\item PairLBT performs similar to \gls{dirlbt} for omnidirectional reception. It is not effective for an omnidirectional reception configuration because the LBT in the opposite direction cannot properly detect all the hidden nodes that are interfering the UE. 
\item \gls{lbtswitch} improves the mean-rate compared to \gls{dirlbt}, \gls{omnilbt}, and \gls{pairlbt}, as shown in Fig.~\ref{fig_res1}.(d). It is able to enhance the fairness of NR-U pairs as compared to \gls{omnilbt}, since more NR-U connections get access to the channel while not affecting negatively the number of WiGig accesses and their average rate. In addition, compared to \gls{dirlbt} and \gls{pairlbt}, since \gls{lbtswitch} is able to properly adapt the type of carrier sense at every gNB as a function of the observed neighboring gNBs/APs density and activity, it provides better performance in such coexistence scenario.
\end{itemize}

\item Receiver-assisted LBT strategies with sensing at gNB and UE side (\gls{omnilbt}-LBR, \gls{dirlbt}-LBR, \gls{pairlbt}-LBR, \gls{lbtswitch}-LBR): In general, sensing at the UE side provides large benefits at unlicensed bands since it overcomes the deficiencies of LBT under beam-based transmissions. This can be observed in the number of connections accessing the channel, their attained mean-rate, and the system sum-rate. We observe that, however, \gls{omnilbt}-LBR is too much conservative and cannot provide the spatial reuse and sum-rate as of \gls{dirlbt}-LBR, \gls{pairlbt}-LBR, and \gls{lbtswitch}-LBR. In general, LBR acts as good neighbor for WiGig nodes as it does not impact the number of WiGig nodes that access the channel and their attained rate, while at the same time NR-U pairs achieve a much larger rate during channel access. Recall that LBR-based techniques get the same access probability than \gls{omnilbt} but, since only the properly selected gNBs access, it provides a higher mean-rate (see Fig.~\ref{fig_res1}.(c)-(d)). 
\end{itemize}

For quasi-omnidirectional reception, as shown in Fig.~\ref{fig_res2}, similar trends are observed but with: 1) lower relative differences in the performance among the different schemes, and 2) larger rates because of the reduced interference levels due to directional reception.
However, few differences are observed for this configuration:
\begin{itemize}
    \item PairLBT with directional reception is able to address the \gls{omnilbt}-\gls{dirlbt} trade-off, since the sensing beam for the opposite direction can be properly adjusted. 
    \item Although LBR-based procedures obtain the largest \gls{qos} (mean-rate), the largest system capacity is given by the no-LBT scheme because excessive interference does not arise due to directional receptions and, thus, the larger the spatial reuse is, the larger the system capacity is.
    \item \gls{lbtswitch} gets a mean-rate similar to LBR-based approaches.
\end{itemize}

We would like to remark that information from UE side is shown to be significantly beneficial to improve the coexistence in unlicensed bands with beam-based transmissions, particularly in the case of omnidirectional reception. This is observed in the performance of \gls{lbtswitch}-LBR.
\gls{lbtswitch}-LBR performs better than the other strategies because it includes sensing at the UE as well as recommendation from UE side regarding the type of carrier sense to be performed at the gNB for LBT. On the other hand, for quasi-omnidirectional reception, one type of UE feedback (either to switch the LBT strategy or to allow/prevent the access through LBR) is sufficient to improve simultaneously the spatial reuse and the \gls{qos}.

\section{Lessons Learned}
\label{sec:learned}
The lessons that we have learned and discussed throughout this article are summarized as follows. 

\begin{itemize}
    \item \textbf{The usage of \gls{pairlbt} and \gls{lbtswitch} in \gls{nru} help in reducing exposed node and hidden node problems, as compared to \gls{omnilbt} and \gls{dirlbt}:} Multiple solutions are available to implement carrier sense at the transmitter side for \gls{lbt} under beam-based transmissions. The two trivial solutions, i.e., \gls{omnilbt} and \gls{dirlbt} have different trade-offs in terms of system performance, fairness, and complexity. It is due to the different types of sensing that accentuate exposed nodes in \gls{omnilbt} and hidden nodes in \gls{dirlbt}. These trade-offs can be addressed by using paired directional sensing at the transmitter side (\gls{pairlbt}), or switching the type of carrier sense at the transmitter as a function of density and activity of neighboring nodes observed from the receiver side (\gls{lbtswitch}). 
    \item \textbf{The efficiency of \gls{pairlbt} and \gls{lbtswitch} is demonstrated in \gls{nru}/WiGig coexistence scenarios:} Results have shown that \gls{pairlbt} is useful for scenarios in which data reception is directional. Otherwise, for omnidirectional data reception, there are hidden node problems that cannot be detected at the transmitter, even with multiple paired sensings. On the other hand, results have shown that \gls{lbtswitch} performs better than \gls{omnilbt}, \gls{dirlbt}, and \gls{pairlbt} because it includes recommendation from the \gls{ue} side regarding the type of carrier sense (omni or dir) to be performed at \gls{gnb}s based on the observed potential interferers. So, information from the \gls{ue} side is beneficial to improve coexistence in beam-based \gls{nru}. 
    \item \textbf{Receiver-assisted \gls{lbt} solutions help in overcoming the deficiencies of sensing only at the transmitter side:} For beam-based communications in the unlicensed band, due to the use of directional antenna arrays, the observed channel status at the transmitter may be different from the perceived interference at the receiver side. Therefore, performing carrier sense at the transmitter side (i.e., \gls{lbt}) may not be sufficient. This can be fixed by using receiver-assisted \gls{lbt} solutions, which provide the receiver (\gls{ue}) an opportunity to sense the shared channel using LBR and assist the transmitter for channel access using a feedback. Indeed, LBR can be combined with different types of sensing at the transmitter side.
   
    \item \textbf{The effectiveness of receiver-assisted \gls{lbt} over \gls{lbt}-based strategies is demonstrated in \gls{nru}/WiGig coexistence scenarios:} Results have shown that sensing at the UE side (LBR) provides large fairness and QoS benefits in \gls{nru}/WiGig coexistence scenarios at mmWave bands. Results confirm that  RTS/CTS-like mechanisms are beneficial to \gls{nru}. Moreover, among the LBT-LBR combinations, it is observed that \gls{lbtswitch}-LBR performs better than \gls{omnilbt}-LBR, \gls{dirlbt}-LBR, and \gls{pairlbt}-LBR. This is due to the fact that, in \gls{lbtswitch}-LBR, the feedback from the UE after performing LBR includes also a recommendation for the type of LBT to be used at the transmitter side. 
    \item \textbf{Coordination of \gls{lbt} processes improves \gls{nru} channel reuse:} Mechanisms to enable frequency reuse among \gls{nru} devices of the same operator are needed to improve the system performance and avoid LBT blocking between devices of the same operator. The potential mechanisms to support intra-\gls{rat} tight frequency reuse are: multi-\gls{ed} strategies, self-defer schemes, and a new mechanism proposed in this paper, i.e., \gls{lbt} coordination, which enables time/frequency coordination of the resource allocation as well as coordination among the \gls{lbt} procedures of different nodes.
    \item \textbf{Sensing at the receiver node is useful to properly update the \gls{lbt} \gls{cws} in beam-based \gls{nru}:} Multiple issues arise when using \gls{harq} feedback to update the \gls{cws} (as done in \gls{laa}) for the case of beam-based transmissions. It is because of the lack of correlation between a collision indicated by a \gls{nack} and the transmit beam, as well as due to the inability to enter in the backoff phase after an incorrect sensing phase. We have proposed a solution to fix these problems by using a receiver-assisted \gls{cws} adjustment that considers paired sensing at the receiver (\gls{ue}) for the \gls{cws} update. It does not use \gls{harq} feedback.
    \item \textbf{Multiple DL/UL switches within the \gls{cot} is beneficial for \gls{nru}:}  Two options are considered for the \gls{cot} structure in \gls{nru}, i.e., a single DL/UL switch and multiple DL/UL switches, each with their pros and cons, as considered in the current discussions for \gls{nru} specification.
    To reduce the end-to-end latency, a \gls{cot} with multiple DL/UL switches is preferred. It is identified that the number of switching points should be further optimized based on the traffic patterns and flow requirements.
    \item \textbf{\gls{ss} block design improvements are needed for initial access in \gls{nru}:} Multiple challenges of the \gls{ss} block design in the unlicensed context arise due to the \gls{lbt} and \gls{ocb} requirements. To reduce the \gls{lbt} impact, multiple occasions for \gls{ss} block transmissions can be used to improve channel access probability. Some \gls{nr} \gls{ss} block patterns need to be redesigned to leave enough time for the sensing phase in between two \gls{ss} block transmissions. To meet the \gls{ocb} requirement in the 60 GHz band, new design solutions for \gls{ss} blocks resource mapping are proposed. It includes frequency-domain \gls{ss} block repetitions, split and/or reordering of the the \gls{ss} block time-frequency structure, and frequency-domain interlaced mapping of the signals that compose the \gls{ss} block.
    \item \textbf{\gls{nr} and eLAA enhancements regarding RACH procedure can be reused for \gls{nru}:} Enhancements to the current four step RACH procedure are needed to reduce the delay associated with it. This can be fixed by increasing the transmit opportunities for each message of the RACH procedure, simplifying the overall RACH procedure (as already contemplated in \gls{nr}), and/or enhancing the \gls{lbt} design for random access. Also, to meet the \gls{ocb} requirements, adaptation in \gls{nr} PRACH preamble formats is needed, as it was done in \gls{elaa}.
    \item \textbf{Paging solutions already defined in \gls{nr} are useful for \gls{nru}:} The uncertainty of channel availability in the unlicensed context complicates the paging procedure in \gls{nru} with standalone and dual-connectivity operations. Multiple opportunities for the paging procedures, for example, using paging message repetitions through the time and/or space domains, have been identified as beneficial for \gls{nru}. Some of such solutions are already being supported in \gls{nr} specification.
    \item \textbf{HARQ procedures defined in eLAA could be reused for \gls{nru}:} Two problems related to the HARQ procedure in \gls{nru} with standalone operation, caused by the usage of the \gls{lbt} requirement, have been identified: HARQ feedback blocking of a DL HARQ process and UL data blocking of an UL HARQ process. To solve the former, the concept of a triggered grant, as per eLAA, can be used. To fix the latter, solutions based on opportunistic and triggered HARQ feedback could be beneficial.
    \item \textbf{There are pros and cons regarding the \gls{lbt} placement in real implementations (\gls{lbt} after or before \gls{mac}):}
    Two implementation-specific solutions for what regards the \gls{lbt} placement versus the scheduling operation are: \gls{lbt} before \gls{mac} processing and \gls{lbt} after \gls{mac} processing. 
    For the \gls{dl} access, pros and cons of each solution are apparent. \gls{lbt} before \gls{mac} processing provides more flexibility at the scheduler, reduces complexity of the scheduler implementation, but it increases the access delay and may require the use of reservation signals. On the other hand, \gls{lbt} after \gls{mac} processing reduces/avoids the need for reservation signals, reduces the access delay if \gls{lbt} success, but requires handling of rescheduling if \gls{lbt} fails. Although this is not discussed in the standardization, and the impact in \gls{nr} may be lower than in \gls{lte} in unlicensed spectrum due to the lower \gls{nr} processing timings, the authors believe that practical implementations should carefully analyze these aspects. 
    \item \textbf{A possible scheduling solution including a specific \gls{lbt} placement is to use spatial replicas:} To address the issues in the \gls{dl} access mentioned in the previous bullet, in this paper we have proposed a new scheduling solution that uses multiple spatial replicas and \gls{lbt} after \gls{mac} processing for \gls{nru} \gls{dl} access. The proposed solution exploits the multi-beam and multi-\gls{trp} deployment in \gls{nr}, while meeting the \gls{lbt} requirement in \gls{dl}, as a way to increase the reliability, to reduce the impact of \gls{lbt} failure on latency, and to reduce the access delay.
    \item \textbf{Alternative \gls{ul} scheduling methods defined in \gls{nr}, FeLAA, and MulteFire are beneficial for \gls{nru}:} \gls{ul} dynamic scheduling in the unlicensed context may incur long delays to UL data transmissions. Scheduling schemes with less dynamic nature, like autonomous \gls{ul} (defined in FeLAA), grant-less \gls{ul} (used in MulteFire), or configured grant (standardized in NR), can be more favorable for \gls{nru} \gls{ul} transmissions in reducing the message exchange overhead and the access delay.
\end{itemize}

\section{Future Perspectives}
\label{sec:future}    
The future perspectives and opportunities for \gls{nru} related research that we envision are:
\begin{itemize}
   \item \textbf{Integration of \gls{mmwave} and sub 7 GHz licensed/unlicensed bands}: Integration of \gls{mmwave} and sub 7 GHz bands has been studied in the \gls{nr} context with licensed bands~\cite{semiari:17,semiari:18}, as well as in the \gls{wigig} context with unlicensed bands~\cite{nitsche:15}. How to potentially reuse and extend them for \gls{nru} by combining licensed/unlicensed/shared paradigms under different operational modes (i.e., carrier aggregation and standalone) is an interesting area for further research~\cite{lu:19}. Also, multi-band and multi-channel selection algorithms in this context could be investigated.
    \item \textbf{\gls{nru} for ultra-reliable and low-latency communications}: The impact of \gls{lbt} on the latency performance of MulteFire has been assessed in~\cite{maldonado:18}, both analytically and through simulations. Extension of the analytic framework and system-level simulations for \gls{nru} are of high interest to understand if \gls{nru} can meet strict low-latency and high-reliability requirements~\cite{TR38913}. If not, then what modifications are required (if any) to support the \gls{urllc} use case.
   \item \textbf{\gls{nru} for future smart factories}: Industry 4.0 has emerged as an important application for \gls{nru} since it requires wireless-connected and privately-owned networks~\cite{8207346,TR22804}. A future research line is to develop theoretical foundations for licensed, unlicensed and shared spectrum paradigms to use \gls{nru} as the \gls{rat} for future smart factories. For example, to accommodate multiple devices with diverse requirements such as extended reality applications, \gls{urllc} devices, sensors, mobile robots, etc. simultaneously. 
   \item \textbf{Improved beam-training for unlicensed-based access}: The impact of \gls{lbt} on the beam training processes needs to be investigated. Recently, authors in~\cite{li:18} proposed a joint directional received-assisted \gls{lbt} (i.e., \gls{dirlbt}/dirLBR) and beam training. It identifies the best beam pair for \gls{nru} communication by taking both channel blocking and channel quality into account. Further research in this line, and the impact on the overall network efficiency should be studied.
   \item \textbf{Beam reciprocity in unlicensed}: Even if \gls{tdd} is used and \gls{dl} and \gls{ul} transmissions are performed within the coherence time interval, it may happen that the best beam for \gls{dl} reception is not the best beam for \gls{ul} transmissions. This is due to \gls{lbt} blocking effects and the differences in the received interference at transmitter and receiver sides, which are accentuated at \gls{mmwave} bands. Therefore, the study of best beam-pair selection, independently for \gls{dl} and \gls{ul}, jointly with the unlicensed band access constraints, could be further investigated.
   \item \textbf{Grant-less \gls{ul} in the unlicensed mmWave bands}: Grant-less \gls{ul} is useful to reduce the scheduling delays and get fast access to the channel at the cost of increased collisions. Therefore, pros and cons of grant-based and grant-free access schemes should be properly evaluated for NR-U beam-based access to unlicensed spectrum. Also, optimization of the access scheme and the number of repetitions for grant-less \gls{ul} to guarantee successful access and decoding while minimizing energy consumption at \gls{ue}s could be investigated.
\end{itemize}

\section{Conclusions}
\label{sec:conc}
In this paper, we highlight the challenges and analyze the potential solutions for \gls{nr}-based access to unlicensed spectrum with beam-based transmissions. We discuss different topics such as channel access, frame structure, initial access, HARQ, and scheduling in the context of \gls{nru}. For the channel access procedures, we review the solutions to support \textit{i}) \gls{lbt} under beam-based transmissions, \textit{ii}) receiver-assisted \gls{lbt} in beam-based transmissions, \textit{iii}) intra-RAT frequency reuse improvement, and \textit{iv}) \gls{cws} adjustment in beam-based transmissions. With the help of simulations, we show that feedback from the receiver significantly improves the performance of coexistence in terms of \gls{qos} and fairness. 
In terms of \gls{cot} structures, slots with multiple \gls{dl}/\gls{ul} switching points within the \gls{cot} are shown to be more suitable for \gls{nru}. For \gls{nru} initial access, we discuss the design consideration for \gls{ss} block design, \gls{rach} procedure, and paging procedure to take \gls{lbt} and \gls{ocb} requirements into account. At the \gls{mac} level, two problems related to the HARQ procedures are identified, for which, we describe the solutions based on self-contained, triggered, and opportunistic HARQ feedbacks. We also discuss the issues related to the dynamic scheduling in \gls{nru}, where, we propose a multiple spatial replicas based solution, and also indicate that the existing scheduling schemes such as grant-less \gls{ul} and configured grant that have less control signaling for \gls{ul} access may be suitable for \gls{nru}. Finally, we provide a summary of all of our main findings as well as future research perspectives for \gls{nru} beam-based transmissions.

\printglossaries

\section{Acknowledgments}
This work was partially funded by Spanish MINECO grant
TEC2017-88373-R (5G-REFINE) and Generalitat de Catalunya grant 2017 SGR 1195. Also, it was supported by InterDigital Communications, Inc.

\bibliography{references}
\bibliographystyle{ieeetr}

\end{document}